\documentclass{aa}
\usepackage{graphicx}
\usepackage{txfonts}
\usepackage{subfigure}
\usepackage{natbib}
\bibpunct{(}{)}{;}{a}{}{,} % to follow the A&A style
\begin{document}
   \title{Towards a physical model of dust tori in Active Galactic Nuclei}

   \subtitle{Radiative transfer calculations for a hydrostatic torus model}

   \author{Marc Schartmann
          \inst{1}\fnmsep\thanks{e-mail: schartmann@mpia.de}
          \and
          Klaus Meisenheimer
          \inst{1}
          \and
          Max Camenzind
          \inst{2}
          \and
          Sebastian Wolf
          \inst{1}
          \and
          Thomas Henning
          \inst{1}
          }

   \offprints{M. Schartmann}

   \institute{Max-Planck-Institut f\"ur Astronomie (MPIA), K\"onigstuhl 17, D-69117 Heidelberg, Germany
         \and
              Landessternwarte Heidelberg, K\"onigstuhl 12, D-69121 Heidelberg, Germany
             }

   %\date{Received / Accepted }

   \abstract{We explore physically self-consistent models of dusty molecular 
tori in Active Galactic Nuclei (AGN) with the goal of interpreting 
VLTI observations and 
fitting high resolution mid-IR spectral energy distributions (SEDs). 
The input dust distribution is 
analytically calculated by assuming hydrostatic equilibrium between 
pressure forces -- due to the turbulent motion of the gas clouds -- and 
gravitational and centrifugal
forces as a result of the contribution of the nuclear stellar distribution and 
the central black hole. For 
a fully three-dimensional treatment of the radiative 
transfer problem
through the tori we employ the Monte Carlo code MC3D. We find that in homogeneous dust
distributions the observed mid-infrared emission is dominated by the
inner funnel of the torus, even when observing 
along the equatorial plane. 
Therefore, the stratification of the distribution of dust grains -- both in
terms of size and composition -- cannot be neglected. In the current study we
only include the effect of different sublimation radii which significantly
alters the SED in comparison to models that assume an average dust grain
property with a common sublimation radius, and suppresses the silicate emission
feature at $9.7\,\muup$m. In this way we are able to fit the mean SED of both
type\,I and type\,II AGN very well. Our fit of special objects for which high
angular resolution observations ($\le$ 0.3\arcsec) are available indicates that the hottest
dust in NGC\,1068 reaches the sublimation temperature while the maximum dust
temperature in the low-luminosity AGN Circinus falls short of 1000\,K.

% not more than 6             
               }

   \maketitle
%
%________________________________________________________________

\section{Introduction}

Active galaxies are among the most energetic objects in the universe. 
With current knowledge, these objects -- although containing 
up to 60 different variations \citep{Ward_03} -- can be described by a small number of
models. The differences between them can be reduced to
geometrical effects. The most common model is called the \emph{Unified Scheme}
\citep{Antonucci_93} of Active Galactic Nuclei (AGN).
According to this scheme, several components can be
distinguished: A supermassive black hole
($10^6$\,-\,$10^{10}\,M_{\sun}$) in the centre surrounded by an accretion disc,
reaching from the marginally stable orbit 
up to several thousands of Schwarzschild radii. By
turbulent processes, material is accreted and the gas in the disc
is heated to several hundred thousand Kelvin. Hence, the spectrum of the emitted radiation
peaks in the UV/optical wavelength range. 
In the outer part, the temperature of the disc drops to about 1000\,K, at 
the start of the larger and geometrically thicker dust reservoir, producing a characteristic 
bump in the IR spectral range (hereafter the {\it IR bump}).
 
The Unified Scheme differentiates between two different types of active galaxies:
The spectral energy distributions (SEDs) of type\,I objects show the so-called blue bump, 
arising from the direct radiation 
of the accretion disc mentioned above. Broad emission lines overlay the continuum 
spectral energy distribution. They come from 
the so-called \emph{Broad Line Region (BLR)} near the centre of the object. Within this area, 
gas clouds are moving at high velocities due to the deep potential well of the central black hole and,
therefore, emit broad spectral lines. 
Type\,II objects do not show a big blue bump and are characterised by narrow spectral lines,  
arising from orbiting gas clouds further away from the gravitational centre.
After broad spectral lines were observed in the polarised light of a type\,II object, it 
was proposed that both AGN types belong to the same class of object with a dust reservoir 
located in a torus which 
blocks the direct view to the centre and the \emph{BLR} 
within the opening of the torus for the case of type\,II objects. Polarised light is 
then produced by electrons and material above the opening angle of the dust torus.
Therefore, the type of object simply depends on the viewing angle towards the torus.

More direct evidence for the existence of an obscuring torus comes from recent observations of
the prototype Seyfert\,II galaxy \object{NGC\,1068} in the mid-infrared by the VLTI with 
MIDI \citep{Leinert_03}. Based on these interferometric data, \citet{Jaffe_04} distinguished two different 
dust components: a hot component in the
centre with a diameter of less than 1\,pc and a temperature of more than 800\,K and
warm dust within an elongated structure perpendicular to the jet axis
with a temperature of approximately 320\,K and a size of 3.4\,pc times 2.1\,pc
perpendicular and parallel to the torus axis, respectively. This means that geometrically thick
tori are consistent with the data. 

The first more detailed radiative transfer simulations for the case of dusty tori were carried out by 
\citet{Pier_92}. They used the most simple dust configuration to describe the
observations available at that time, consisting only of spectral energy distributions -- due to the
large distances (several tens of Mpc) and small sizes (less than 100\,pc) of these objects: 
a cylindrical shape with a cylindrical hole
in the middle. Although pointing out that dust -- in order to avoid destruction by hot
gas -- must be contained in clumps, they applied a homogeneous dust
distribution.
Characteristics of their modelling are the small sizes of a few pc in diameter and
very high dust densities.
The small amount of cold dust leads to a too narrow dust temperature range, resulting 
in spectra insufficiently extended towards longer wavelengths, compared to observations.          
\citet{Granato_94} preferred larger wedges (tens to hundreds of pc in diameter) 
and thereby solved the problem of the too narrow IR bumps compared to
observations.
Further difficulties when comparing calculations with data emerge from the 
so-called silicate feature problem. 
The silicate feature is a resonance in the spectral energy distributions at $9.7\,\muup$m, arising
from the stretching modes in silicate tetrahedra. For optically thin configurations, this feature is seen in 
emission. Looking at optically thick tori, two cases have to be distinguished: The feature appears in emission, 
if the temperature decreases along the line of sight away from the observer and an absorption feature 
is seen for the case of increasing temperature along the line of sight.
Between those two cases, at an optical depth $\tau_{9.7\,\muup \mathrm{m}}\,\approx\,1$, a mixture of both --
so-called self-absorption -- is visible \citep{Henning_83}.
Therefore, for typical torus geometries, the feature is expected to arise in absorption for type\,II objects and 
in emission in type\,I sources. Up to now, no observation of a type\,I object
has shown the feature 
in emission.   
In both studies mentioned, a significant reduction of the silicate emission-feature could be found 
for the case of 
type\,I galaxies. However, a lot of fine tuning of the
parameters was needed.
\citet{Granato_94} point out that small silicate grains are selectively destroyed by radiation pressure 
induced shocks in the 
inner part of the torus close to the central source. Therefore, they set up a depletion radius 
for small silicate grains and are able to reduce the silicate emission at $9.7 \muup$m. 
\citet{Manske_98} succeeded in avoiding the feature by using a toroidal density distribution 
comparable to the one of \citet{Granato_94} with 
a large optical depth in combination with a strong anisotropic radiation source, which 
is a more physical approach to describe the accretion disc.
Another more realistic way to cope with the silicate feature problem was proposed by
\citet{Nenkova_02}. They introduce a clumpy structure again within a flared disc geometry. 
Their radiative
transfer simulations show that within a large range of values of their parameters, a
reduction of the emission feature is feasible.  
Another possibility to investigate AGN torus models is the simulation of polarisation maps and 
their comparison with observations. For a first approach see \citet{Wolf_99b}.

In this first paper of a series, we present radiative transfer calculations for a hydrostatic model of dusty tori.
We use the so-called \emph{Turbulent Torus Model}, introduced by \citet{Camenzind_95}. 
The great advantage of this model arises
from the fact that the dust density 
distribution and the geometrical shape of the torus do not have to be set up
arbitrarily and independent of each other, but both result from 
physically reasonable assumptions. 
Our calculations combine structure modelling and fully
three-dimensional Monte Carlo radiative transfer simulations. 
The following chapter will summarise the structure modelling together with all ingredients needed to 
perform the radiative transfer simulations. 

In a further step towards more physical models of AGN dust tori, the homogeneous dust distribution 
will be replaced by a clumpy dust model. 
Later, we will start with hydrodynamic simulations in combination with radiative transfer calculations,
in order to gain information about the temporal evolution of spectral energy distributions and surface 
brightness maps of dusty tori.

Chapter \ref{chap:Result} deals with
the results we obtained with the first approach. Spectral energy distributions and surface
brightness maps are calculated and discussed for part of
the simulation series we conducted.  
In chapter \ref{chap:Comparison} we
discuss comparisons of our model results with large aperture data as well as with very high 
spatial resolution data.

\section{The model}
\label{chap:Model}

\subsection{The Turbulent Torus Model}
\label{sec:TTM-model}

It is known that many spiral galaxies harbor young nuclear star clusters
\citep[e.g.][]{Gallimore_03,Walcher_04,Boeker_02,Boeker_04}. 
Especially during late stages of stellar evolution, stars release large
amounts of gas through stellar winds and the ejection of planetary nebulae. 
With increasing distance from the star, dust forms from the gas phase. 
Due to the origin of the dust from ejection processes of single stars, we
expect the dust torus
to consist of a cloudy structure. 
As even current interferometric instruments are not able to resolve single clouds
of the dust distribution, 
they are assumed to be small.  For the sake of simplicity, we treat 
the torus as a continuous medium in this paper.
The dust clouds are ejected by the stars and therefore take over the stellar 
velocity dispersion $\sigma_*$. As we neglect interaction effects between
these clouds, we can assume that they possess the same velocity as the stars, called
the turbulent velocity $v_{\mathrm{t}}$ ($v_{\mathrm{t}} \approx \sigma_*$). 

We are dealing with stationary models only. Therefore, the presence of continuous 
energy feedback by supernovae explosions as well as continuously 
injected mass by
stellar winds is not taken into account. The dust clouds are embedded
into hot gas, which leads to the destruction of dust grains by 
sputtering effects. Therefore, dust has a finite lifetime and it is 
likely that the composition differs from the one found in interstellar 
dust within our galaxy (see Sect.~\ref{sec:dust}). The effects of magnetic fields on charged
dust particles are neglected.
The radiation pressure of the extreme radiation field of the central 
AGN exerted on the dust grains only affects the very innermost part 
of the dust density distribution and is therefore not taken into account 
within our simulations.

Due to their environment, the dusty clouds are subject to gravitational forces.
Together with centrifugal forces -- because of the rotation of the central star 
cluster -- an effective potential can be formed. In the following, the three main
components will be discussed:
First, the gravitational potential of the supermassive black hole. As the torus extends to  
several hundred thousands of Schwarzschild radii from the centre, we can treat the black hole as
a point-like source of gravitation with a Newtonian potential, 
given in cylindrical coordinates:
      \begin{eqnarray}
        \phi_{\mathrm{BH}}(R,z) = -\frac{G \, M_{\mathrm{BH}}}{\sqrt{R^2+z^2}}.
      \end{eqnarray}
The second contribution we take into account is the gravitational potential of the nuclear star
cluster, dominating the contribution by stars in the galaxy core. 
From high resolution observations of the centres of galactic nuclei \citep[e.\,g.\,][]{Gebhardt_96} 
a surface brightness distribution profile 
can be derived. This can be well fitted by a two-component power law, the so-called Nuker law, 
given by the following equation \citep{Lauer_95,Byun_96,Gebhardt_96,Faber_97}: 
      \begin{eqnarray}
        I(R,z) & = & 2^{\frac{\beta-\delta}{\alpha}} \, I_{\mathrm{c}} \, \left( \frac{R_{\mathrm{c}}}{\sqrt{R^2+z^2}} \right) ^{\delta} \nonumber\\
 & & \left[ 1+ \left(\frac{\sqrt{R^2+z^2}}{R_{\mathrm{c}}} \right)^{\alpha} \right]^{\frac{\delta - \beta}{\alpha}},
      \end{eqnarray}
where $R_{\mathrm{c}}$ is the core radius (or break radius), $\beta$ the slope of the surface brightness 
distribution outside $R_{\mathrm{c}}$,
$\delta$ the slope inside $R_{\mathrm{c}}$, $\alpha$ characterises the width of the transition region and 
$I_{\mathrm{c}}$ is the surface brightness at the location of $R_{\mathrm{c}}$.

When assuming a constant mass-to-light ratio, this distribution mirrors the
projected mass density distribution directly. 
By applying a least-square fitting procedure, one can obtain the 3D density profile, for which an analytical expression
can be given. This is the so-called Zhao density profile family \citep{Zhao_97}: 
      \begin{eqnarray}
        \rho_*(R,z) = \frac{M_*}{4 \, \pi \, R_{\mathrm{c}}^3} \, \frac{C \left(\tilde{\alpha},\tilde{\beta},\tilde{\delta}\right)}{\left(R^2+z^2\right)^{\frac{\tilde{\delta}}{2}} \, \left(1+\left(R^2+z^2\right)^{\frac{\tilde{\alpha}}{2}}\right)^{\frac{\tilde{\beta} - \tilde{\delta}}{\tilde{\alpha}}}},
      \end{eqnarray}
where the parameters $\tilde{\alpha}$, $\tilde{\beta}$ and $\tilde{\gamma}$ are
complicated functions of the parameters $\alpha$, $\beta$ 
and $\gamma$ of the Nuker law. The quantity $C$ is a constant, depending on these parameters. 

By solving the Poisson equation, the gravitational potential of the nuclear star cluster can be obtained.
Throughout this paper, all simulations use the special case of
the Hernquist profile \citep{Hernquist_90}. According to observed profiles of
cores of spiral galaxies, the exponent of the
density distribution within the core radius is
close to 1 in most cases (corresponding to the
Hernquist profile). Recent simulations of dark matter halos by \citet{Williams_04} -- which belong to the
same family of stellar systems -- also give an upper limit for the exponent of the
density distribution within the core radius between  
1.5 and 1.7 in the inner part.\\
With the parameters $\tilde{\alpha}$\,=\,$1$, $\tilde{\beta}$\,=\,$4$ und $\tilde{\delta}$\,=\,$1$, we finally get the 
needed gravitational potential:
\begin{eqnarray}
  \phi_{\mathrm{Hernquist}}(R,z) = - \frac{G \, M_*}{\sqrt{R^2+z^2}+R_{\mathrm{c}}}. 
\end{eqnarray}   
The third part of the effective potential is the centrifugal potential. Apart from 
the turbulent velocity, the stars -- and therefore the dust clouds -- possess an additional velocity 
component from the rotational motion around the gravitational centre. 
Due to the lack of observational data, 
we adopt the following distribution of specific angular momentum for such a 
stellar system:
\begin{eqnarray}
  j_{\mathrm{spec}} = \sqrt{G \, (M_{\mathrm{BH}} \, + M_*(R_{\mathrm{T}})) \,
 R_{\mathrm{T}}}
 \, \left( \frac{R}{R_{\mathrm{T}}}  \right)^{\gamma}.
\end{eqnarray} 
Here, $R_{\mathrm{T}}$ is the torus radius, defined as the location, where the
effective potential reaches its minimal value. This corresponds to an
equilibrium of centrifugal and gravitational forces in our solution,
where the angular momentum attains the Keplerian value. Therefore, a
distribution of the specific angular momentum is needed, which fixes the value at
the location of the torus radius to the Keplerian value. We used 
a less steep decline of the specific angular momentum beyond the
torus radius than for the Keplerian case, which is in agreement with 
expectations for elliptical star
clusters. Such toroidal gas distributions are expected to be unstable on 
time scales larger than the dynamical time at the torus radius. The torus 
is however steadily refilled so that a quasi-stationary configuration results.\\
\begin{figure}[t!]
  \resizebox{\hsize}{!}{\includegraphics{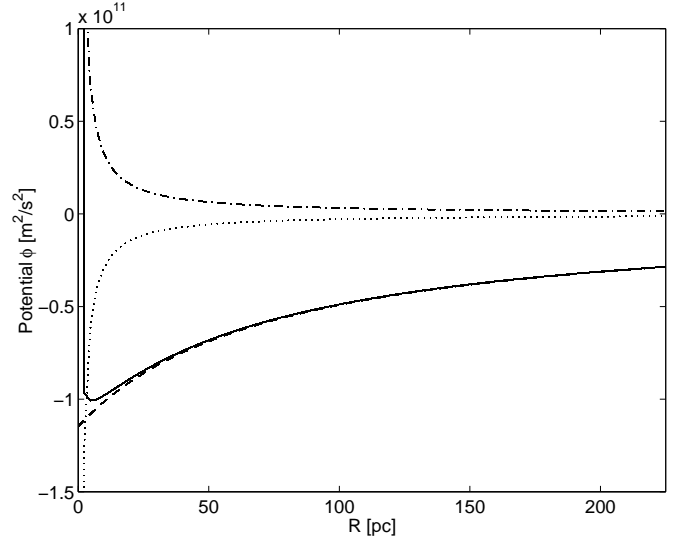}} 
  \caption{Comparison of the contributions of the three components of the
  effective potential for the parameters of our standard model in the equatorial plane. The solid line
  corresponds to the effective potential, the dotted line to the BH-potential,
  the dashed line to the potential of the nuclear star cluster and the
  dashed-dotted line to the centrifugal potential.} 
  \label{fig:tmean001_potentialvergleich} 
\end{figure}
By integrating over the centrifugal force -- under the assumption of constant angular velocity on 
cylinders -- we obtain the centrifugal potential:
\begin{eqnarray}
\label{equ:zfpotential}
 \phi_{\mathrm{CF}}(R,z) = 
 \frac{G\,(M_{\mathrm{BH}} \, + M_*(R_{\mathrm{T}}))}{2\,R_{\mathrm{T}}\,(1-\gamma)}\, \left( \frac{R}{R_{\mathrm{T}}} \right)^{2\,(\gamma-1)}.
\end{eqnarray} 
The effective potential and its three single components are
shown in Fig.~\ref{fig:tmean001_potentialvergleich} for the case of our standard model (described
in Sect.~\ref{sec:param_model}) in the equatorial plane. 
As can be seen, the effective potential (given by the solid line) in the outer part of the torus is 
dominated by the potential of the
nuclear star cluster (dashed line).
Moving closer to the centre, the black hole potential (dotted line) wins,
which leads to a further bending down of the potential, before the
centrifugal potential (dashed-dotted line) dominates. The latter is the only repulsive term
and leads to a centrifugal barrier close to the centre, preventing the models
to be gravitationally unstable. \\

The force given by the effective potential has to be balanced by the turbulent pressure in hydrostatic 
equilibrium. This pressure results from the turbulent motion of 
the dust clouds and can be expressed by the following 
(isothermal) equation of state:
 \begin{eqnarray}
   P(\rho_{\mathrm{d}})=\rho_{\mathrm{d}} \, v_{\mathrm{t}}^2.
 \end{eqnarray} 
After inserting this into the equation for the hydrostatic equilibrium and 
solving the differential equation under the assumption that 
$v_{\mathrm{t}}$ is constant, we obtain the dust density distribution:
 \begin{eqnarray}
\label{equ:staubdichte}
    \rho_{\mathrm{d}}  =   \rho_{\mathrm{d}}^0 \,
    \exp \left[\frac{G\,M_*}{v_{\mathrm{t}}^2\,R_{\mathrm{c}}} \,
    \left\{\frac{M_{\mathrm{BH}}/{M_*}}{\sqrt{\frac{R^2+z^2}{R_{\mathrm{c}}^2}}}
    + \frac{1}{1+\sqrt{\frac{R^2+z^2}{R_{\mathrm{c}}^2}}} \right. \right. \nonumber\\
  \left. \left. {} - \frac{R_{\mathrm{c}}\,(M_{\mathrm{BH}} \,+
    M_*(R_{\mathrm{T}}))}{2\,M_*\,R_{\mathrm{T}}(1-\gamma)}\, \left(
    \frac{R}{R_{\mathrm{T}}} \right)^{2\,(\gamma-1)} \right\}\right]. 
 \end{eqnarray}
Here, $\rho_{\mathrm{d}}^0$ is a measure for the total dust mass.
From this equation, some basic features of the model can be derived:
First of all, equipotential surfaces have the same shape as surfaces of constant
density, so-called isopynic surfaces. 
The shape of these surfaces only depends on the reduced parameter $M_{\mathrm{BH}}/{M_*}$ and the 
reduced coordinates $R/{R_{\mathrm{c}}}$, $R/{R_{\mathrm{T}}}$ and 
$z/{R_{\mathrm{c}}}$ and the exponent of the angular
momentum distribution $\gamma$.
As mentioned above, the torus radius is connected with the minimum of the
potential, the maximum of the density and the maximum of pressure and lies
within the equatorial plane, due to symmetry considerations.

\subsection{Parameters of the model}
\label{sec:param_model}

\begin{figure*}
\centering
\mbox{
  \subfigure{\includegraphics[width=8.5cm]{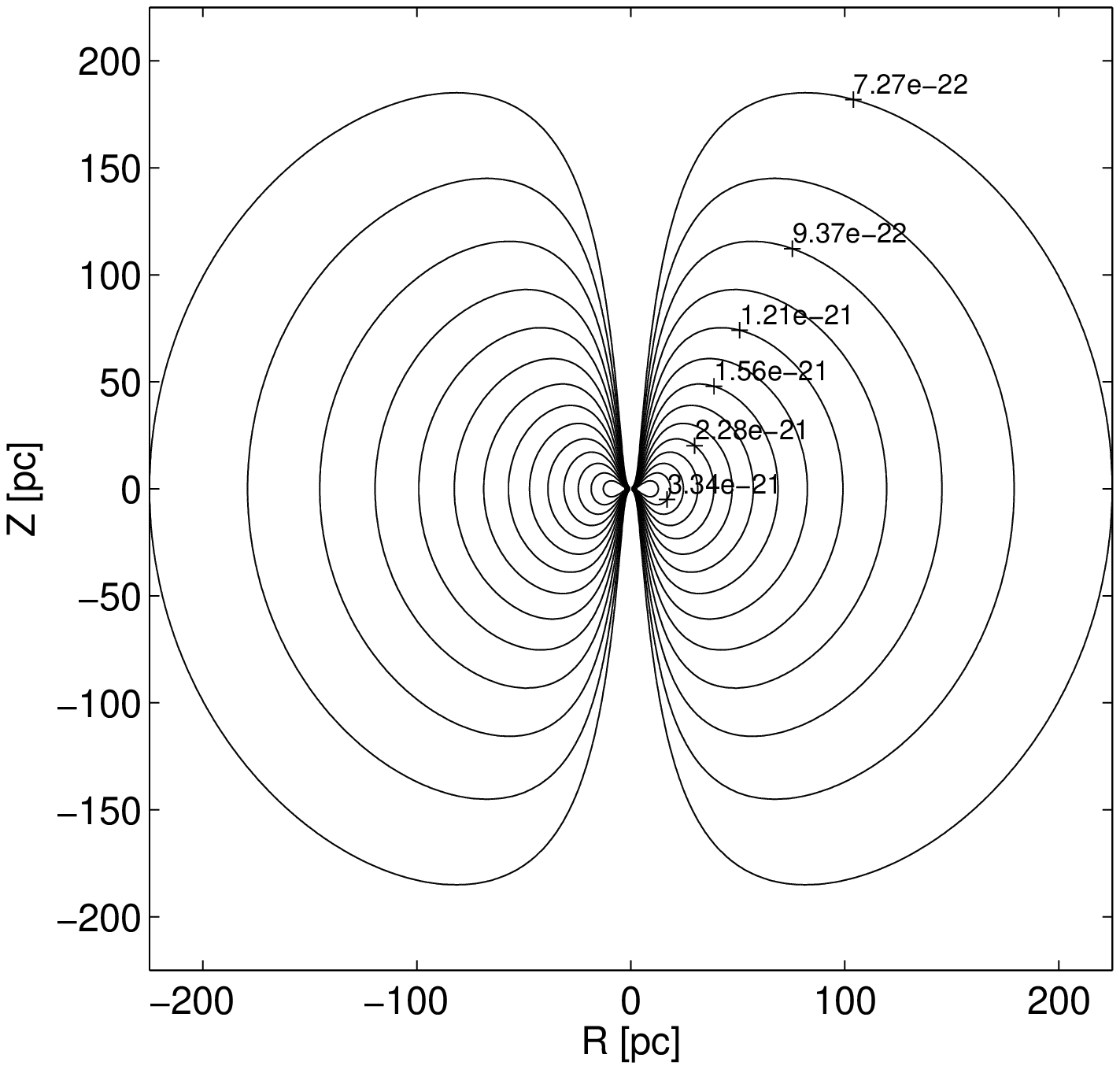}}
  \subfigure{\includegraphics[width=8.5cm,angle=90]{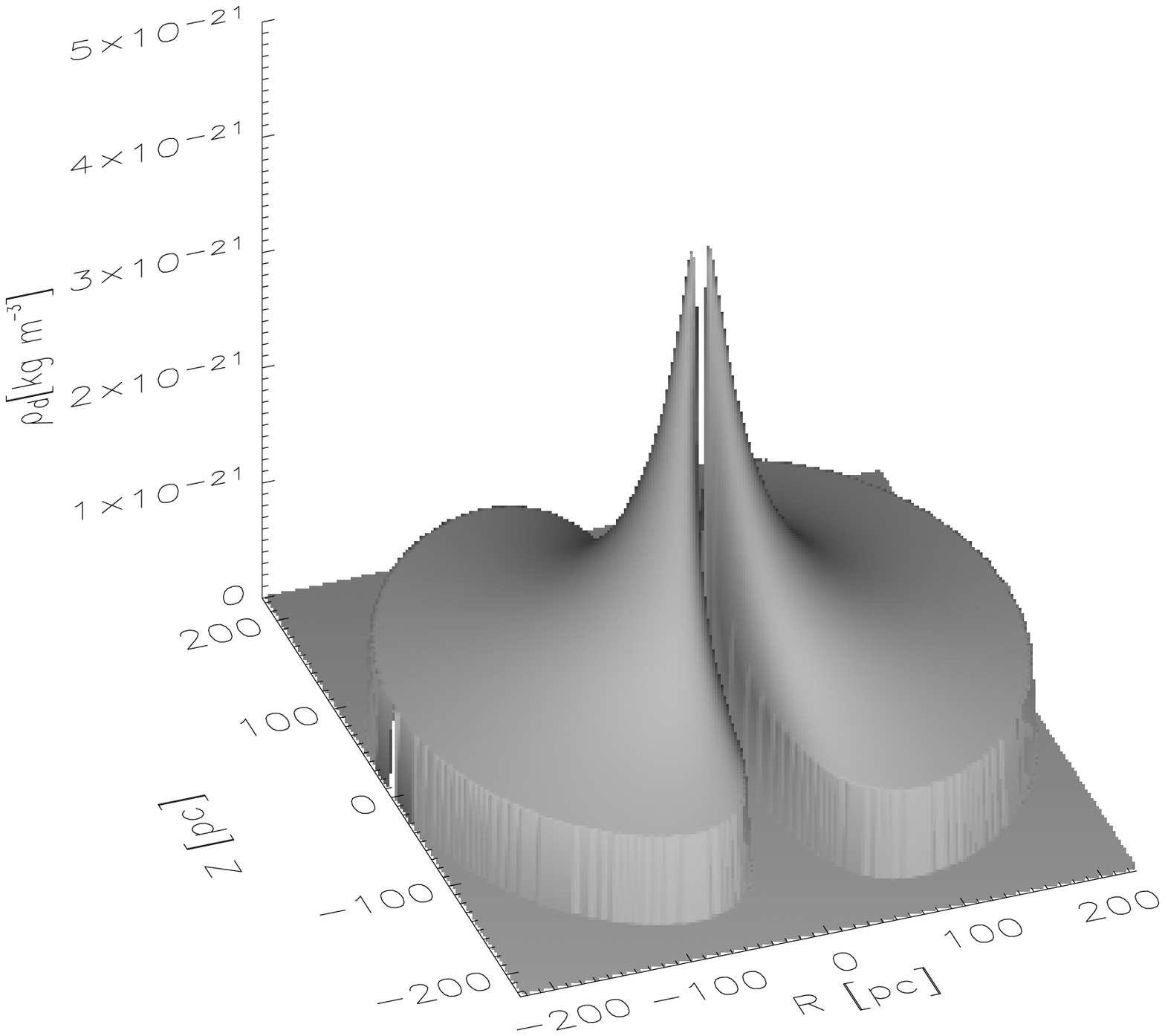}}
}
  \caption{Visualisation of our standard model: {\bf a)} Isopynic line plot (logarithmically equidistant) 
           in a meridional plane of the torus. The numbers give the dust density in kg\,$\mathrm{m}^{-3}$ 
           of the corresponding isopynic line. {\bf b)} Dust density distribution in a meridional
           plane. \label{fig:standardmodel_torus_dichte}} 
\end{figure*}

The introduced standard model will be used to
represent the core of a typical Seyfert galaxy. Therefore, we calculated the mean black
hole mass and mean luminosity of a large sample of Seyfert galaxies,
given in \citet{Woo_2002}. We find mean values of $\left<M_{\mathrm{BH}}\right>$\,=\,
$6.6 \cdot 10^7 \, \mathrm{M}_{\odot}$
and $\left<L_{\mathrm{disc}}\right>$\,=\,$1.2\cdot 10^{11} \, \mathrm{L}_{\odot}$.
After calculating a set of simulations with different dust density distributions, leading to
different density contrasts between the maximum density and the outer part of
the distribution, we found that a relatively shallow density distribution is
needed in order to obtain SEDs extending to wavelengths longer than 10-20$\,\muup$m. 
To achieve this, a relatively steep radial distribution of specific
angular momentum is required ($\gamma$\,=\,$0.5$ leads to the steepest -- but
still stable -- 
distribution). 
But this gives only a fairly small effect. It is more efficient to increase 
the core radius, which leads to a broadening of the density
distribution, due to the flattening of the potential of the star cluster in
the outer part of the torus. A change of the total 
mass of the stars mainly scales the
potential of the stellar cluster and therefore finally the density
distribution. 
By taking this into account, we chose the 
core radius for our mean model to be $75\,$pc with a mass of 
the nuclear star cluster of $2 \cdot 10^9\,\mathrm{M}_{\odot}$.

To constrain the other parameters,  one can use the fact that not all of them are
independent from each other. Comparable to the fundamental plane of
global galaxy parameters, the cores of ellipticals and bulges of 
spiral galaxies span the so-called core fundamental plane, relating the
parameters core radius $R_c$, the central surface brightness $\mu_0$ and the
central velocity dispersion $\sigma_*$ \citep{Kormendy_1987,Kormendy_1997}.
These studies give
several relations between relevant parameters.
The relation between the core radius
and the central velocity dispersion of the stars is interesting for us:  
\begin{eqnarray}
  \label{eqn:sigma_Rc}
  \sigma_* \approx 201 \, \left( \frac{R_{\mathrm{c}}}{75\,\mathrm{pc}} \right)^{0.18} \,
  \frac{\mathrm{km}}{\mathrm{s}}.
\end{eqnarray}
\begin{table}[!t]
\centering
\caption{\label{tab:param_meanseyf} Parameters used for the simulation of our
  mean Seyfert model.}
\begin{tabular}{lcc}
\hline
\hline
Parameter & Value \\
\hline
$M_{\mathrm{BH}}$   & $6.64\cdot10^7 \, M_{\sun}$           \\
$R_{\mathrm{c}}$    &  75\,pc                               \\
$M_{*}$             & $2.0\cdot 10^9 \, M_{\sun}$          \\
$L_{\mathrm{disc}}$          & $1.23\cdot 10^{11} \, L_{\sun}$      \\
$L_{\mathrm{disc}}/{L_{\mathrm{Edd}}}$ & 6\% & \\
$R_{\mathrm{T}}$    & 5\,pc                                 \\
$M_{\mathrm{dust}}$ & $5.79\cdot10^5 \, M_{\sun}$           \\ 
$\tau_{9.7\,\muup \mathrm{m}}$              & 2.0     \\
$R_{\mathrm{out}}$  & 225\,pc                                \\
$v_{\mathrm{t}}$    & $201\,\mathrm{km}\,\mathrm{s}^{-1}$ \\
$\gamma$            & 0.5                                  \\ 
$d$                 & 45\,Mpc                               \\
\hline
\end{tabular}
\end{table}
As already mentioned above, the dusty clouds are produced by the stars and 
take over their
velocity dispersion. Therefore, we choose the turbulent velocity of the clouds
$v_{\mathrm{t}}$ to be equal to the velocity dispersion of the stars in the central region of
the galaxy. According to equation (\ref{eqn:sigma_Rc}) the value $v_{\mathrm{t}}$\,=\,$201 \,
\mathrm{km}\,\mathrm{s}^{-1}$ is obtained.  
Interactions between the different phases (dust, molecular and hot gas) within
the torus and
between other dust clouds, which may alter their velocity distribution are
not taken into account at present.
The outer radius of the torus is chosen to be three times the core radius of
the nuclear star cluster, where the density distribution has already reached an
approximately constant value. 
An upper limit for the total mass of the nuclear star cluster is determined by the Jeans
equation, which relates structural with dynamical parameters for an isothermal
model of the central stellar distribution under the assumption of a constant
velocity dispersion. 
The remaining two parameters $R_{\mathrm{T}}$ and $\gamma$ are chosen such 
that the dust density
distribution remains flat enough to obtain IR bumps, which are broad enough to
compare with data and
to obtain a torus model with an inner radius small enough to allow the
majority of the grain species to reach their sublimation temperatures. 
Finally, the mass of the dust, enclosed into the torus, is determined by the depth
of the silicate feature at $9.7\,\muup$m. We chose 
$M_{\mathrm{dust}}$\,=\,$5.79\cdot10^5 \, M_{\sun}$ (corresponding to an optical 
depth of $\tau_{9.7\,\muup \mathrm{m}}$\,=\,$2.0$), which yields depths 
comparable to observed spectra.

The parameters we assumed for our standard model are summarised in Table \ref{tab:param_meanseyf} and the 
resulting model is shown in Fig.~\ref{fig:standardmodel_torus_dichte}.
As can be seen, the torus has a relatively dense core and the density
distribution gets shallow in the outer part. The maximum is reached in the
equatorial plane (due to symmetry considerations) at the torus radius $R_{\mathrm{T}}$ 
and the models possess a more
or less pronounced cusp at the inner end of the torus. The dust-free cone with
quite steep walls will be called funnel in the following.
The turbulent velocity always tends to increase the height of the torus, while
the angular momentum tends to flatten it.

\subsection{Dust properties}
\label{sec:dust}

Although the very extreme physical conditions near the central source make it
likely that the dust composition may be altered compared to interstellar dust
in our own galaxy, we assume -- for the sake of simplicity and comparability
-- a typical dust
extinction curve of interstellar dust. We use the model of
\citet{Mathis_77} which fits well the observed extinction curve of the interstellar
medium.

\begin{figure*}
\centering
\mbox{
  \subfigure{\includegraphics[angle=90,width=8.5cm]{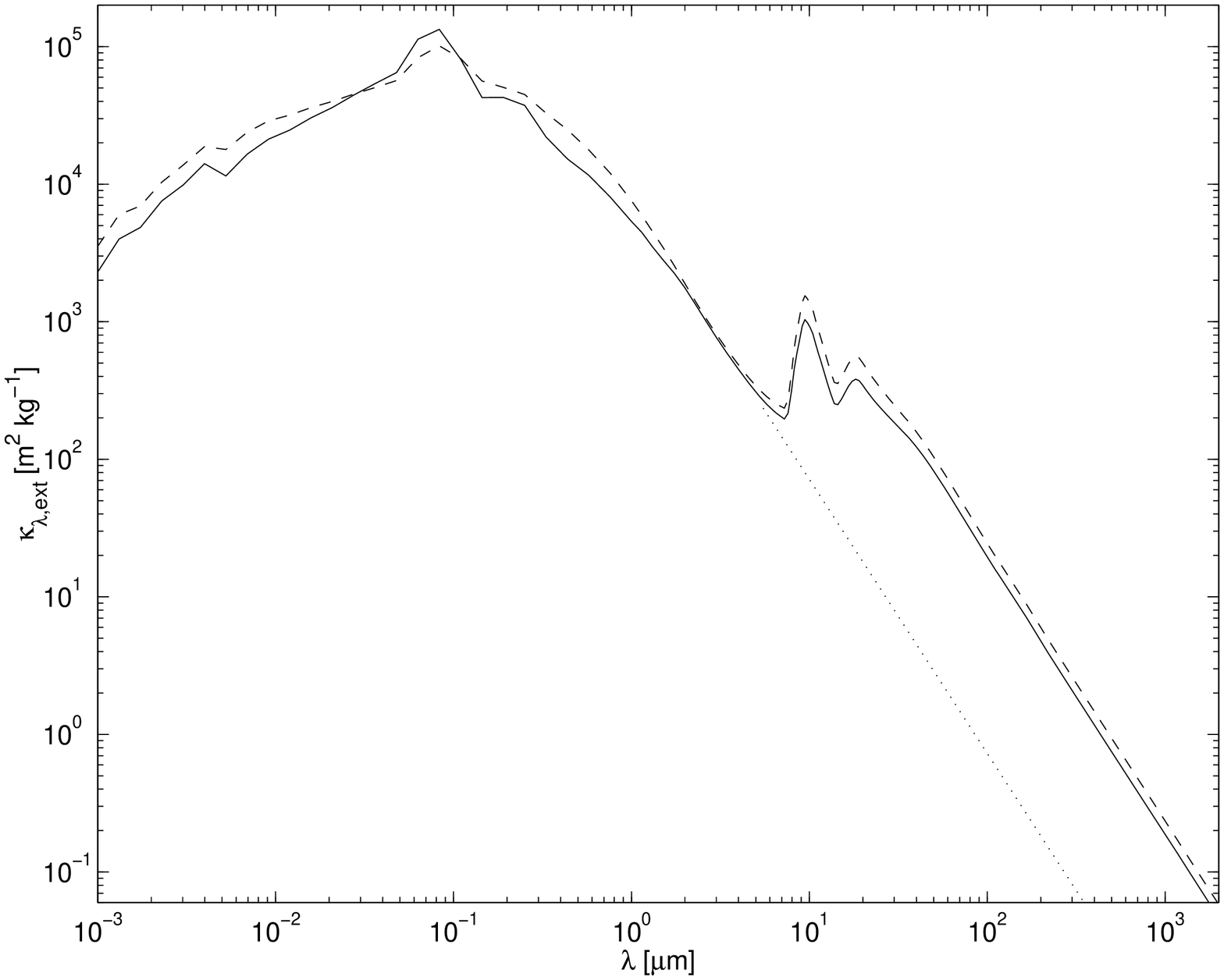}}
  \subfigure{\includegraphics[angle=90,width=8.5cm]{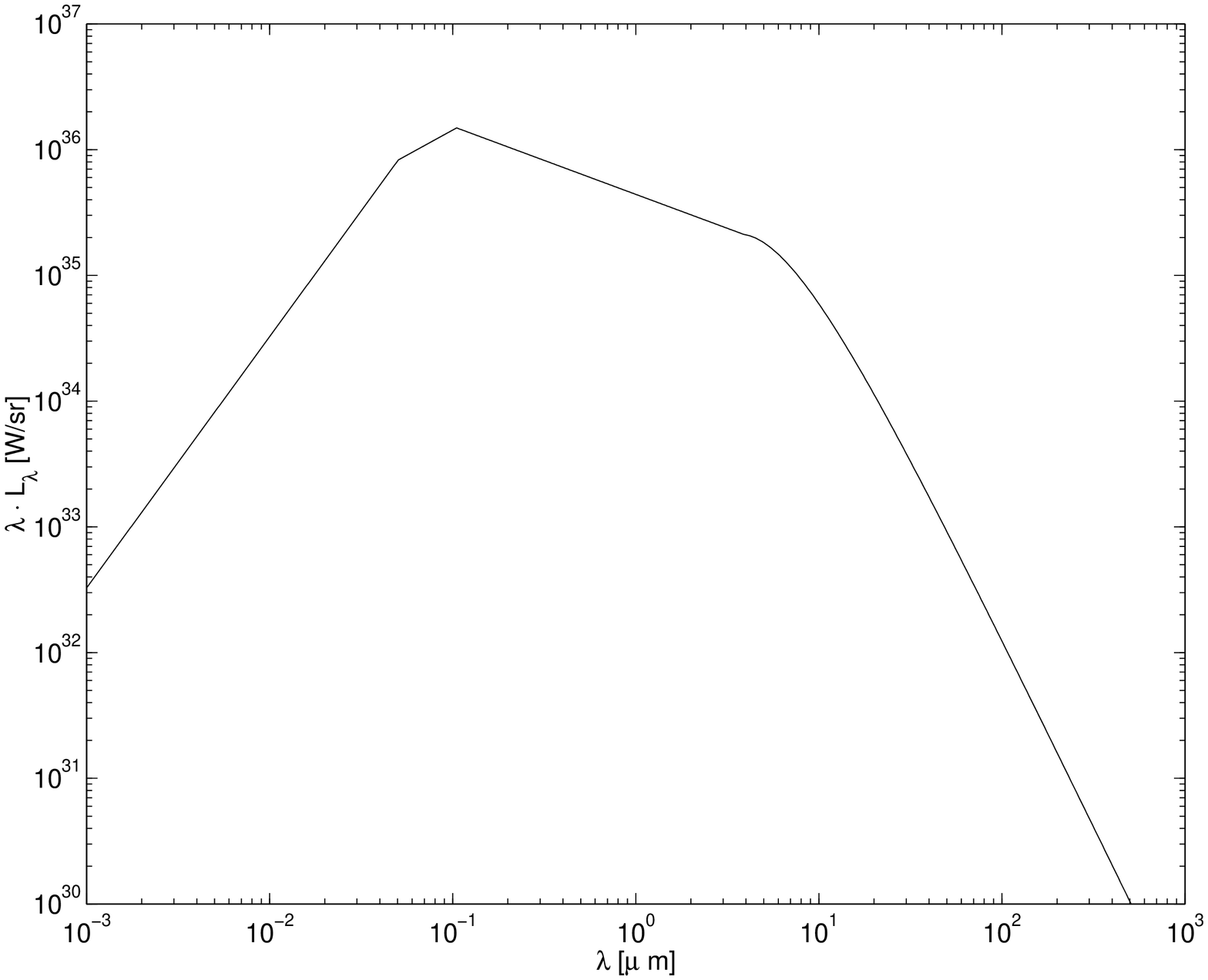}}
}
  \caption{Input parameters of our simulations: {\bf a)} Course of the mass extinction coefficient 
           plotted against the wavelength for the averaged dust model (dashed line) and single grains
           (solid line): composition of silicate and graphite grains with varying grain radii (for details
           see text). The dotted straight line should clarify the offset in the continuum extinction 
           curve. {\bf b)} SED of the central accretion disc, used as the primary energy source.} 
  \label{fig:input} 
\end{figure*}

They found a number density distribution (MRN-distribution hereafter), which is proportional to the grain
radius to the power 
of -\,3.5 and a range of grain radii between $0.005\,\muup$m and $0.25\,\muup$m. We
use 15 different grains\footnote{The number of dust grains increases
  the computational time for simulating the temperature
  distribution, therefore we tried to minimise the number of
  grains. Simulations, where we doubled the number of grains showed no
  significant differences in the spectral
  energy distributions.}, containing a mixture of 5
silicate and 10 graphite grains -- each species with different grain radii according 
to the MRN grain size distribution -- and a mass fraction of 62.5\% silicate and
37.5\% graphite. Because of the anisotropic behaviour of graphite, two
different sets of optical constants are necessary: one where
the electric field vector oscillates parallel to the crystal axis of the
grain (5 grains) and another one, where it oscillates perpendicular to it (5 grains). For them, the standard 
$\frac{1}{3}\,-\,\frac{2}{3}$ approximation is used, which is valid exactly in the Rayleigh limit and 
reproduces extinction curves reasonably well, as shown by \citet{Draine_93}.  
Optical data (refraction indices) are taken from \citet{Draine_84}, \citet{Laor_93} and 
\citet{Weingartner_01}
and the scattering matrix elements and coefficients are calculated under the 
assumption of spherical grains
using the routines of \citet{Bohren_83} within the radiative transfer code MC3D, 
which will be discussed in Sect.~\ref{sec:mc3d}.

In addition, a comparison was made between a split into 15 different
grains and the usage of only 
one grain, but with weighted mean values for the extinction, absorption and scattering 
cross sections and the Stokes
parameters, which describe the intensity and the polarisation state of the photon package.
The weights result from the abundance and the size distribution of the 
respective material (for details about the averaging see \citet{Wolf_03b}). 
The resulting extinction curves of these two cases are shown in
Fig.~\ref{fig:input}a, where the solid line corresponds to the splitted
model and the dashed line to the model with mean characteristics. 
For every dust species (type of material and grain size) the temperature distribution
is determined separately. 
In Fig.~\ref{fig:input}a one can clearly see the two silicate features at
$9.7\,\muup \mathrm{m}$ and $18.5\,\muup \mathrm{m}$. The extinction curves drop down quite steeply
at longer wavelengths and an offset can be recognised around the
wavelengths of the two features between the continuum spectra shortward of $10\,\muup$m and longward
of $20\,\muup$m, emphasised by the dotted line fragment.  

\subsection{Primary source}

In our modelling, the dust in the torus is solely heated by an accretion disc
in the centre of the model space. In most of the
simulations we use a point-like, isotropically radiating 
source with a luminosity of $1.2\cdot 10^{11}\, \mathrm{L}_{\odot}$ in our standard
model,  which corresponds to roughly
$6\%$ of the Eddington luminosity. The spectral energy distribution (SED) is
composed of different power laws and a Planck curve as follows. In the
ultraviolet to optical wavelength regime, we use the composite mean quasar spectrum
from \citet{Manners_02}. For wavelengths longer than Lyman alpha,
it includes data from 2200 quasars of the Sloan Digital Sky Survey \citep{Berk_01}, 
which results in a power law with a spectral index of 
$\alpha_{\nu}\,$=$\,-0.46$ \footnote{$\alpha_{\nu}$ is defined as $F_\nu \propto \nu^{\alpha_{\nu}}$}. Shortwards 
of Lyman alpha, data from radio-quiet
quasars \citep{Zheng_1997}, taken with the HST Faint Object Spectrograph, is used.
The data leads to a spectral index of $\alpha_{\nu}$\,=\,$-1.8$.
For wavelengths longer than $10\,\muup$m, we use a decline of the spectrum according to
the Rayleigh-Jeans branch of a Planck curve with an effective temperature of
1000\,K, reflecting the smallest temperatures expected from the thermal emission
of the accretion disc. At shorter wavelengths, the spectrum decays with 
a spectral index of $\alpha_{\nu}$\,=\,$-3$. Due to the lack of observational data in
this wavelength range, the index is chosen in agreement 
with theoretical modelling of accretion disc spectra performed by \citet{Hubeny_2000}.     
Taking this shape and normalising to the
chosen bolometric luminosity of the accretion disc yields the input spectrum shown
in Fig.~\ref{fig:input}b.

\subsection{The radiative transfer code MC3D}	     
\label{sec:mc3d}

MC3D is a three-dimensional continuum radiative transfer code based on the Monte Carlo
approach \citep{Fischer_94, Wolf_99a}. It is able to manage arbitrary
three-dimensional dust
and electron configurations and different primary sources of radiation. 
Diverse shapes and species of dust grains can be implemented
as well.  
Working in the particle picture, the bolometric luminosity of the
central source is divided into photon packages, so-called weighted photons. 
They
are fully described by their wavelength $\lambda$ and their Stokes vector,
describing the intensity and the polarisation state of the particle.
According to a provided radiation characteristic and spectral energy
distribution of the source, such a weighted photon is emitted. Within the dust or
electron configuration, absorption and scattering take place, where Thomson
scattering, Rayleigh and Mie scattering is applied to the weighted photon. 
In a last step, these photons are detected after leaving the convex model space.

According to this procedure, temperature distributions as well as -- in a second
step -- spectral energy distributions and surface brightness distributions can
be calculated. 
Temperature distributions have to be calculated using real radiative transfer. 
In order to minimise CPU time and get smoother surface
brightness distributions, a raytracer (also inherent in MC3D) can be used to
obtain SEDs and maps.

Details about MC3D and the built-in features can be found in
\citet{Wolf_99a}, \citet{Wolf_00}, \citet{Wolf_01} and \citet{Wolf_03}.

\citet{Pascucci_04} tested MC3D for 2D structures against various other continuum radiative transfer
codes, where they found good agreement.
We also performed a direct comparison for the special case of AGN dust tori
with the simulations of \citet{Granato_94}, one of the standard torus models to compare to, 
calculated with his grid based code.
\begin{figure}
  \resizebox{\hsize}{!}{\includegraphics[angle=90]{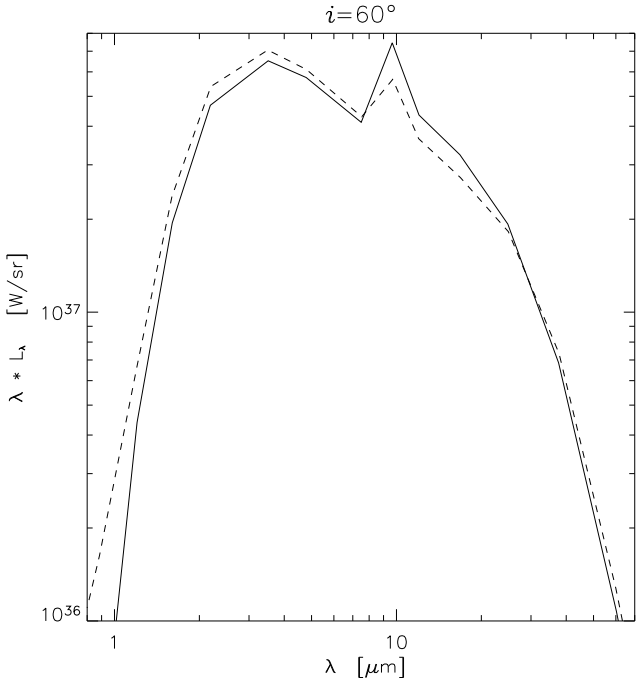}} 
  \caption{Comparison of MC3D (solid line) with the code of \citet{Granato_94}
  (dashed line).}   
  \label{fig:comp_granato} 
\end{figure}
The results are
displayed in Fig.~\ref{fig:comp_granato}, where the solid line corresponds to
the calculation with MC3D and the dashed line to the original spectrum of
Granato.   
The underlying model is a wedge geometry with dust of constant
density. The ratio between outer and inner radius is set to 100 with a half
opening angle of the dust-free cone of 40\,\degr\, and an optical depth of 0.3 at $9.7\,\muup
\mathrm{m}$ (for details of the model setup see \citet{Granato_94}). The original
spectrum of Granato's code is taken from the library of SEDs
described in \citet{Galliano_03}.
   
As can be seen in Fig.~\ref{fig:comp_granato}, good agreement between the two codes is found with
deviations smaller than 30\% within the relevant wavelength range.
The reason for the remaining differences between the two SEDs is due to the fact that only
standard calibrations of the code-parameters were used and no
fine tuning was performed. Minor differences of the model setup are responsible for the 
remaining discrepancies among the SEDs.

\section{Results and discussion}
\label{chap:Result}

First, the use of single dust grains is motivated, before we 
introduce our model by presenting temperature distributions and 
an inclination study. After that, several parameters of our
standard model have been varied and the resulting effects will be discussed.

If not noted otherwise, all of the SEDs shown are pure dust re-emission 
spectra. The direct radiation
from the central source is always omitted, as we are only interested
in the mid-infrared wavelength range, where the torus emission dominates 
the SED.
Optical depths are given for lines of sight within the equatorial
plane from the centre outwards.

\subsection{Averaged dust mixture versus single grains}
\label{sec:av_dust}

\begin{figure*} 
\centering
\mbox{ 
   \subfigure{
     \includegraphics[width=8.5cm,angle=90]{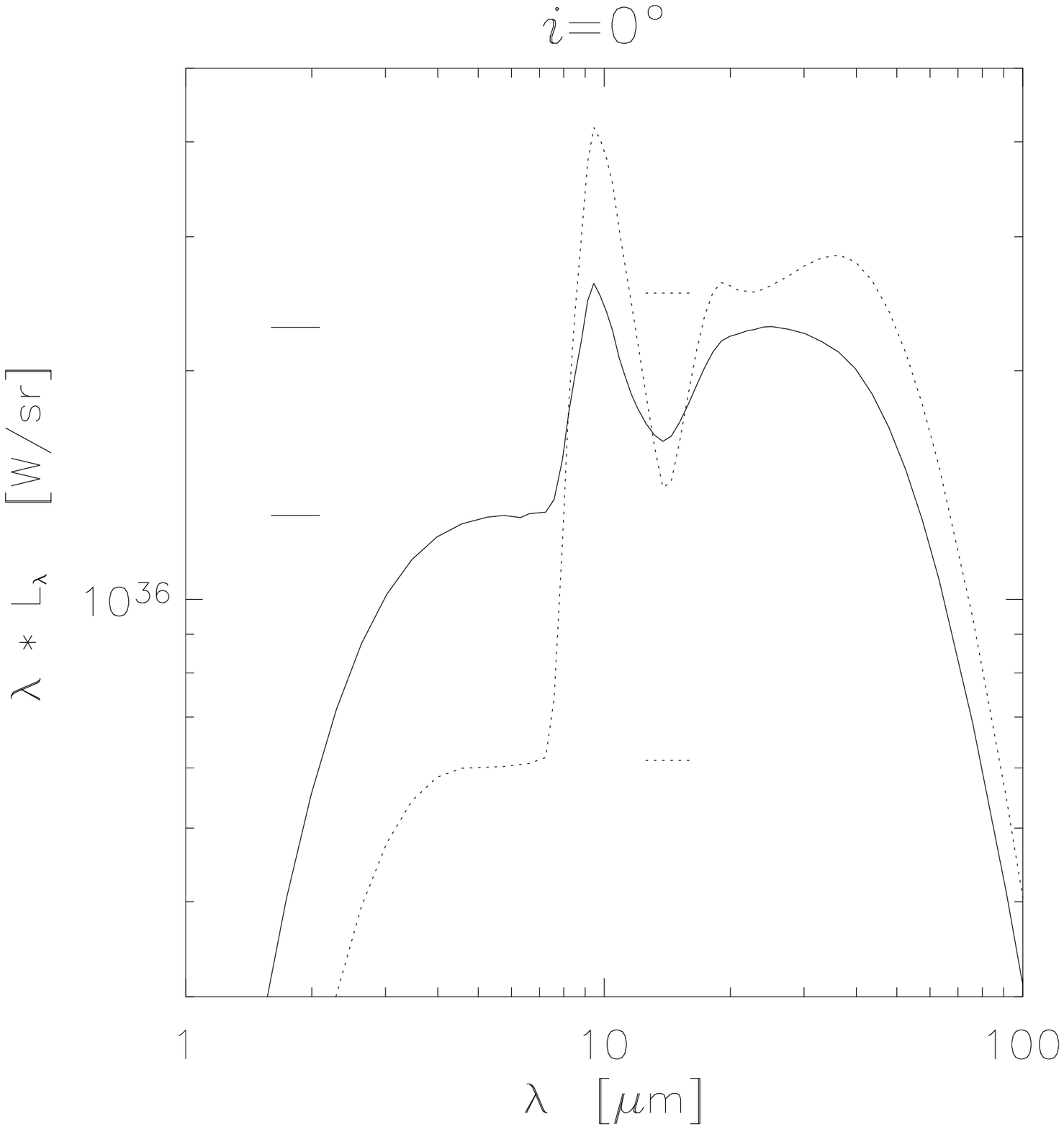}
   }
   \subfigure{
     \includegraphics[width=8.5cm,angle=90]{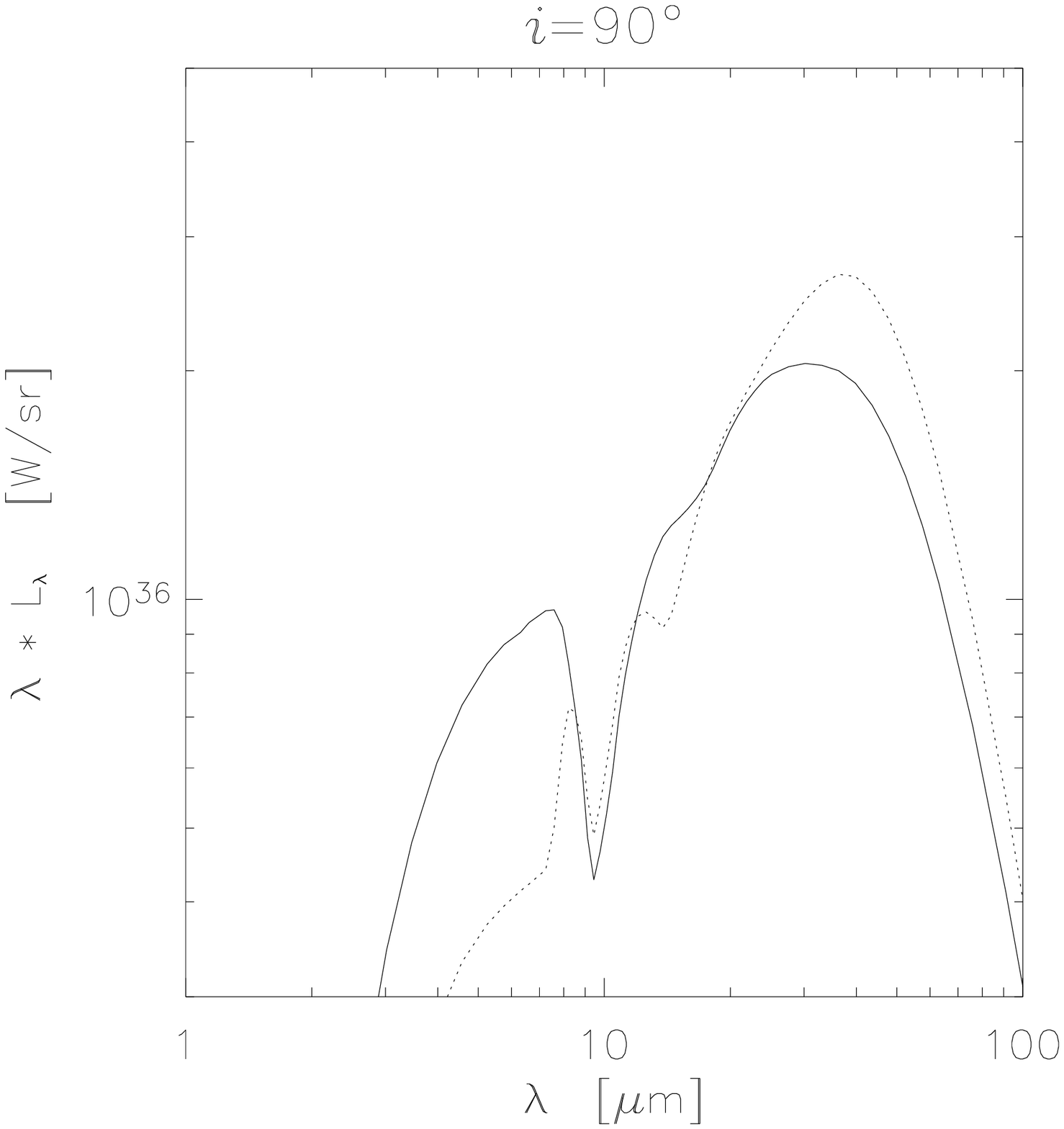} 
   }
}
\caption{SEDs at inclination angles of 0\,\degr\, and 90\,\degr\, for the case
  of averaged dust properties (dotted line) and dust splitted in grains of
  different species and sizes (solid line). The horizontal line segments in the left
  panel indicate the offset of the continuum SEDs of the individual curves.} 
\label{fig:dust_av_single}
\end{figure*}

Detailed Monte Carlo simulations are always very time consuming.
In order to reduce the CPU time, an averaged dust grain mixture
\citep{Wolf_03b} is often
considered instead of a real mixture of single grains. However, temperature distributions
have to be computed for every single grain. 
Therefore, computation time
strongly scales with the number of dust grains considered. 
As \citet{Wolf_03b} showed for the case of protoplanetary discs and we also confirmed for
AGN tori, an averaged dust model works perfectly well, if the inner radius of the disc/torus is larger
than the sublimation radii of the grains (see discussion later). 
But for most of our models, this is not the case, as the inner radii of the
tori are often determined by sublimation themselves. 
These sublimation radii follow an approximative formula
for the local thermodynamic equilibrium. It only takes emission of the central 
source and extinction by dust species with smaller inner radii into account.  
Although the resulting maximal temperatures for the single grains differ by up to 
10\% relative to the assumed sublimation temperatures of silicate and graphite grains, 
this seems to be an adequate 
procedure, as sublimation temperatures are not very well known.
With this new dust model, the situation looks
much different, as shown in Fig.~\ref{fig:dust_av_single}, although the
overall shape of the extinction curve looks similar compared to the one of the 
averaged grain model (see Fig.~\ref{fig:input}a).
The typical shape of the resulting SEDs will be described in more detail in
Sect.~\ref{sec:inc_sed}. In the following, we address just
the major differences between these two cases.
The solid line corresponds to our standard model (see Fig.~\ref{fig:dust_av_single}) with a dust
mixture containing 15 different grains (3 different grain species -- silicate 
and two orientations of graphite grains -- with 5
varying grain radii each), while the dotted line is calculated for the same
torus parameters, but the dust model is given by a single grain type with
averaged properties concerning chemical composition and grain size.
Using the averaged grain model, the silicate features are much more pronounced
compared to the other case. The same happens with the offset in the SED
shortwards of the $9.7\,\muup \mathrm{m}$ feature. The reason for this is the varying
sublimation temperatures of the different dust species. Silicate grains
sublimate at about 1000\,K, while graphite grains sublimate around 1500\,K.
Smaller particles are heated more efficiently than larger grains.
Therefore, smaller
grains reach the sublimation temperature earlier and the inner radius of their
density distribution must be larger than the one for the distribution of large
grains. This effect leads to a layering of grain species and sizes, where silicate
grains are further out (in the radial direction) than graphite grains and the
distribution of smaller grains is shifted outwards compared to the
distribution of larger grains. 
Neglecting this layering and using averaged properties within one grain leads
to the effects shown in Fig.~\ref{fig:dust_av_single}. The reason for these effects is that
some of the dust species, represented by the grain, are heated to temperatures higher than 
their sublimation temperatures. This is an unphysically treatment of these components. 
Especially affected are the smallest silicate grains, which are the most abundant ones in our dust
model. 
The effects are mainly more explicitly visible characteristics of the dust
extinction curve properties of silicate grains -- especially the offset in the continuum spectrum
near the feature (demonstrated by the horizontal lines in Fig.~\ref{fig:dust_av_single}) 
and the strength of the feature itself. Reaching only the
permitted smaller temperatures, these effects are partially balanced by the
contribution of the other grain types. Using the minimum of sublimation
temperatures of the mean grain components leads to a Wien branch shifted to
longer wavelengths. 

Being mainly interested in the mid-infrared wavelength range, 
we decided to split the dust model in single grains for all of the models shown in this
paper. All
species obey the same density distribution, according to physical reasoning
(see Sect.~\ref{sec:TTM-model}) and the mass fraction of the individual species,
derived from the power law number density distribution.

The inner radius of the distribution of every dust species is calculated using the formula for
the local energy equilibrium and taking extinction by all other dust species into
account. Spheres with these radii around the central source are left vacant concerning these particular 
dust species. The rest of the calculated equipotential surfaces are filled with dust, according 
to the dust model used.

\subsection{Temperature distribution}
\label{sec:temp_dis}

\begin{figure*}
\centering
\mbox{ 
   \subfigure{\includegraphics[width=8.0cm,angle=90]{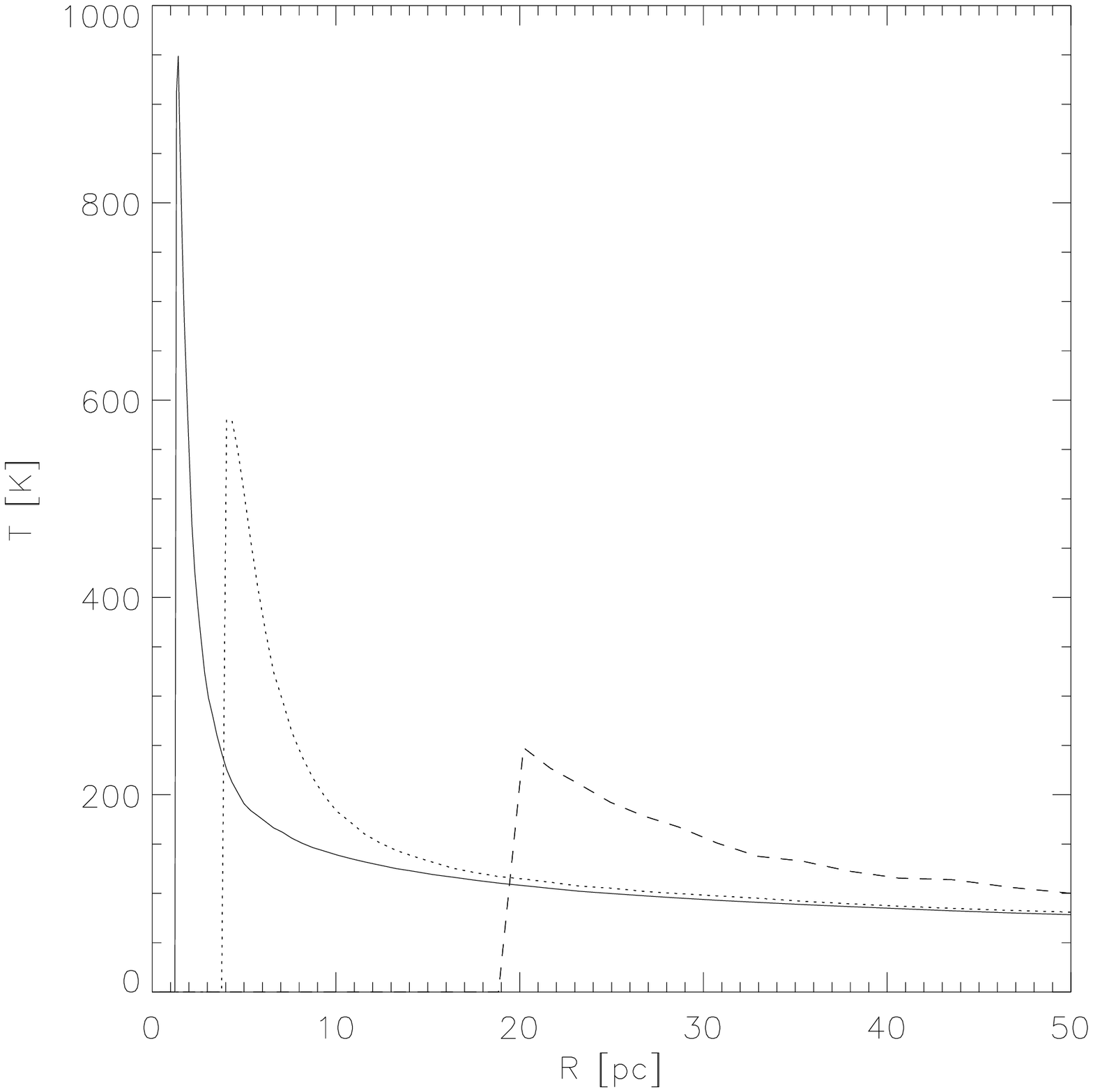}}
   \subfigure{\includegraphics[width=8.0cm,angle=90]{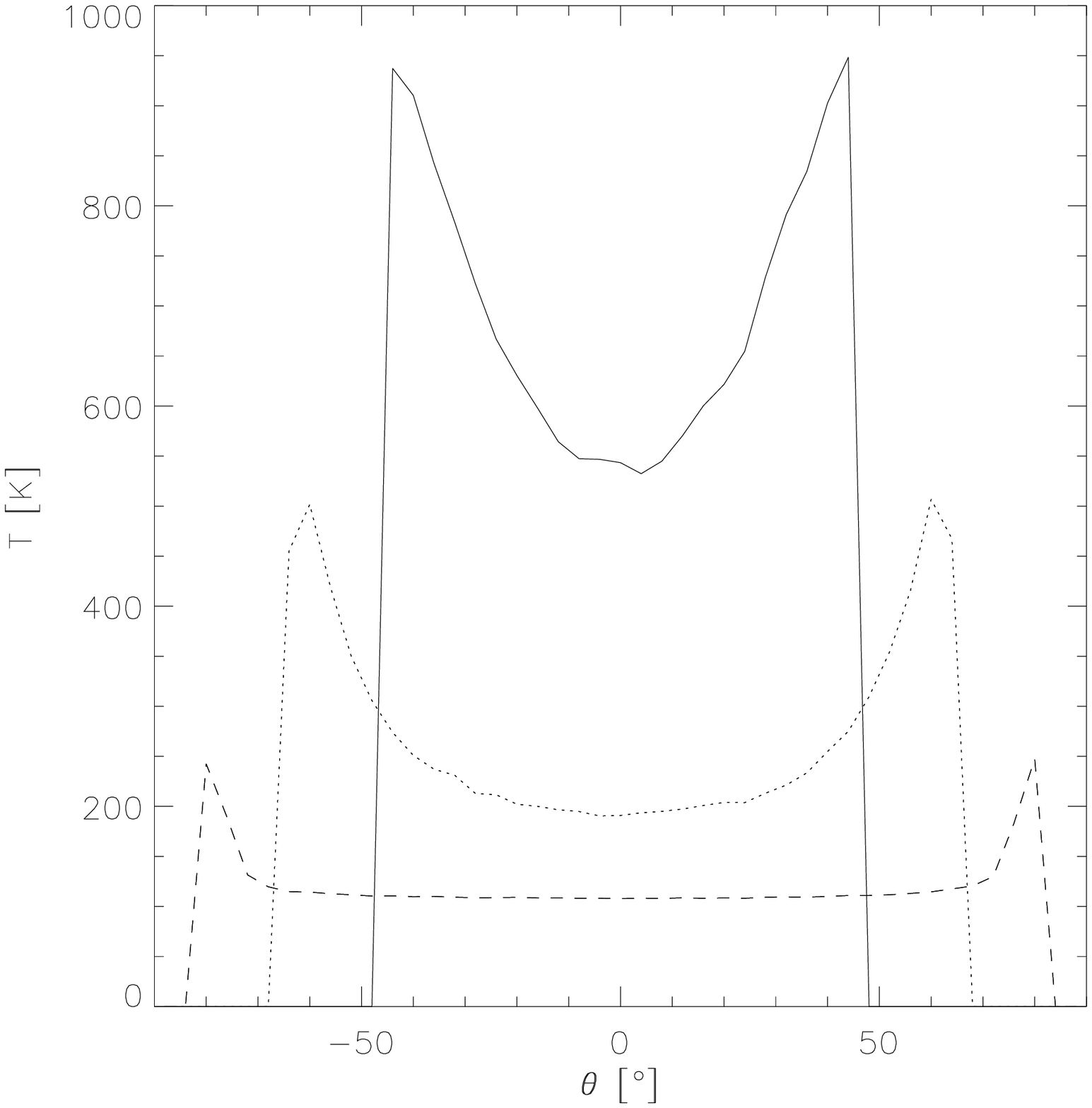}}
}
\caption{Temperature distributions of the smallest silicate grain component: {\bf a)} Radial temperature 
         distributions. The solid line corresponds to an equatorial line of sight,
         the dotted line to an inclination angle of $i=30\,\degr$ and the dashed line to 
         $i=20\,\degr$. {\bf b)} Temperature distributions in $\theta$-direction. 
	 The solid line corresponds to a radial
         distance to the centre of 2\,pc, the dotted line to 5\,pc and the dashed line 
         to 20\,pc.\label{fig:temp_small}}
\end{figure*}

\begin{figure*}
\centering
\mbox{ 
   \subfigure{
     \includegraphics[width=8.0cm,angle=90]{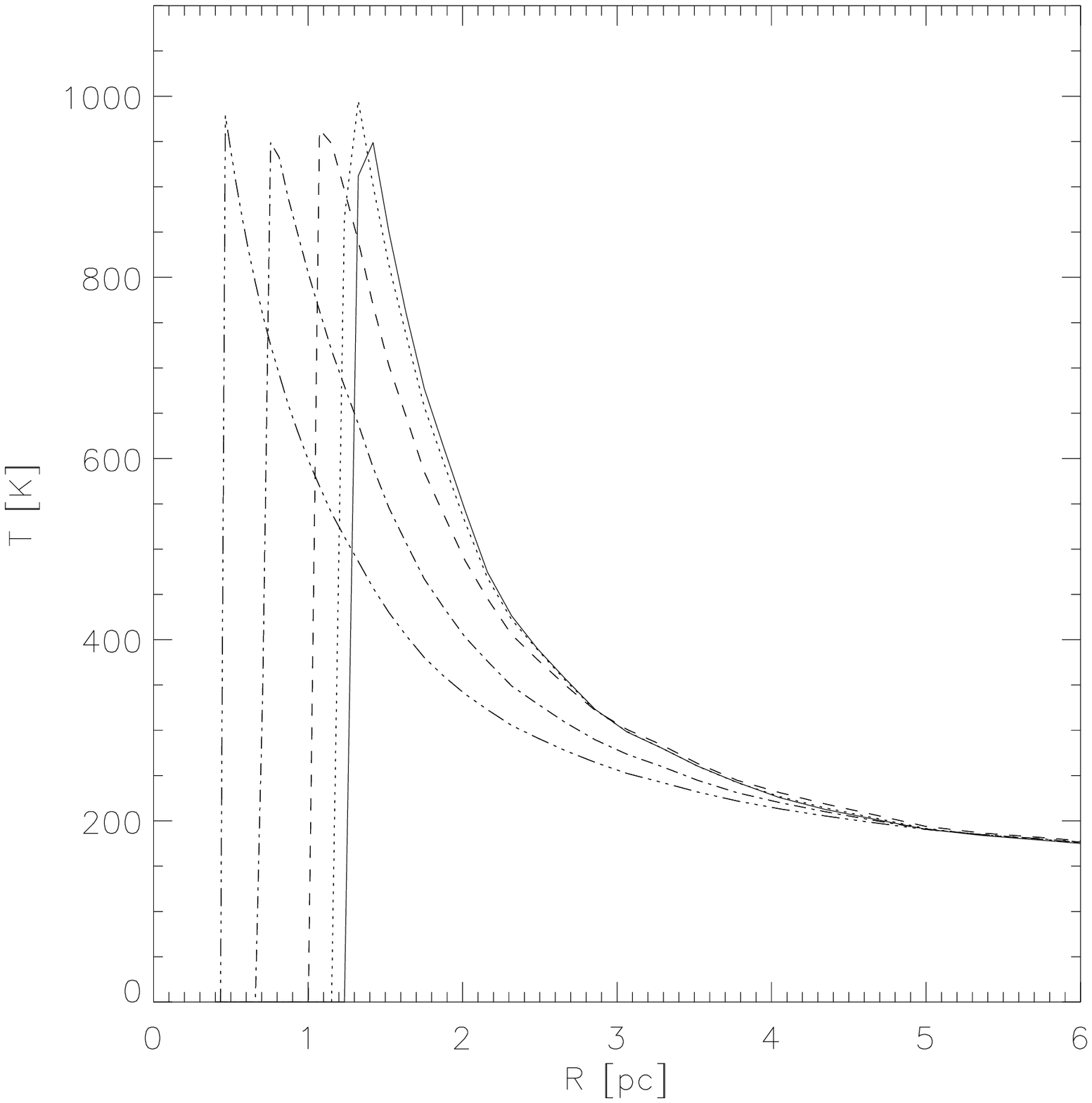}}
   \subfigure{
     \includegraphics[width=8.0cm,angle=90]{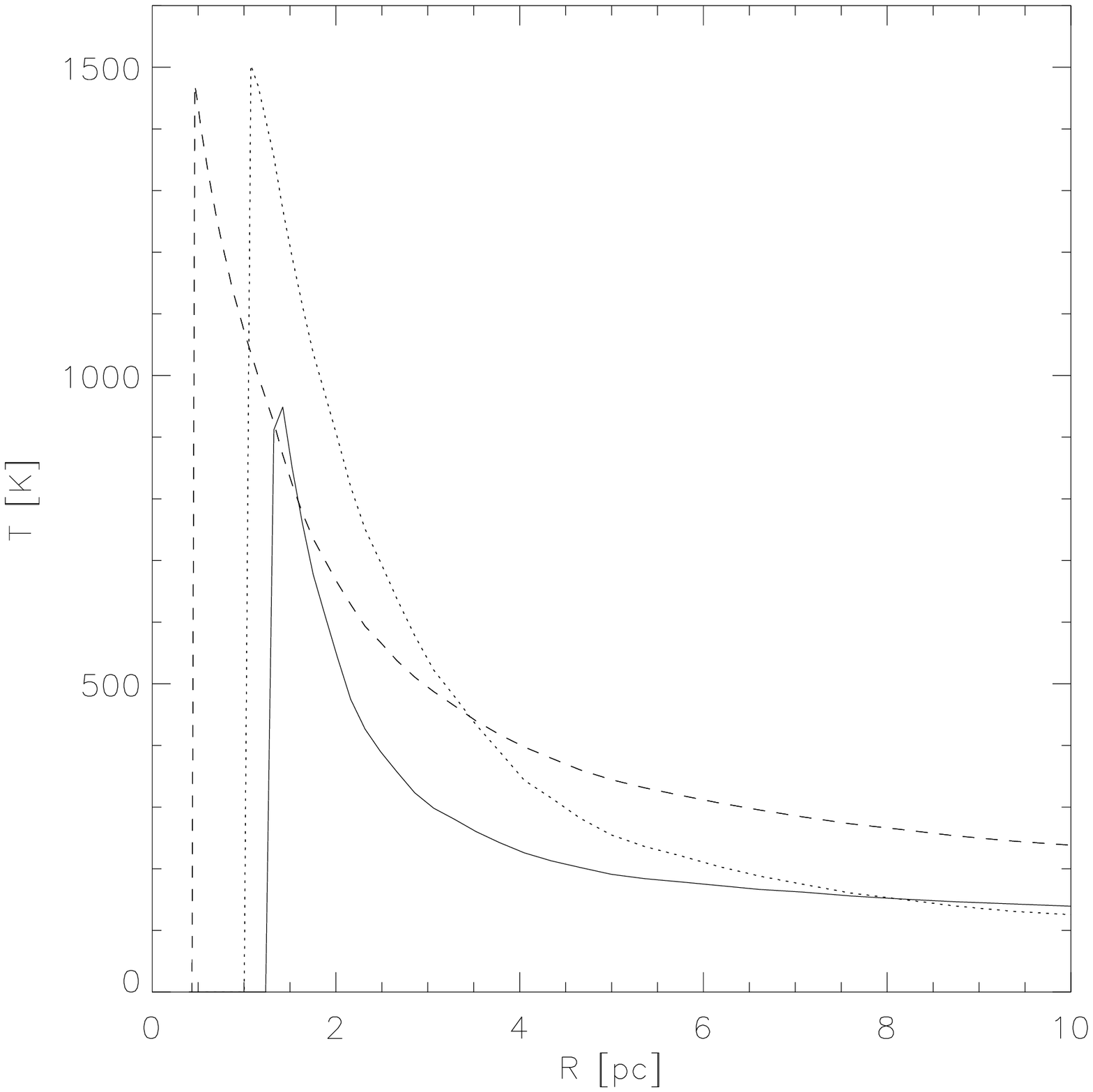}}
}
\caption{Comparison of the temperature distributions of different dust species within the equatorial plane:
         {\bf a)} Temperature distributions for different sized silicate grains: $a=0.005\,\muup$m (solid line), 
              $a=0.013\,\muup$m (dotted line), $a=0.035\,\muup$m (dashed line), $a=0.094\,\muup$m (dashed-dotted line), 
              $a=0.25\,\muup$m (dashed-dotted-dotted-dotted line). {\bf b)} Temperature distributions for 
	      different dust species: the smallest silicate component (solid line),
              the smallest graphite component with E $\parallel$ c (dotted line) and the smallest graphite component with E $\perp$ c
              (dashed line). \label{fig:dust_size}}
\end{figure*}

Fig.~\ref{fig:temp_small} shows the temperature distribution of our standard Seyfert model 
for the smallest silicate grain component, which is also the most abundant one 
(see Sect.~\ref{sec:dust}). In the left panel, radial temperature distributions for 
an equatorial line of sight (solid line), an inclination angle of $i\,=\,30\,\degr$ (dotted line) and 
$i\,=\,20\,\degr$ (dashed line) are plotted. With increasing inclination angle, 
the intersection of the respective 
line of sight with the torus moves to larger distances from the central source (compare to 
Fig.~\ref{fig:inclinationstudy}b). 
Therefore, the maxima of the temperature distributions 
move further out and the maximum temperature decreases as well.
The smaller the inclination the smaller the dust density at the point of the intersection of
the line of sight with the torus. This causes shallower temperature distributions with 
decreasing inclination.

The right panel of Fig.~\ref{fig:temp_small} shows various cuts in $\theta$-direction through 
the temperature distribution at radial distances from the centre of 2\,pc, 5\,pc and 20\,pc. 
A minimum curve is expected because of the direct illuminated -- and
therefore hotter -- funnel walls. In the outer part of the torus, the temperature contour lines 
get spherical symmetric, as can be seen by the flattening of the curve at 20\,pc distance from
the centre.

Fig.~\ref{fig:dust_size} displays the temperature distributions within the equatorial plane for 
different kinds of grains. In the left panel, the layering of different sized silicate grains 
can be read off. The smallest grains (given by the solid line in Fig.~\ref{fig:dust_size}a) 
are heated more efficiently and, therefore, their distribution 
possesses the largest sublimation radius. Fig.~\ref{fig:dust_size}b compares the temperature
distributions within the equatorial plane for the smallest grains of the three dust species. 
The solid line corresponds to silicate grains, the dotted and dashed curve to graphite with 
the two orientations of the polarisation vector relative to the optical axis. Silicate grains
are heated more efficiently than graphite grains.

\subsection{Inclination angle study for SEDs}
\label{sec:inc_sed}

\begin{figure*} 
\centering
\mbox{ 
   \subfigure{
     \includegraphics[width=8.5cm,angle=90]{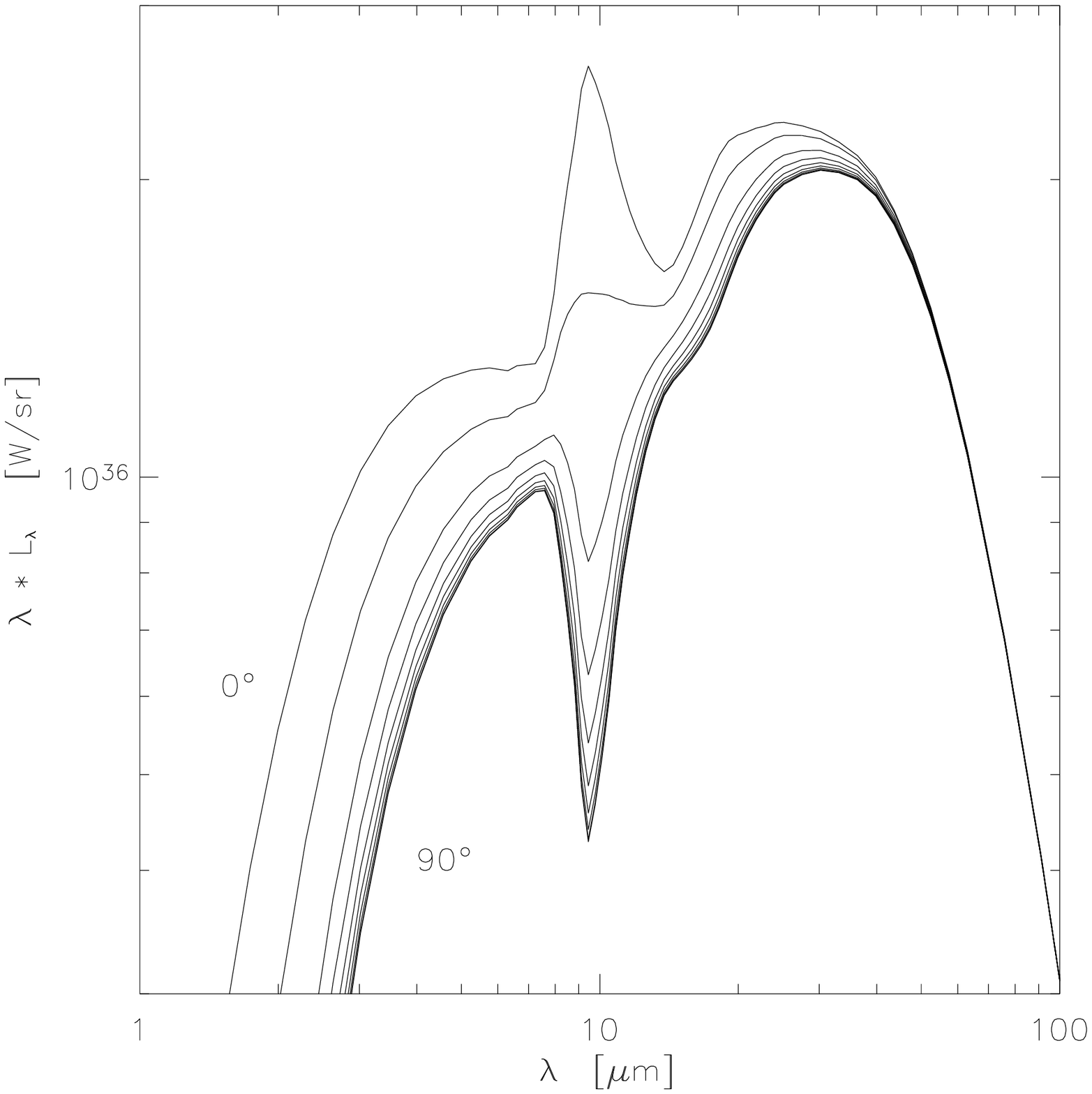}}
   \subfigure{\includegraphics[width=8.5cm]{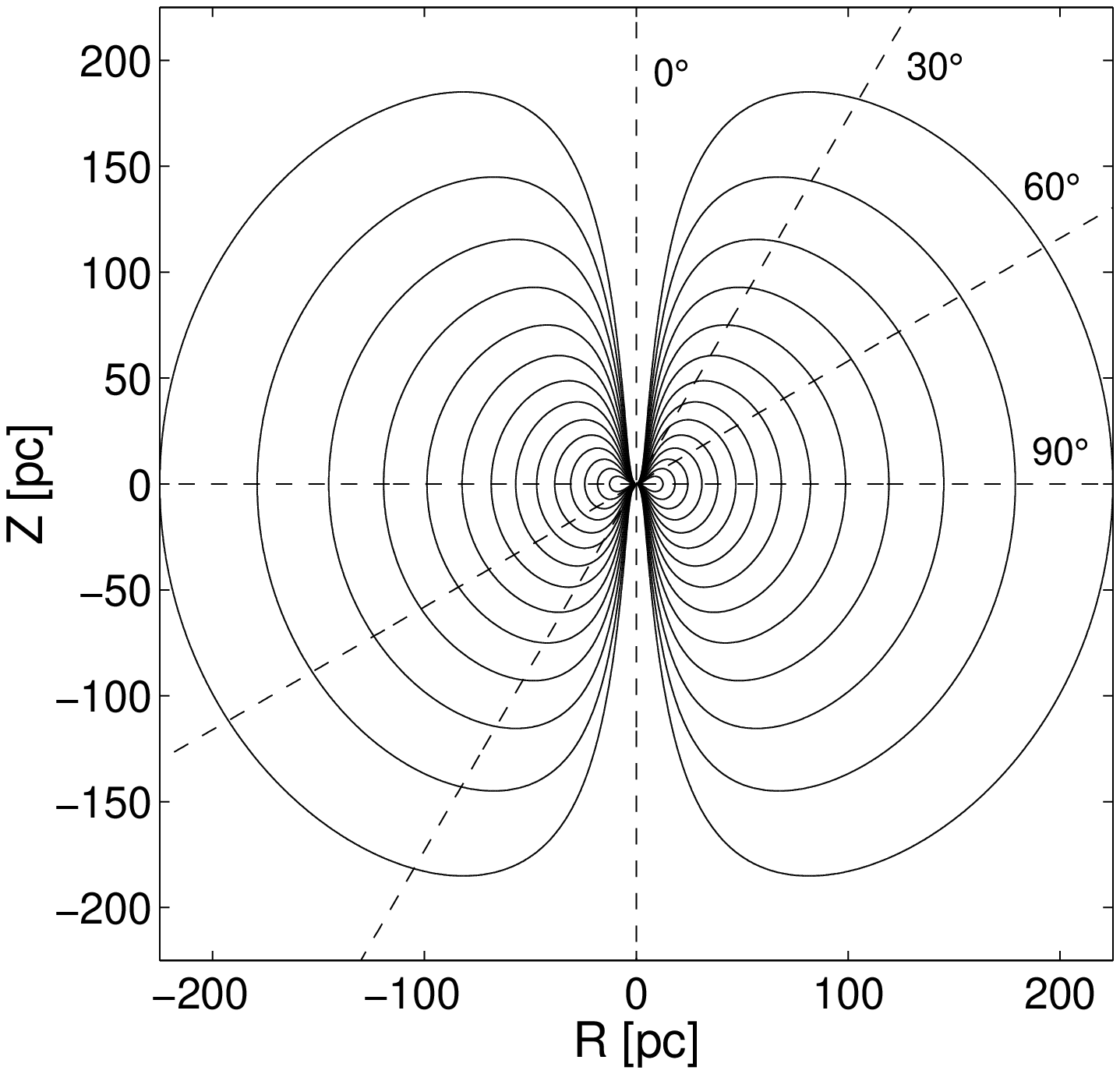}}
}
\caption{Inclination angle study for our standard model: {\bf a)} SEDs of the dust re-emission. 
         The inclination varies from 0\,\degr\, (upper curve) to 90\,\degr\, (lowermost curve) 
	 in steps of 10\,\degr\,. {\bf b)} Illustration of some inclination angles of lines of 
	 sight frequently used for our standard model.\label{fig:inclinationstudy}}
\end{figure*}

In Fig.~\ref{fig:inclinationstudy}a we show an inclination study for
SEDs. The uppermost curve belongs to an angle of 0\,\degr, which
corresponds to an extreme Seyfert\,I case, where we see the torus face-on. Typical 
inclination angles of Seyfert\,I galaxies are between 10\,\degr and 20\,\degr. Then, the
inclination angle is increased up to 90\,\degr, the edge-on case, in steps of
10\,\degr. Some of the inclinations are visualised in Fig.~\ref{fig:inclinationstudy}b for the case
of our standard model. 
Only dust re-emission SEDs are shown, which result in the so-called IR bump. It arises from dust at different 
temperatures within the toroidal distribution.
The two characteristic silicate features at $9.7\,\muup$m and $18.5\,\muup$m are present in the SED. 
The $18.5\,\muup$m feature is less pronounced and partially hidden in the global maximum of the IR bump.
Just looking at the
continuum spectrum and neglecting the silicate bands for a second, an offset in
the flux is visible. This offset (explained in Sect.~\ref{sec:av_dust} and visualised in 
Fig.~\ref{fig:dust_av_single}) is again due to the offset in the extinction
curve (see Fig.~\ref{fig:input}a) and especially visible for the case of silicate grains. 

For $i=0\,\degr$, we directly see the torus
and the primary source and, therefore, expect the silicate feature in emission. 
Moving to higher inclination angles, the extinction
to occur along the line of sight, mainly caused by cold dust in the outer part of the torus,
increases. Less and less of the hotter, directly illuminated parts of the torus
can be seen directly. Therefore, the silicate feature changes from emission to absorption. 
With rising extinction along the line of sight,
the increase of the SEDs at small wavelengths is shifted to longer
wavelengths and the IR bump gets more and more narrow. The increase at small
wavelengths can be explained by a Planck curve with the highest occuring dust 
temperatures (at the very inner end of the cusp) altered by rising extinction along the line of sight.
For our modelling, the torus gets optically thin between 40 and
$50\,\muup \mathrm{m}$ (due to the steep decrease of the extinction curve, compare to 
Fig.~\ref{fig:input}a). Therefore, all curves coincide and the dust configuration
emits radiation isotropically for longer wavelengths.

When we look at the extinction in the visual wavelength range plotted against
the inclination angle, $A_{\mathrm{V}}$ reaches about 5 at an
inclination angle of 10\degr. This means that the opening angle of our torus model is
too small compared to simple estimates of the torus opening angle using the ratio between 
Seyfert I and II galaxies, e.\,g.\,\citet{Osterbrock_88} find a half opening angle 
between 29\degr and 39\degr.

\subsection{Inclination angle study for surface brightness distributions}
\label{sec:inc_map}

\begin{figure*} 
\centering
\mbox{ 
   \subfigure{
     \includegraphics[width=4.0cm,angle=90]{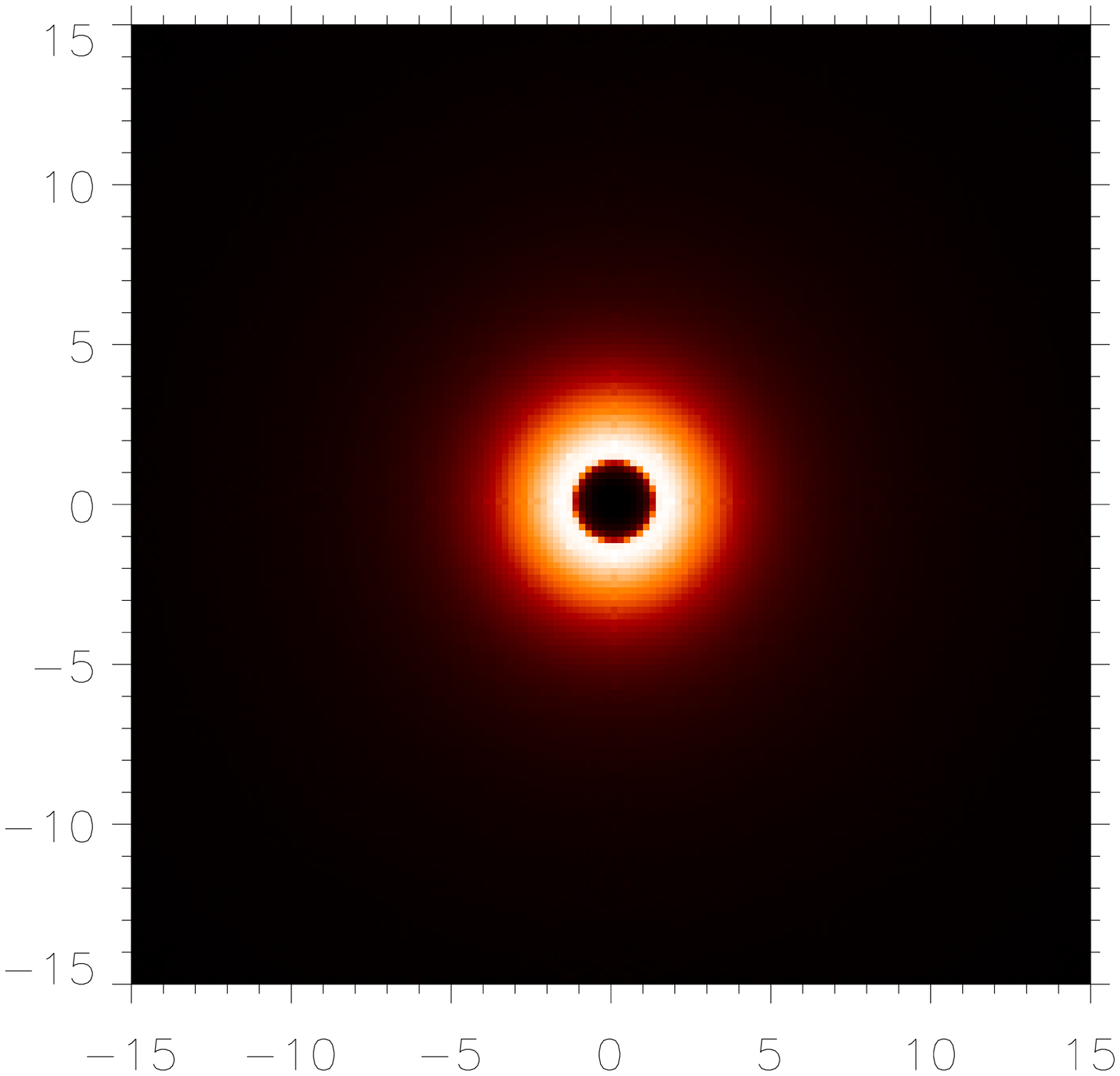}
   }
   \subfigure{
     \includegraphics[width=4.0cm,angle=90]{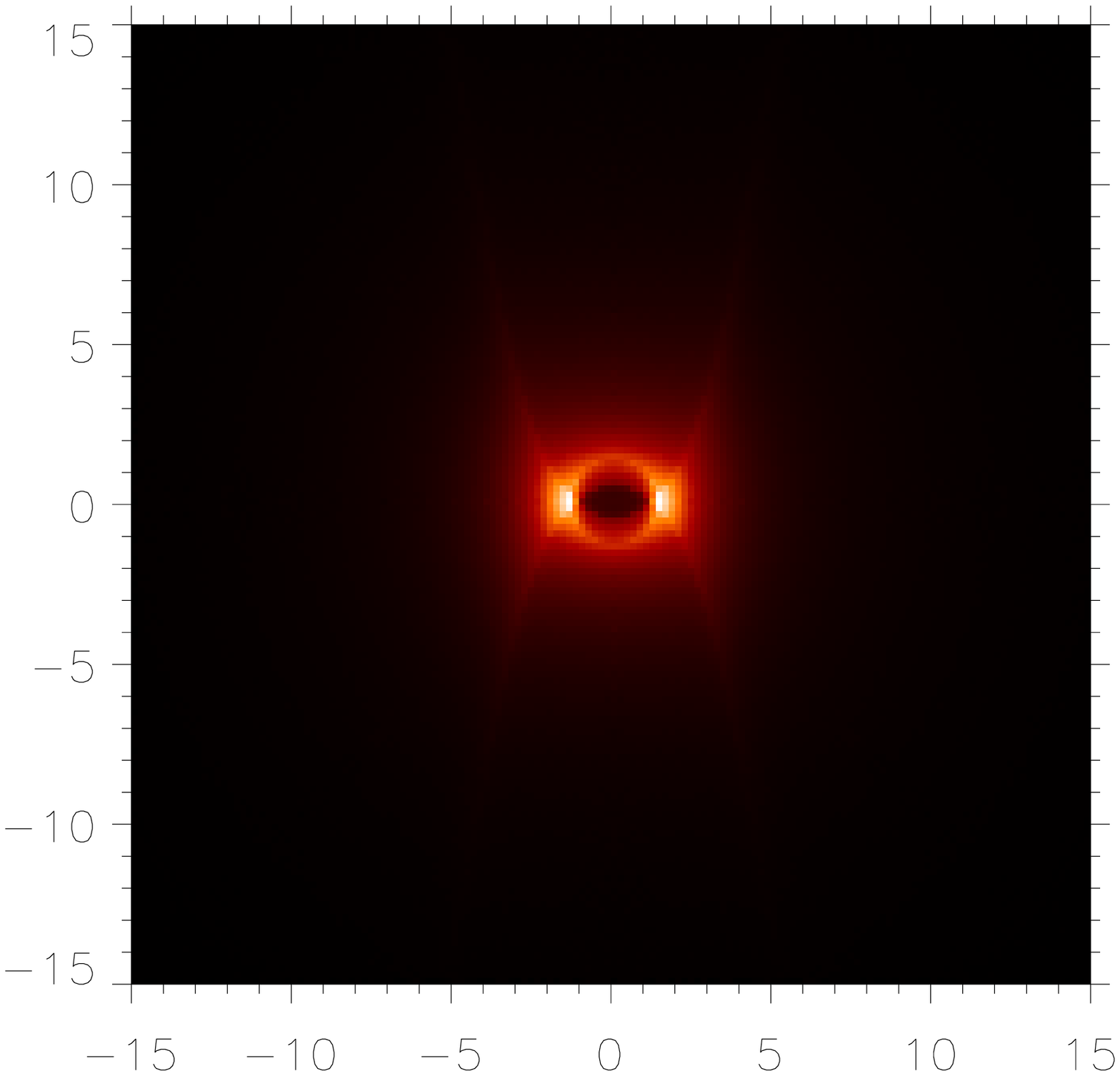} 
   }
   \subfigure{
     \includegraphics[width=4.0cm,angle=90]{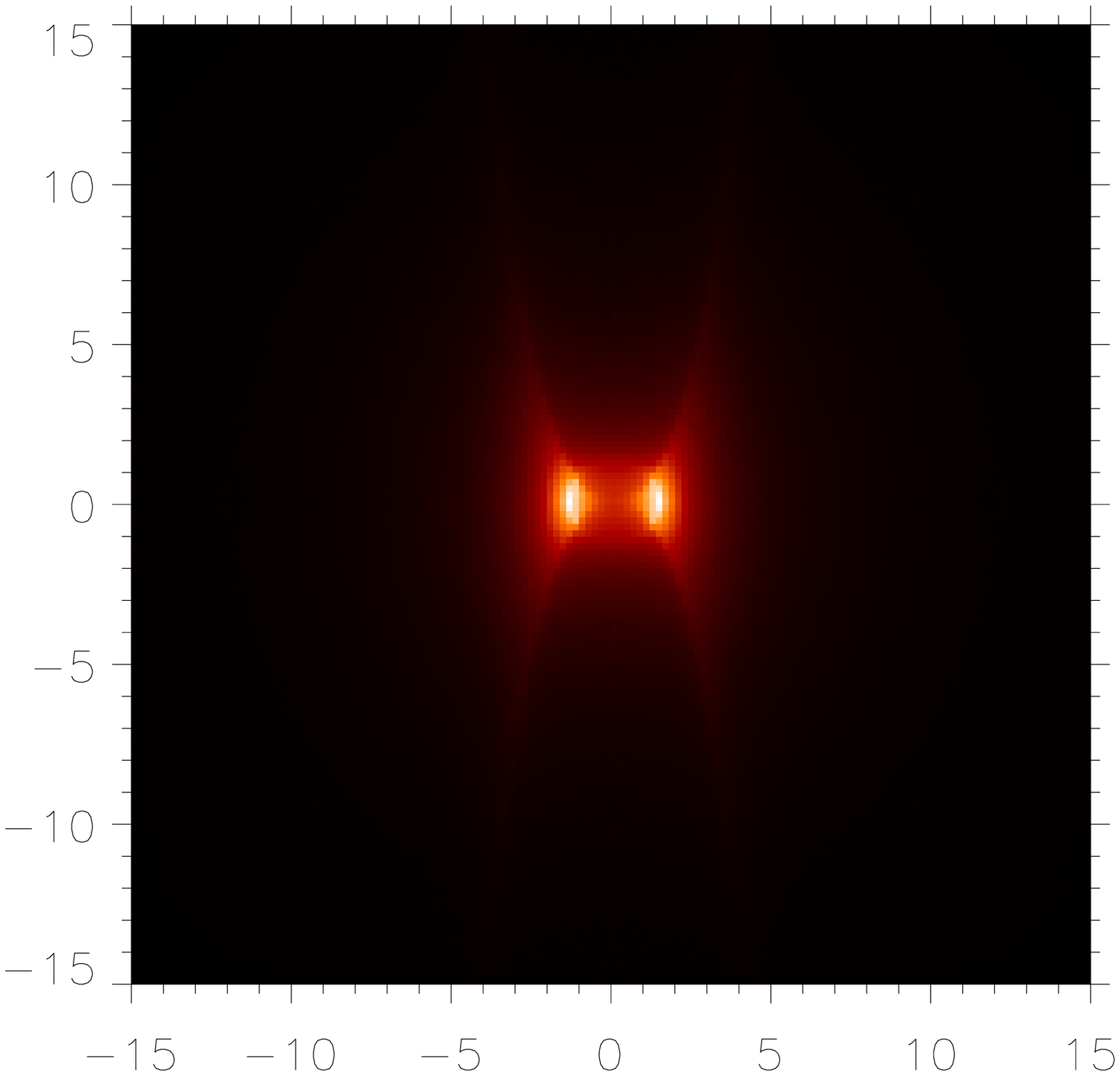}
   }
   \subfigure{
     \includegraphics[width=4.0cm,angle=90]{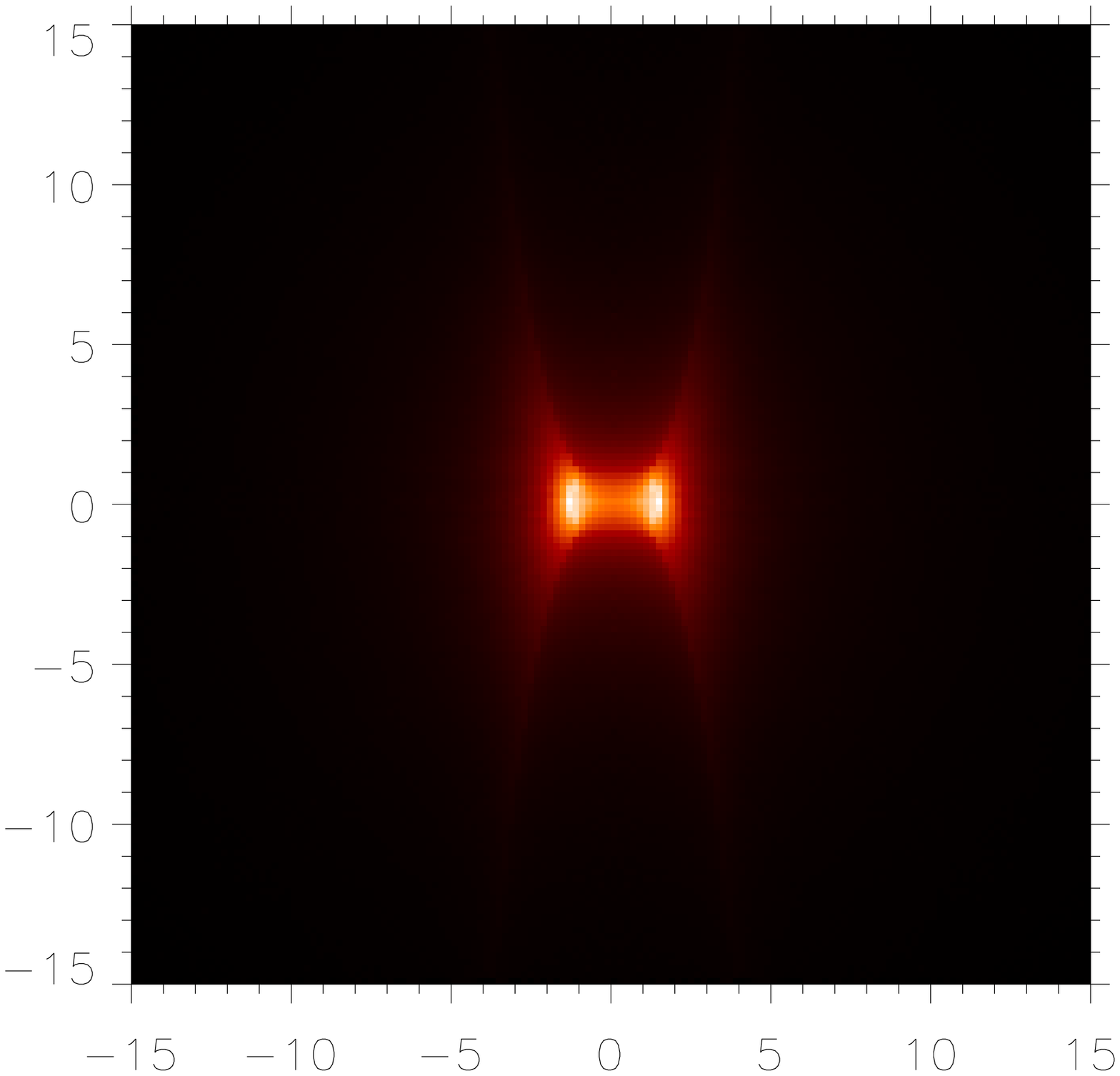} 
   }
}
\mbox{ 
   \subfigure{
     \includegraphics[width=4.0cm,angle=90]{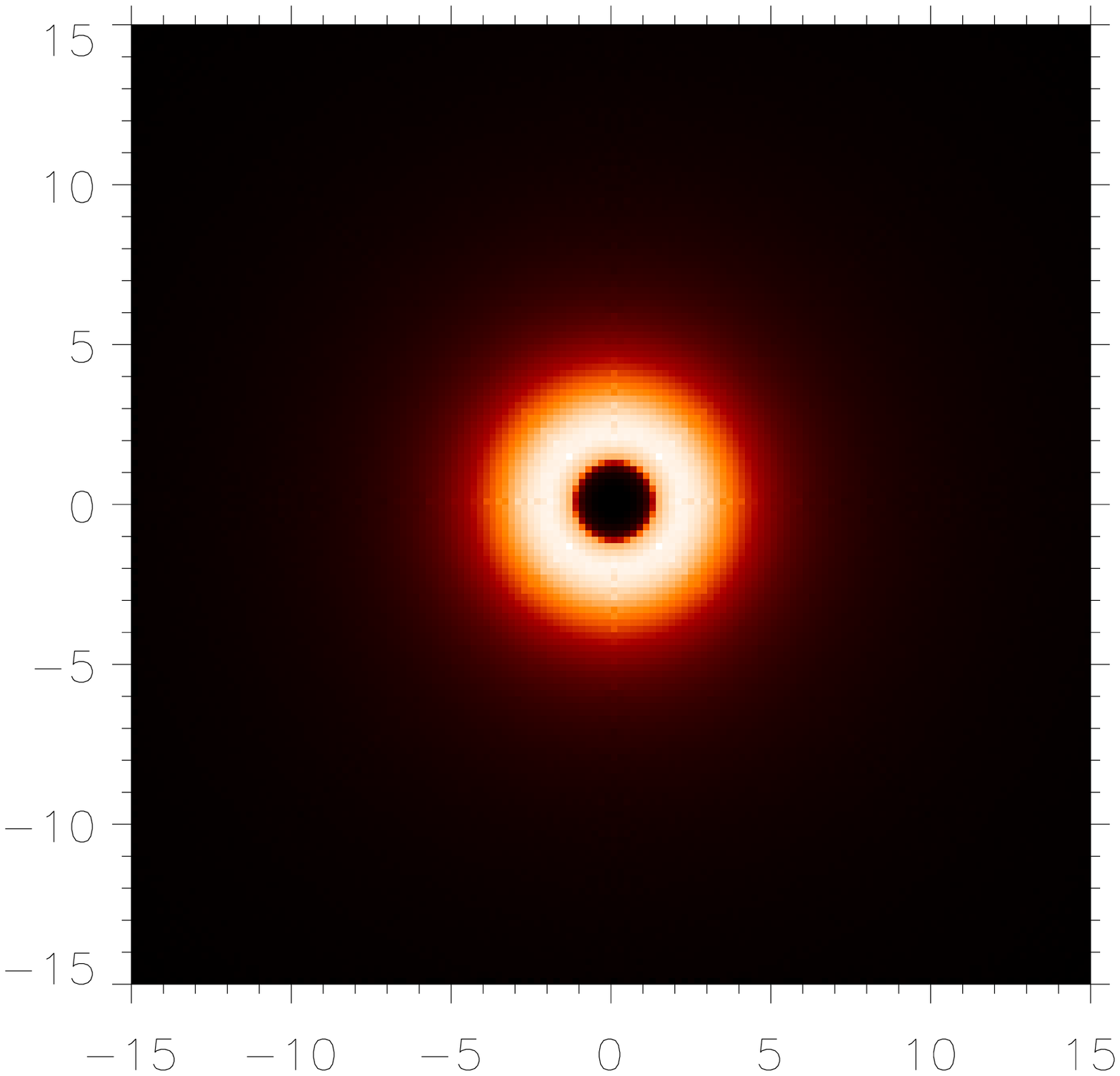}
   }
   \subfigure{
     \includegraphics[width=4.0cm,angle=90]{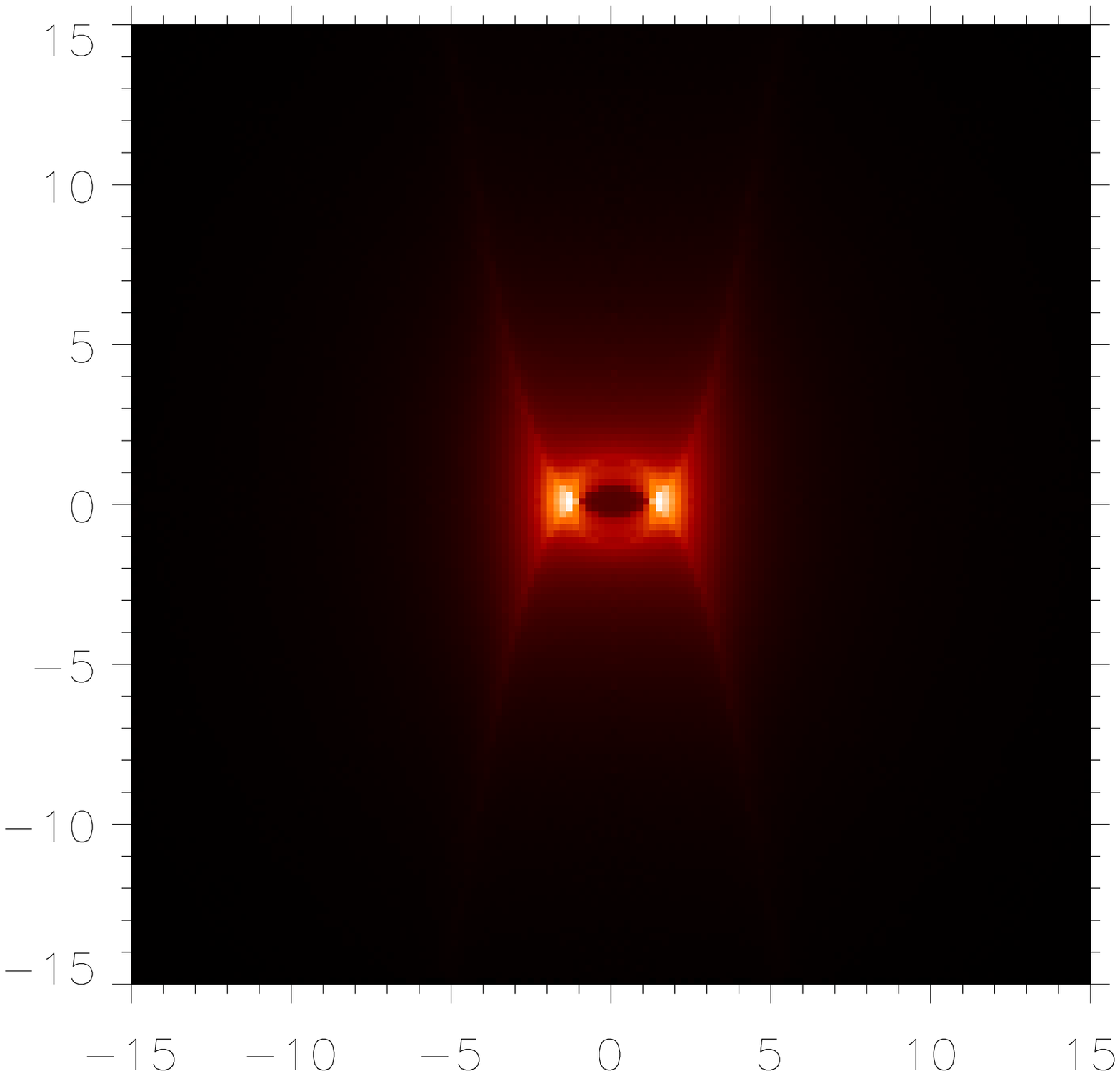} 
   }
   \subfigure{
     \includegraphics[width=4.0cm,angle=90]{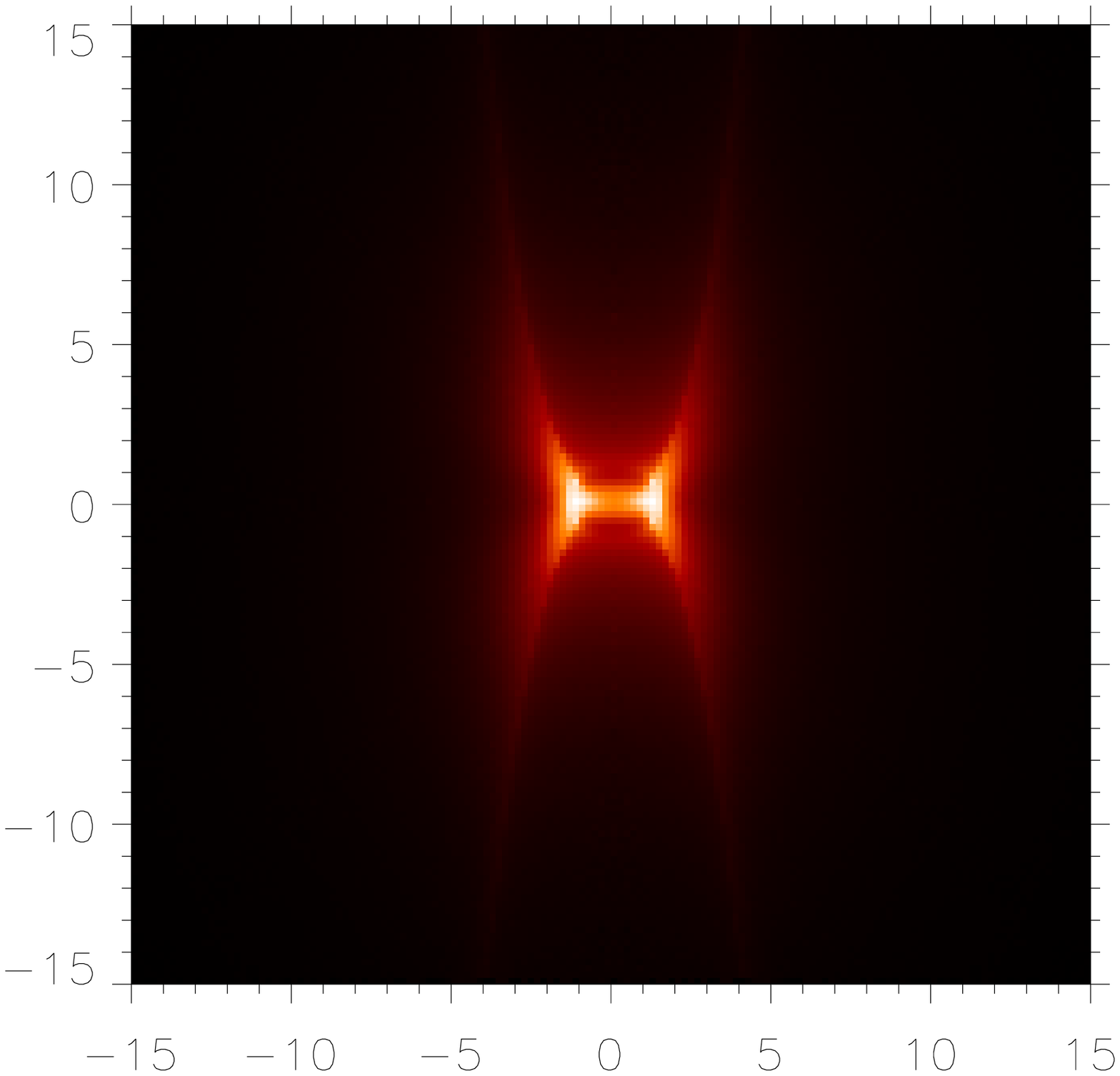}
   }
   \subfigure{
     \includegraphics[width=4.0cm,angle=90]{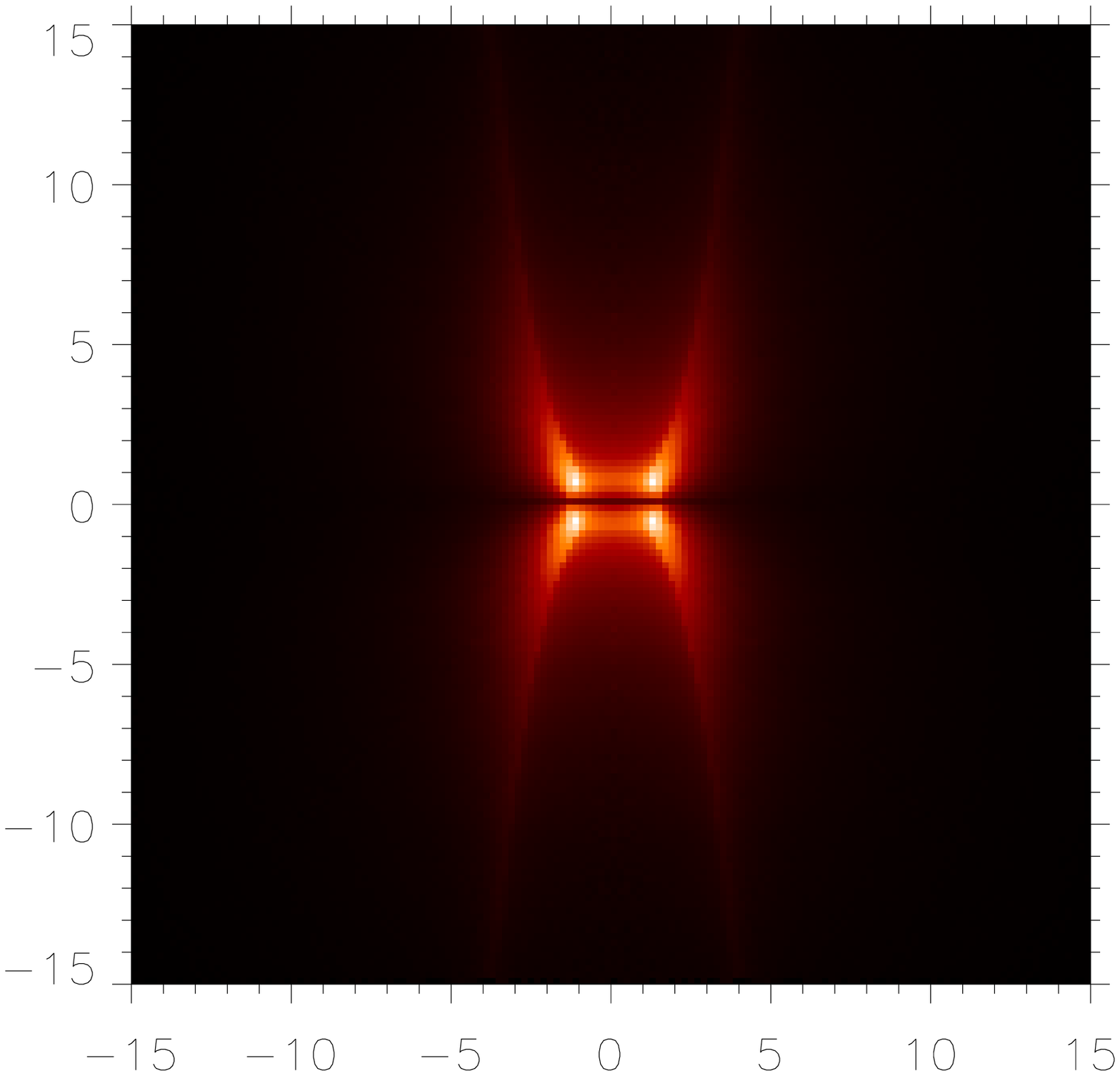} 
   }
}
\caption{Surface brightness distributions of our standard model with an
  isotropically radiating central source (first row)
  and with a $\left|\cos\theta\right|$ - radiation characteristic (second row) at inclination
  angles of 0\,\degr, 30\,\degr, 60\,\degr\, and 90\,\degr\, given with a
  linear colour
  scale ranging from 0 to the maximum value of the respective image.} 
\label{fig:inc_map}
\end{figure*}

In the following section, an inclination study for surface brightness
distributions is presented. The underlying model is our standard model introduced earlier. 
As wavelength we chose $\lambda$\,=\,$13.18\,\muup \mathrm{m}$, which lies in the
continuum spectrum outside the silicate feature and still belongs to the
wavelength range covered by MIDI \citep{Leinert_03}. Using the Wien law as a rough estimate of the
temperature, where the dust emits the maximal radiation at this wavelength, we
get about 230\,K. Looking at the steep radial temperature distributions of the various
grains (see Fig.~\ref{fig:dust_size}b), we expect the radiation to come from the direct vicinity of the
inner radius of the torus. The resulting surface brightness distributions for inclination
angles of 0\,\degr, 30\,\degr,
60\,\degr\, and 90\,\degr\, are shown in the upper row of Fig.~\ref{fig:inc_map}. The images are given in a linear
colour scale emphasising areas of maximum surface brightness. The axis labeling denotes 
the distance to the centre of the AGN in pc. When we consider flux scales, the object is
always assumed to be at a distance of 45\,Mpc, which is a typical value for Seyfert galaxies, 
which will be observed with MIDI.

Looking at the image with an inclination angle of 90\,\degr (the last image in the first row of
Fig.~\ref{fig:inc_map}), the most central part of the
torus funnel is visible. As already mentioned above, the dust
species sublimate at different distances from the central source. Within these
inner radii, no such grains can survive. Therefore, a layering of dust grains
exists in the vicinity of the cusp. This is partly visible here. While the
very inner end of the torus appears quite faint due to the depletion of many
grain species (especially the most abundant small silicate grains and also
the smallest graphite grains),   
a semilunar feature is visible further out. At this point, the dust density
increases strongly, because the sublimation radii of small silicate grains are
reached. They provide the highest mass fraction of all dust species (for this
dust model, 
the two smallest silicate grain populations carry about 43\% of the total mass). 
Apart from this, an x-shaped emission area appears. Here we see the directly
illuminated funnel walls. As the dust density peaks further out at a distance
of 5\,pc in the equatorial plane, the dust density next to the funnel (outer
equipotential lines) is quite low. Hence, only an x-shaped feature remains from
the emission of the walls,
which arises from a summation effect along the lines tangential to these
walls of the funnel. 
The two maxima are connected via an emission band, which gets fainter and more
narrow towards the centre, again due to a summation effect, which is stronger
for the case of a line of sight tangential to the ring than perpendicular.\\ 
At an inclination angle of 60\,\degr, the same features remain, but the two
semilunar areas of maximal emission get connected via thin and faint bows, due
to the different line of sight to the ring-like structure. 
This effect further increases at $i$\,=\,$30$\,\degr. Now, the whole ring of the
main dust grain contributors is visible. We see here the intersection of the
sublimation sphere of these kind of grains with the torus. 

In the face-on case ($i$\,=\,$0$\,\degr), we get the expected ring-like structure. In
the central area no emission is present, because in our modelling the funnel of
the dusty torus itself is free of any material. The
maximum surface brightness does not result from the innermost part of the
torus -- although these dust grains possess the highest temperatures -- 
because of the summation effect already mentioned above. Only very little
dust can exist here because of the density model itself and because the
sublimation radii of the main dust contributors lie further out. Summing up
along the line of sight results in the shown distribution of surface
brightness.

\subsection{Wavelength study for surface brightness distributions}

%alles mit tmean009 \\
%Projekte:  - mmean013 (7.244 mu) --> nr. 42 (Kontinuum, moeglichst nahe bei 8mu) \\ 
%           - mmean014 (9.772 mu)         50 \\
%           - mmean017 (13.18 mu)         58  hinter feature (Minimum)\\
%           - mmean016 (30.2  mu)         74  Maximum fuer den Fall groesserer Inklinationen\\

%Max(0 degrees) = 0.0076933616    \\
%Min(0 degrees) = 0.0076933616e-5 \\
%Max(90 degrees) = 0.0021265470 \\
%Min(90 degrees) = 0.0021265470e-5 \\

\begin{figure*} 
\centering
\mbox{ 
   \subfigure{
     \includegraphics[width=4.0cm,angle=90]{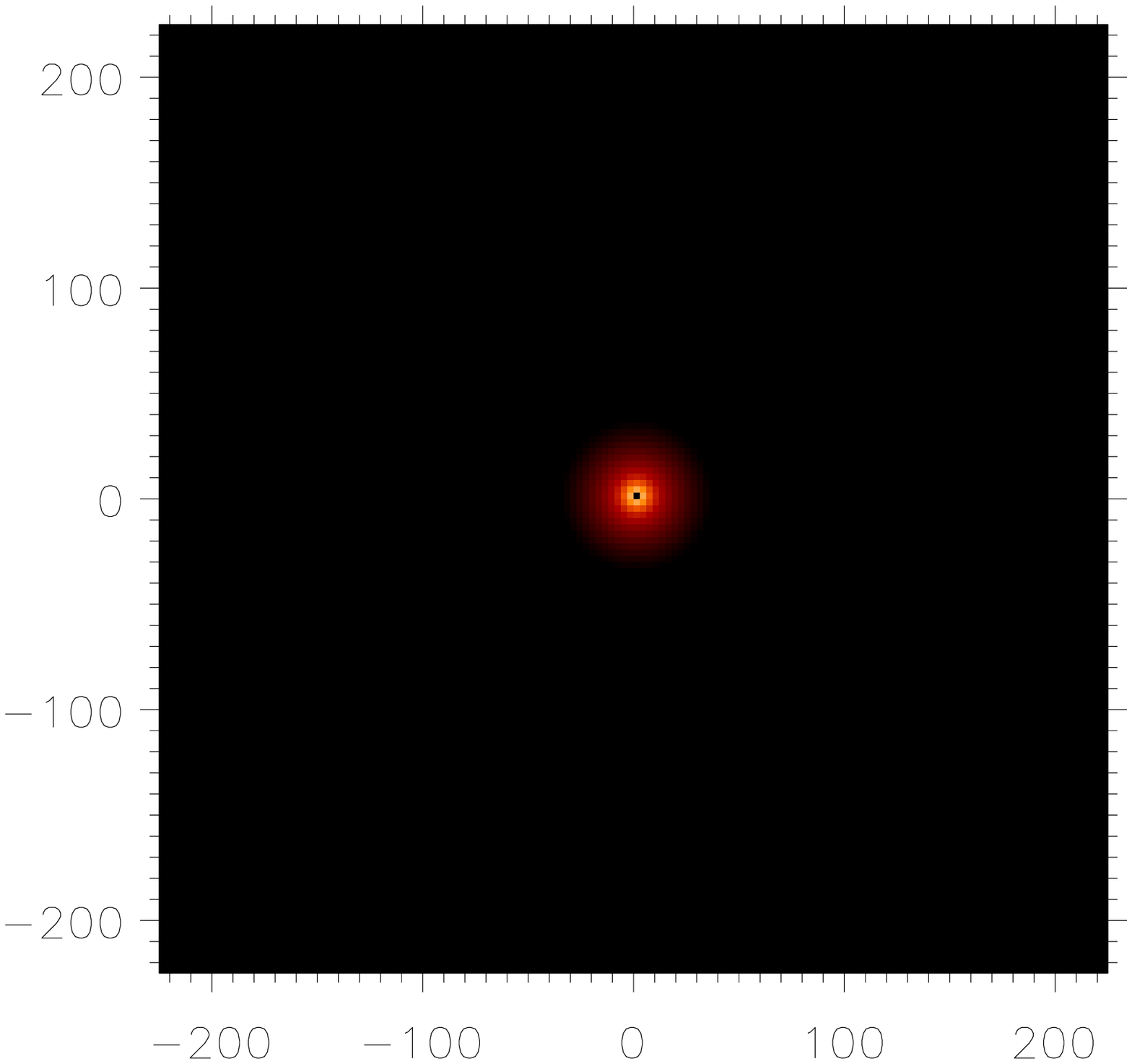}
   }
   \subfigure{
     \includegraphics[width=4.0cm,angle=90]{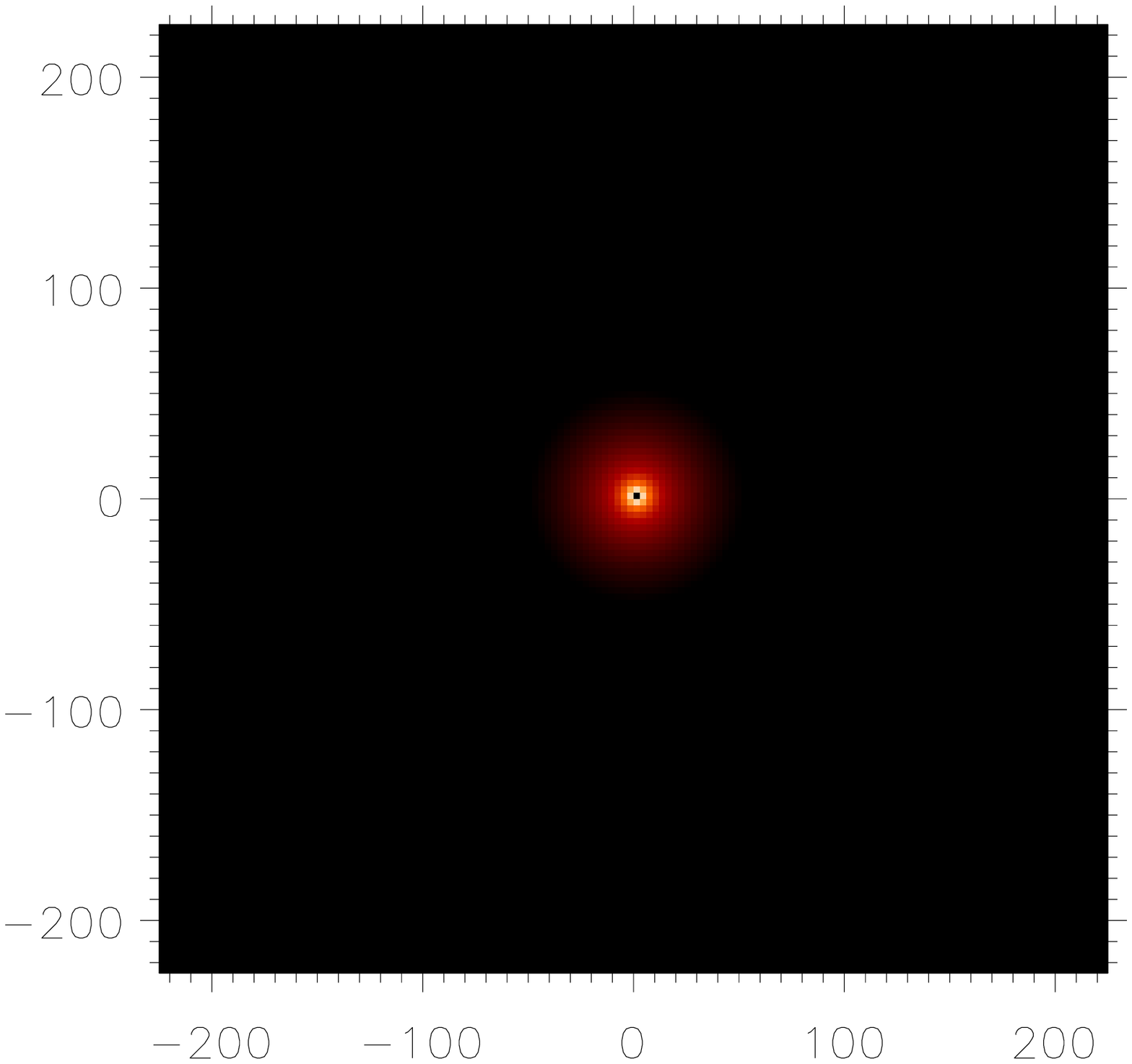} 
   }
   \subfigure{
     \includegraphics[width=4.0cm,angle=90]{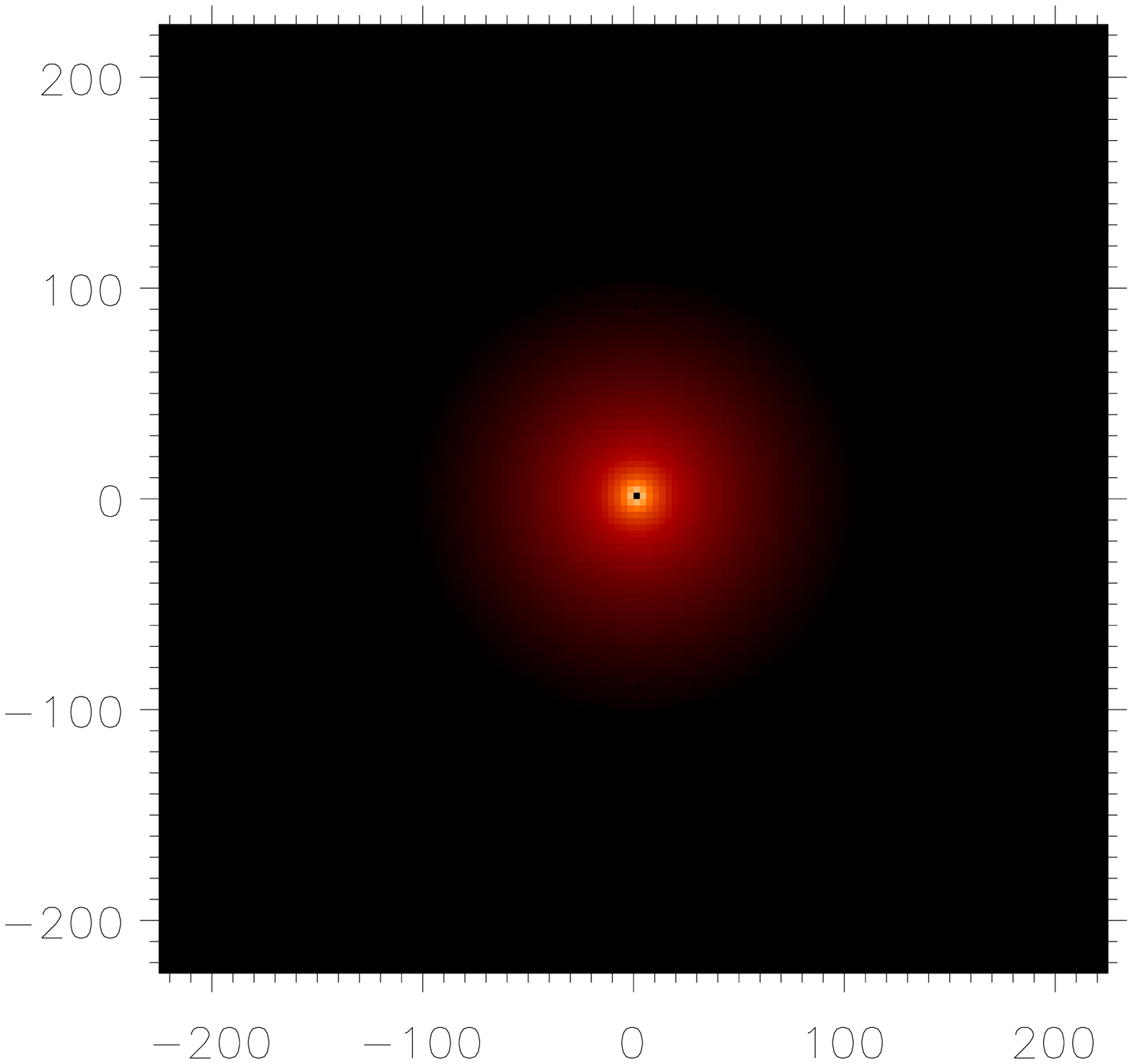} 
   }
   \subfigure{
     \includegraphics[width=4.0cm,angle=90]{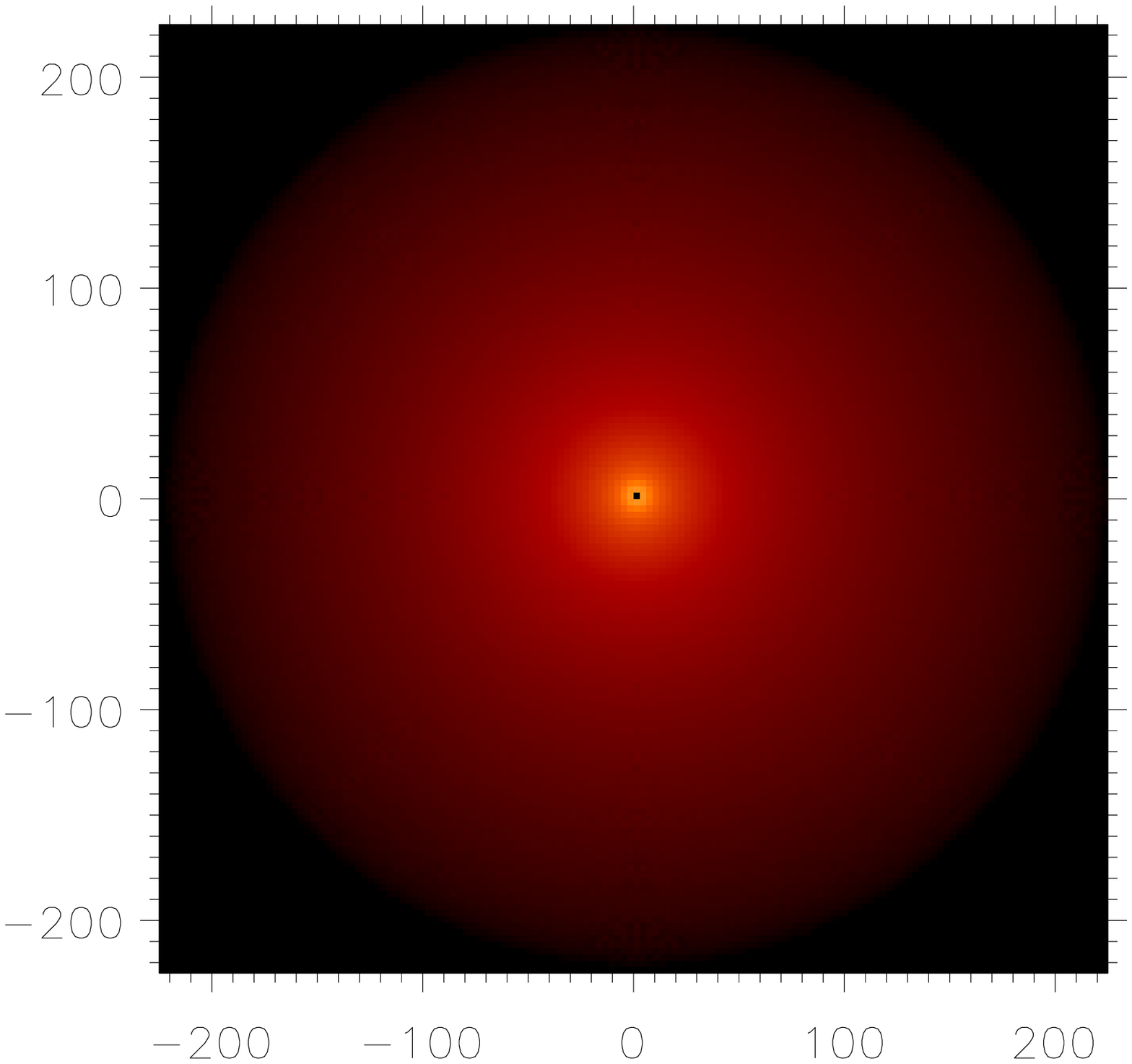} 
   }
}
\mbox{ 
   \subfigure{
     \includegraphics[width=4.0cm,angle=90]{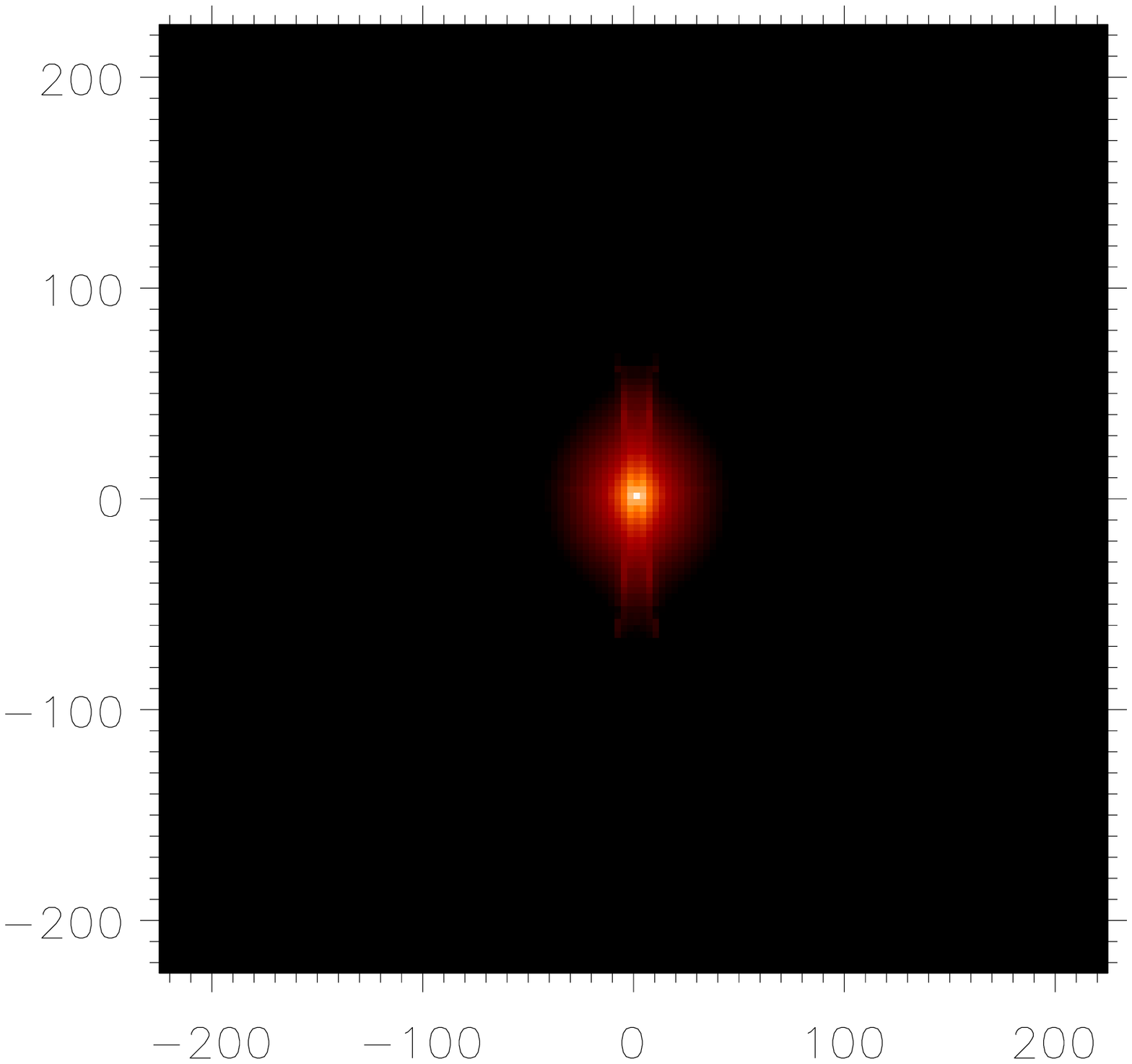}
   }
   \subfigure{
     \includegraphics[width=4.0cm,angle=90]{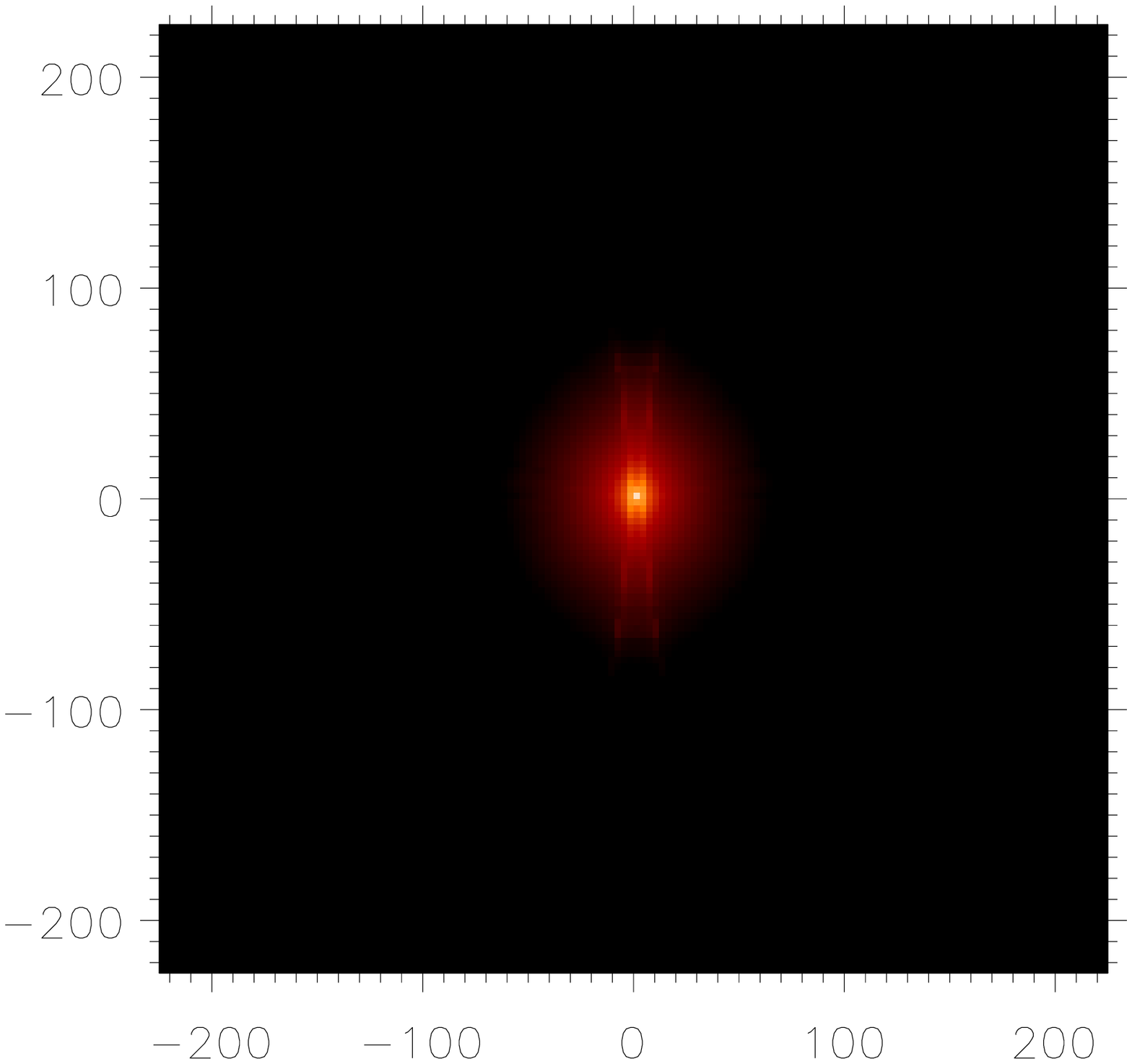} 
   }
   \subfigure{
     \includegraphics[width=4.0cm,angle=90]{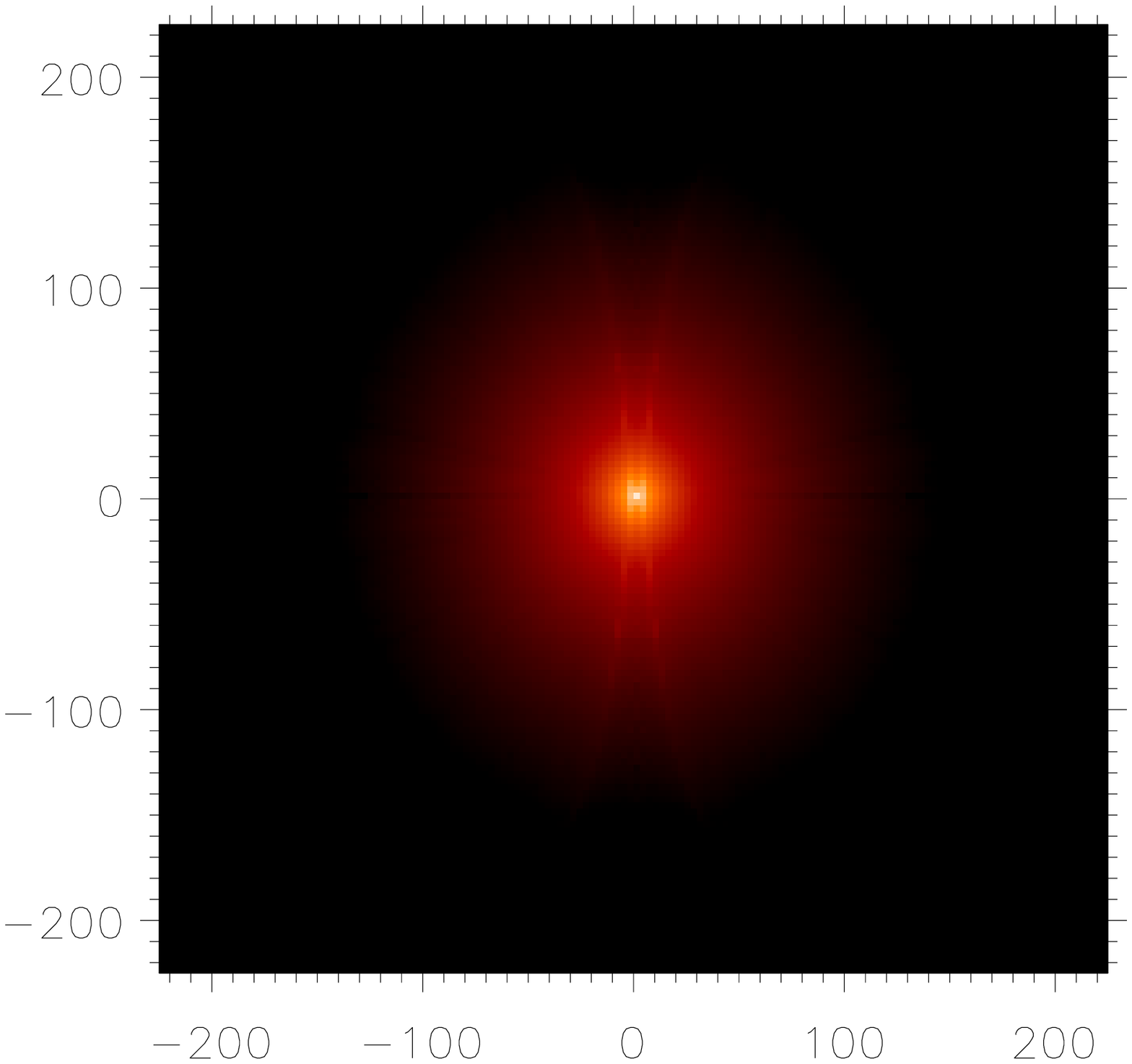} 
   }
   \subfigure{
     \includegraphics[width=4.0cm,angle=90]{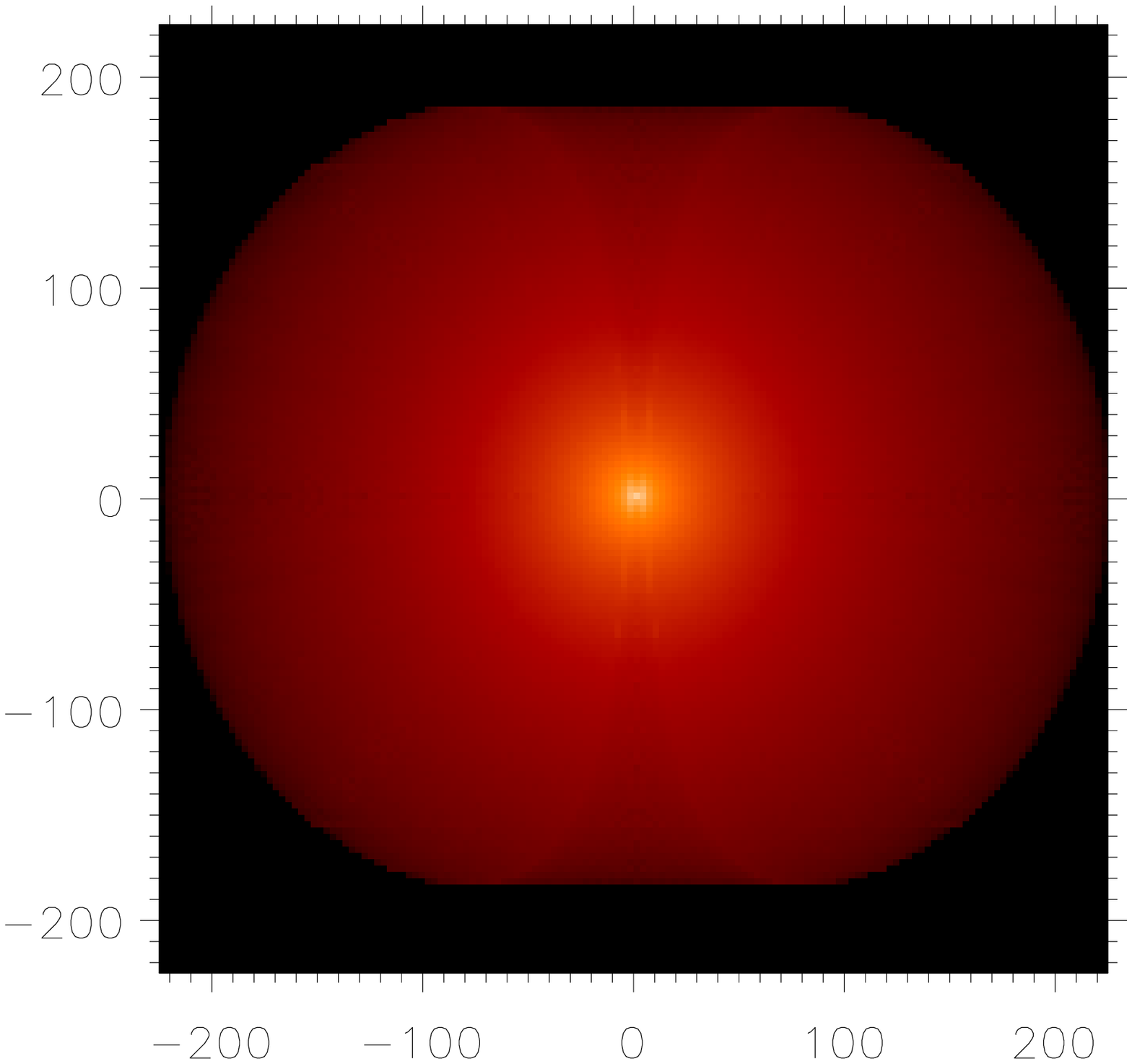} 
   }
}
\caption{Surface brightness distributions of our standard model for different
  wavelengths from left to right ($7.22\,\muup \mathrm{m},\,9.7\,\muup \mathrm{m}, \,13.18\,\muup
  \mathrm{m}, \,30.2\,\muup \mathrm{m}$). Maps are
  given in logarithmic scale for inclination angles of $i$\,=\,$0\,\degr$ (upper row)
  and $i$\,=\,$90\,\degr$ 
  (lower row). See text for a detailed description.} 
\label{fig:wavel_map}
\end{figure*}

In Fig.~\ref{fig:wavel_map} we show surface brightness distributions at
different wavelengths for the face-on and the edge-on view on our mean Seyfert 
torus model. Images for the same inclination angles are given with the same 
logarithmic colour scale and a dynamic range between the
global maximum at the respective inclination angle and $10^{-3}\%$ of this value.
Again, only re-emission spectra of the dust configuration are shown. The
direct radiation of the energy source is neglected. When we would take this
into account, the centre would always yield the highest brightness in the surface brightness
distribution.
The wavelengths are chosen as follows:
We start with $7.22\,\muup \mathrm{m}$, located outside the silicate feature at $9.7\,
\muup \mathrm{m}$, but closest to $8\,\muup \mathrm{m}$, where the wavelength range of the MIDI 
\citep{Leinert_03}
instrument starts. The next is the wavelength of the feature itself, then
between the two silicate features (at $13.18\,\muup \mathrm{m}$) at the upper end of the 
MIDI wavelength range and
finally we chose the overall maximum of the SEDs ($30.2\,\muup$m).

At wavelengths smaller than approximately $2\,\muup \mathrm{m}$, re-emission of dust can be
neglected and pure emission of the central source remains, attenuated by the dust. 
The surface brightness distributions at varying wavelengths highlight
explicitly the different temperature domains. According to {\sc Wien}'s law, the
maximum surface brightness at a wavelength of about $7\,\muup \mathrm{m}$ is emitted by dust of a
temperature of around $400\,$K. This area is concentrated to the innermost part
of the dust distribution, caused by the very steep radial temperature
curve there. Dust heated to about $100\,$K -- which is reached in
the flattening part of the temperature distribution -- has its maximum radiation at a wavelength of
around $30\,\muup \mathrm{m}$. Therefore, emission can be seen from a large fraction of the torus
volume. The same behaviour is also visible in the case of an inclination of
$90\,\degr$. While for small wavelengths basically the x-shaped structure
is visible, at longer wavelengths one can see the whole torus. This is also
due to the fact that the extinction curve drops off very steeply and,
therefore, we expect emission from all over the torus body.

\subsection{Implementation of a radiation characteristic}

\begin{figure}
  \resizebox{\hsize}{!}{\includegraphics[angle=90]{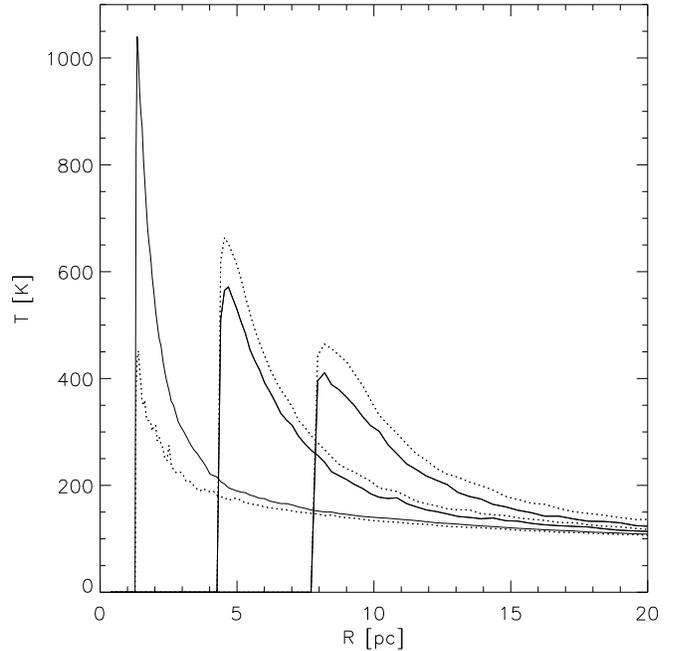}} 
  \caption{Comparison of the temperature distribution of our standard model (solid lines) 
  with the model with an anisotropically radiating accretion disc (dotted lines). 
  The respective pairs of radial temperature distributions for the smallest silicate 
  grain component are shown.
  From left to right: within the equatorial plane, for an inclination angle of 30\degr and
  an inclination angle of 20\degr.}   
  \label{fig:radchar_temp} 
\end{figure}

\begin{figure*}
\centering
\mbox{ 
   \subfigure{
     \includegraphics[width=8.5cm,angle=90]{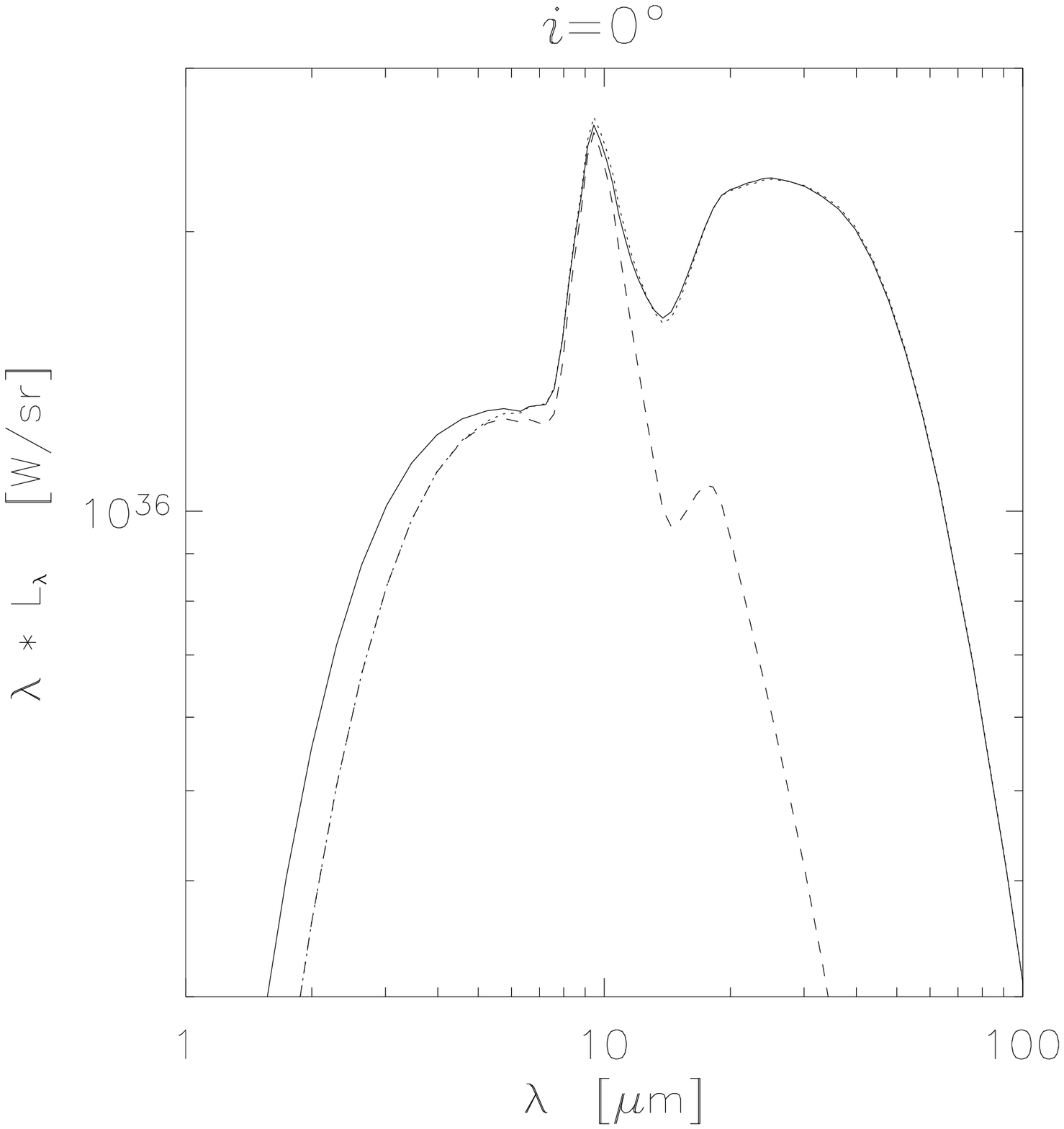}
   }
   \subfigure{
     \includegraphics[width=8.5cm,angle=90]{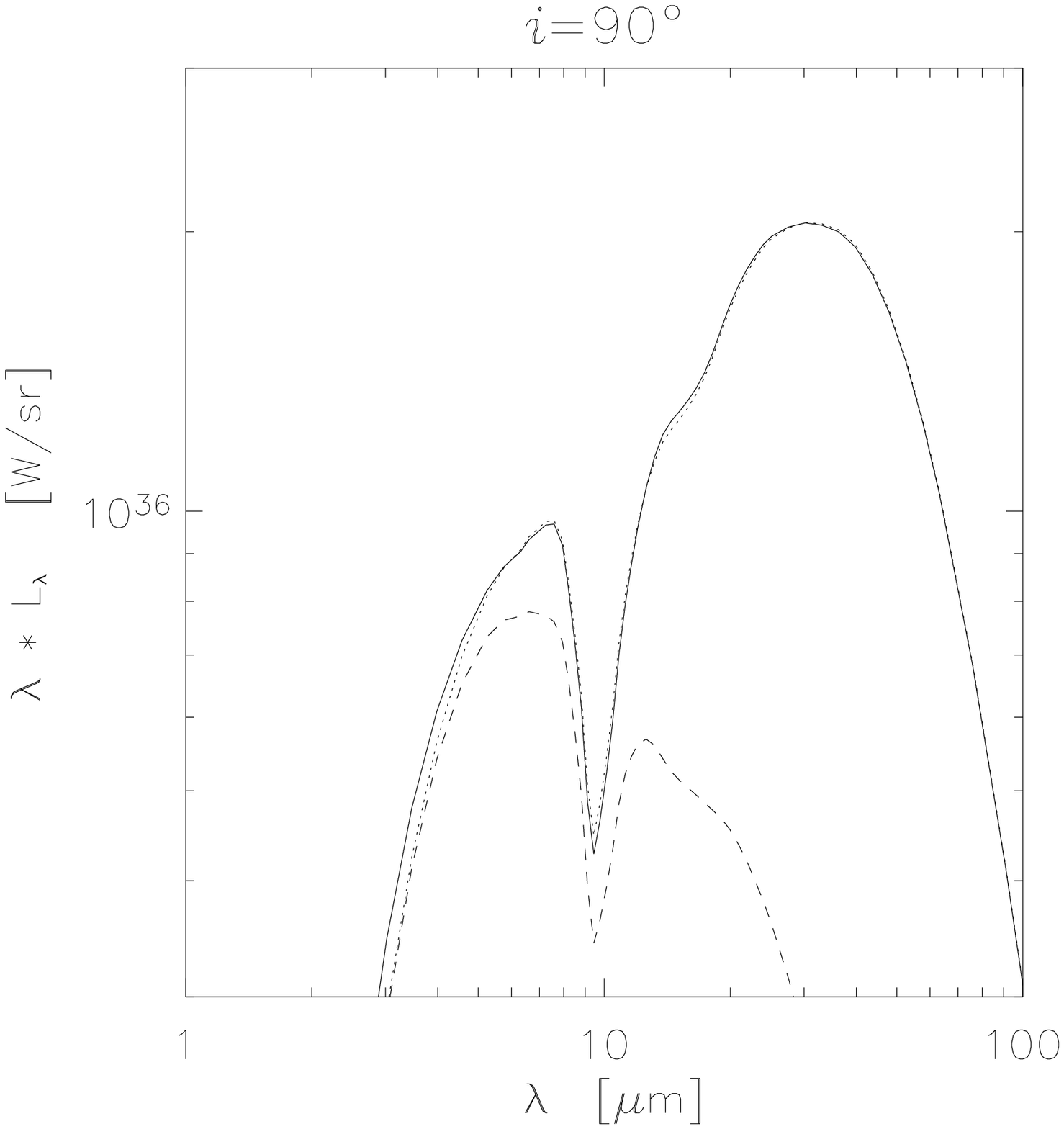} 
   }
}
\caption{SEDs at inclination angles of 0\,\degr\, and 90\,\degr\, for our standard
  model with an isotropically radiating accretion disc (given by the solid
  line) and for the same model but with an accretion disc with a
  $|cos(\theta)|$ - radiation characteristic, without zoom (dotted line) and
  zooming into the inner 10\,pc in radius (dashed line).} 
\label{fig:radchar_SED}
\end{figure*}

A further step towards a more physical model was the introduction of a radiation
characteristic of the accretion disc (see also \citet{Manske_98}). We wanted to check if this
yields less steep radial temperature distributions, helping us to
explain the increasing depth of the silicate feature seen in correlated flux
measurements performed by MIDI \citep{Jaffe_04}. 
In an extended
optically thick accretion disc, a significant part of the radiation
arises from the surface. Therefore, in a simple approach, we
implemented a radiation characteristic following the same $|\cos(\theta)|$ law
for all wavelengths. 
After
emission of a photon package from the central source, 
the direction of emission is chosen by chance, but in accordance with this 
radiation characteristic. 

In Fig.~\ref{fig:radchar_temp}, the
differences in the temperature distributions of the smallest silicate dust
component are visualised. 
Here, one can directly
see that compared to the isotropic radiating case, the $|\cos(\theta)|$
characteristic for the photon emission leads to hotter areas near the funnel
walls, while the temperature within the equatorial plane gets lower. This is
understandable, because the midplane is mainly heated by indirect reradiation of
the dust in this case. 

In Fig.~\ref{fig:radchar_SED}, SEDs
of our standard model with an isotropically radiating accretion disc
(given by the solid line) are compared to the case of an accretion disc with a 
$\left| \cos \theta \right|$ - radiation characteristic (dotted line). The dashed 
line corresponds to the model with a radiation characteristic, but zoomed into the
central 10\,pc in radius and will be discussed in Sect.~\ref{sec:zoomin}. 
As the maximum dust temperature is smaller when implementing a radiation characteristic, 
the left branch of the IR bump is slightly shifted to longer wavelengths. 
The dotted line coincides here with the dashed line. 
With increasing inclination, this effect gets less important due to increasing extinction 
along the line of sight. 
In disagreement with \citet{Manske_98}, we cannot find any change
of the depth of the silicate feature, when comparing the SED with an isotropically radiating
accretion disc with the anisotropic case. This is caused by the different geometry in the 
innermost part of the torus and other values of the
optical depth of our modelling compared to their approach. The cusp of the torus, which is 
only present in our modelling leads to a more pronounced silicate feature in emission for
small inclination angles.\\
In the second row of Fig.~\ref{fig:inc_map}, surface brightness distributions at different inclination
angles are shown for the case of our standard torus model, illuminated by the
accretion disc described above. Here, one can directly see that the radiation characteristic leads to
an enhanced heating of the funnel walls. This is best seen when comparing
the images at an inclination angle of $0\,\degr$. The area of maximum emission
increases for the case of an implemented radiation characteristic, 
because more energy is emitted into the relevant solid angle
and can, therefore, heat the funnel efficiently up to a larger height. 
In contrast to this, the equatorial plane can only be heated by reemitted
photons and, therefore, only smaller temperatures are reached. This effect is
visible directly when looking at the image for the edge-on view to the
torus. The midplane appears as a narrow absorption band.

\subsection{Dust mass variation study for SEDs}
\label{sec:duma}

\begin{figure*} 
\centering
\mbox{ 
   \subfigure{
     \includegraphics[width=8.5cm,angle=90]{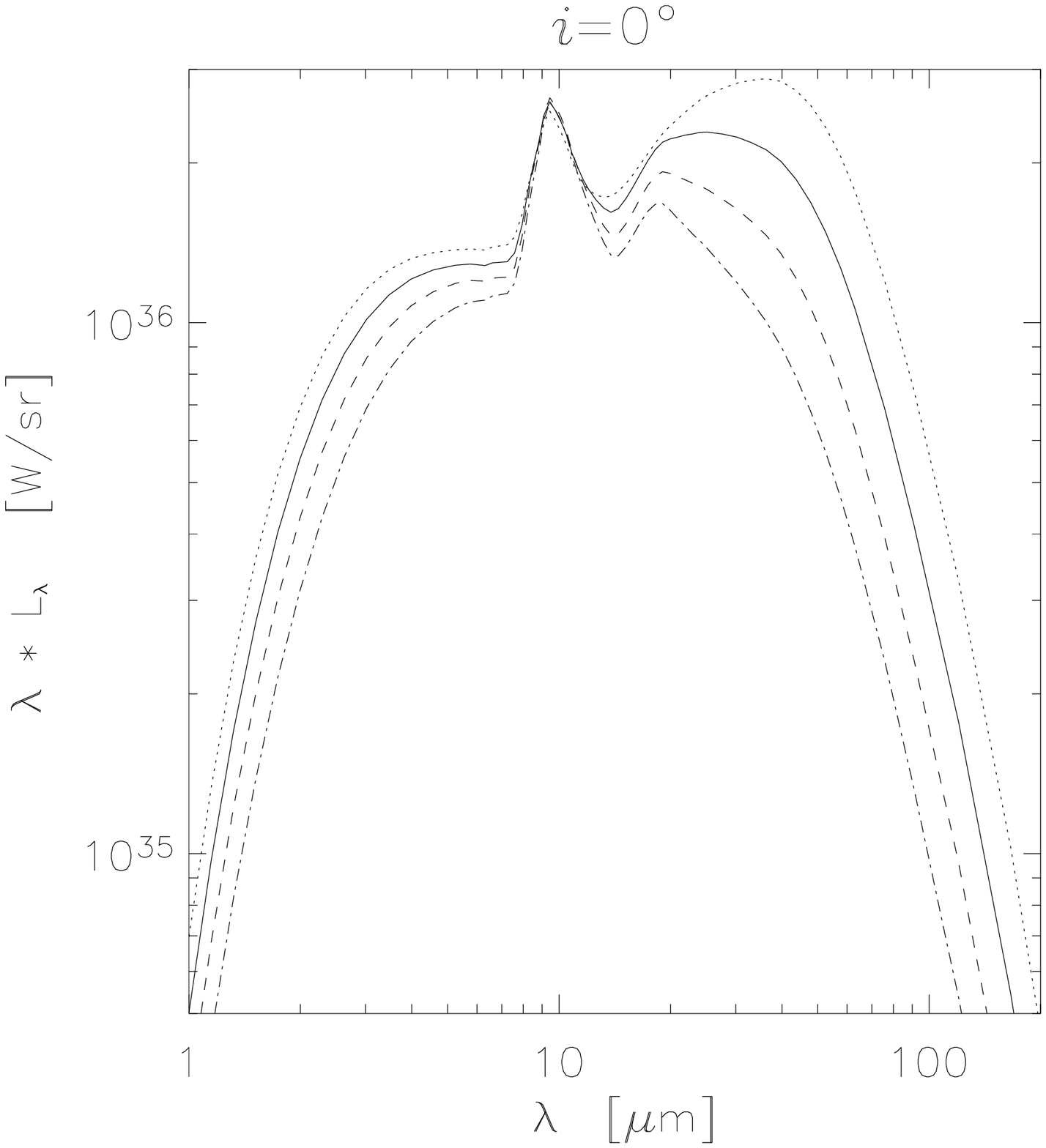}
   }
   \subfigure{
     \includegraphics[width=8.5cm,angle=90]{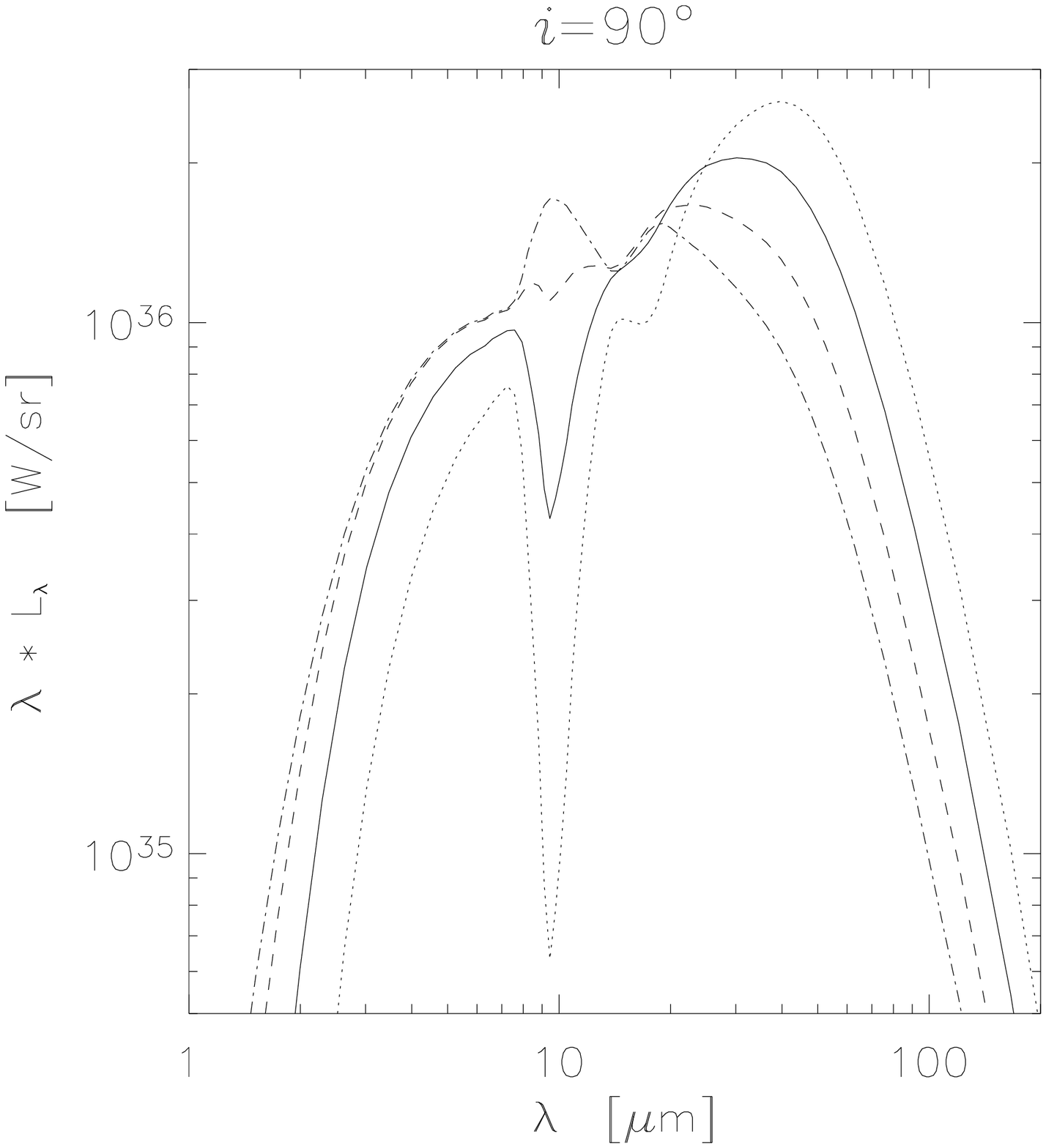} 
   }
}
\caption{SEDs at inclination angles of 0\,\degr\, and 90\,\degr\, for different
  optical depths along an equatorial line of sight:
  $\tau_{9.7\,\muup \mathrm{m}}$\,=\,$4.0$ (dotted line), $\tau_{9.7\,\muup \mathrm{m}}$\,=\,$2.0$ 
(solid line -- our standard model ), $\tau_{9.7\,\muup \mathrm{m}}$\,=\,$1.0$ (dashed line) and 
$\tau_{9.7\,\muup \mathrm{m}}$\,=\,$0.5$ (dashed-dotted line).} 
\label{fig:tau_SED}
\end{figure*}

To study the impact of changing the dust mass, which is enclosed in the torus, 
we started with our standard model (given by the solid line in
Fig.~\ref{fig:tau_SED}) with an optical depth of $\tau_{9.7\,\muup \mathrm{m}}$\,=\,$2.0$,
doubled the mass (dotted line, $\tau_{9.7\,\muup \mathrm{m}}$\,=\,$4.0$) and halved it twice
(dashed line with $\tau_{9.7\,\muup \mathrm{m}}$\,=\,$1.0$ and
dashed-dotted line with $\tau_{9.7\,\muup \mathrm{m}}$\,=\,$0.5$). The
resulting SEDs are shown in Fig.~\ref{fig:tau_SED} for the Seyfert\,I and Seyfert\,II case
with inclination angles of 0\,\degr\, and 90\,\degr\, respectively. 
As the optical depth is given by the integral of the product of the dust
density and the mass extinction coefficient along the line of sight, it scales
linearly with the enclosed dust mass:
\begin{eqnarray}
  \tau_{\mathrm{ext}} = \int_{\zeta}\limits \rho_{\mathrm{d}} \, \kappa_{\mathrm{ext}} \, ds.
\end{eqnarray}
When we choose a line of sight within the dust-free hole of the torus --
corresponding to the Seyfert\,I case -- the whole temperature range of the dust
in the torus can be seen. 
With increasing optical depth, more and more photons are absorbed within a decreasing 
volume near the funnel of the torus and reemitted in the infrared wavelength range.
This leads to a rise of the
temperature of dust near the funnel. But due to the high optical depths, less
photons reach the outer part of the torus and, therefore, the temperature
drops. Altogether, a steeper temperature distribution in radial direction
results. Smaller minimum temperatures lead to a shift of the Rayleigh-Jeans
branch to longer wavelengths and larger maximum temperatures lead to a shift
of the Wien branch to shorter wavelengths. Altogether, the IR bump gets
broader. The higher temperatures in the inner region also cause more flux at
short wavelengths, while the larger amount of cold dust produces more
flux at longer wavelengths. The global maximum of the IR bump is shifted to higher
wavelengths because of the smaller temperatures at the point of the maximum of the
dust density. 
  
For an inclination angle of 90\,\degr, which corresponds to a Seyfert\,II-like
case, there is always dust on the line of sight. Only the outer and,
therefore, the coldest part
of the torus can be seen directly. That is why we can use  
the reasoning
for the Seyfert\,I case for the long wavelength region. 
The innermost (and therefore hottest) part of the torus can only be seen
through an increasing amount of dust. The higher the extinction
the less we can look into the torus and the less hot areas of the torus
can be seen. Therefore, the left branch of the
IR bump is shifted to higher wavelengths when the extinction increases.
The $9.7\,\muup \mathrm{m}$ silicate feature changes from emission over partial absorption
to total absorption, comparable to large inclination angles in the study shown
in Sect.~\ref{sec:inc_sed}. As discussed above, increasing dust mass affects the
optical depth linearly. When $\tau$\,=\,$1$ is reached, the torus gets optically
thick, and, therefore, the feature changes from emission to
absorption. Between these two cases, a smooth transition region is visible, where
the feature can be seen in partial absorption and emission, called self-absorption \citep{Henning_83}.
Intersecting the $\tau$\,=\,$1$ line with the optical depth plotted
against the wavelength though displays the expected behavior of the
silicate feature in the spectrum. The same holds for the second silicate
feature at $18.5\,\muup \mathrm{m}$ but in alleviated form as it is much weaker and
partially covered by the global maximum of the IR bump.\\
The study shows that the SEDs are very sensitive to changes in the dust
mass. Already a change by a factor of four changes the feature from
emission to complete absorption (see solid line compared to dashed-dotted line
in Fig.~\ref{fig:tau_SED}). Therefore, the silicate feature is a quite
sensitive tool to distinguish the optical depth along the line of sight.

\subsection{Dust mass variation study for surface brightness distributions}

\begin{figure*} 
\centering
\mbox{ 
   \subfigure{
     \includegraphics[width=4.0cm,angle=90]{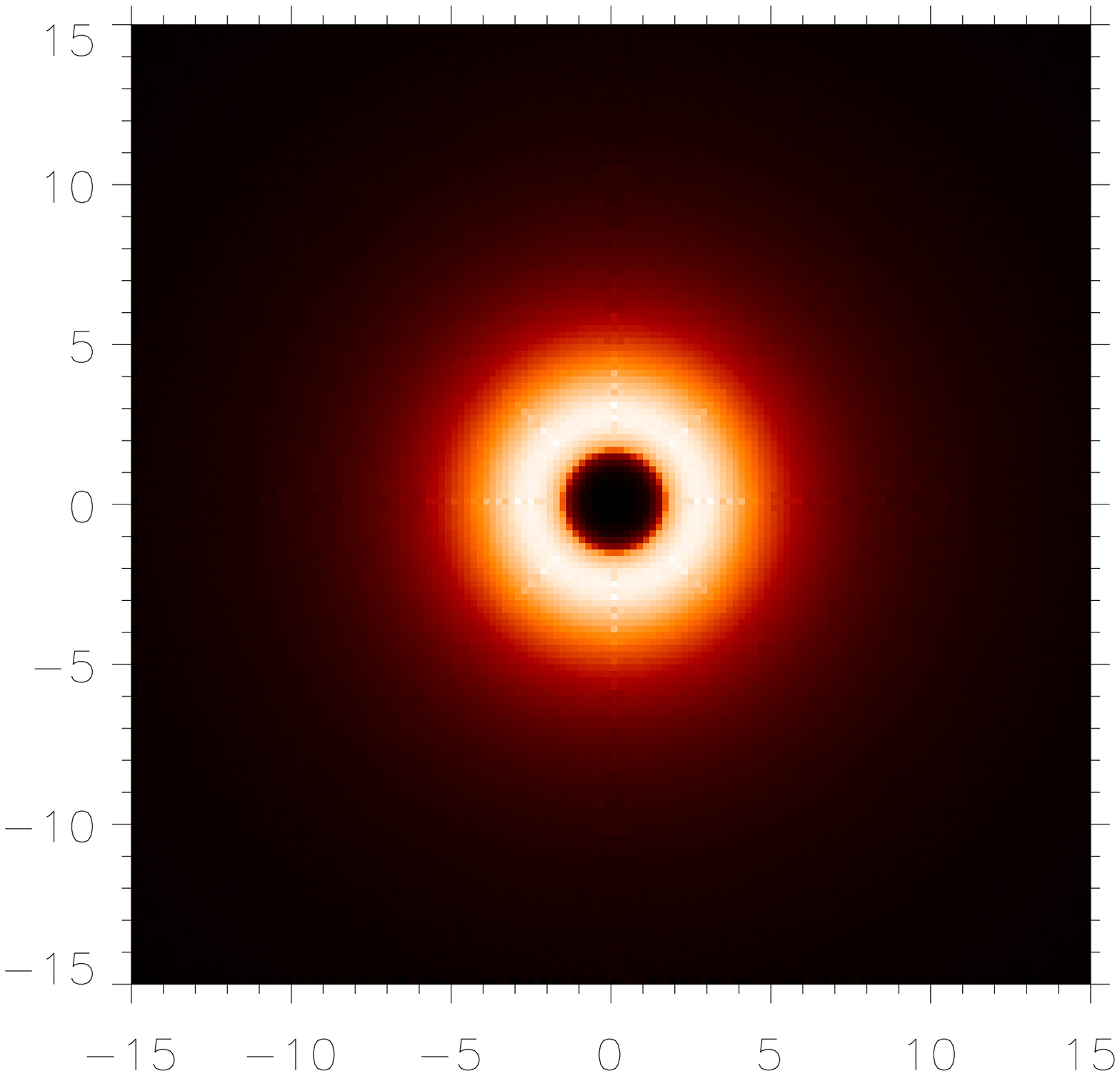}
   }
   \subfigure{
     \includegraphics[width=4.0cm,angle=90]{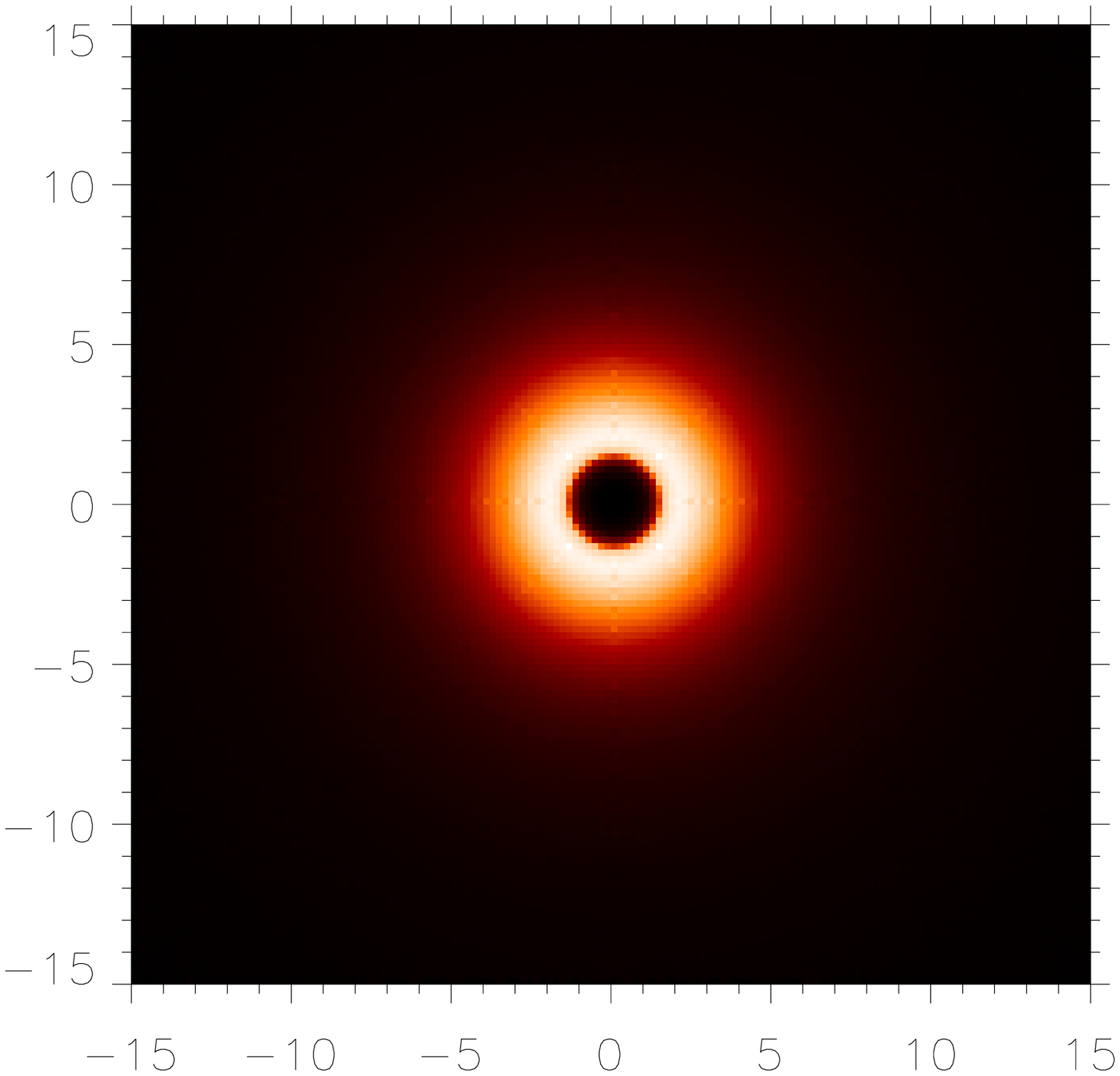} 
   }
   \subfigure{
     \includegraphics[width=4.0cm,angle=90]{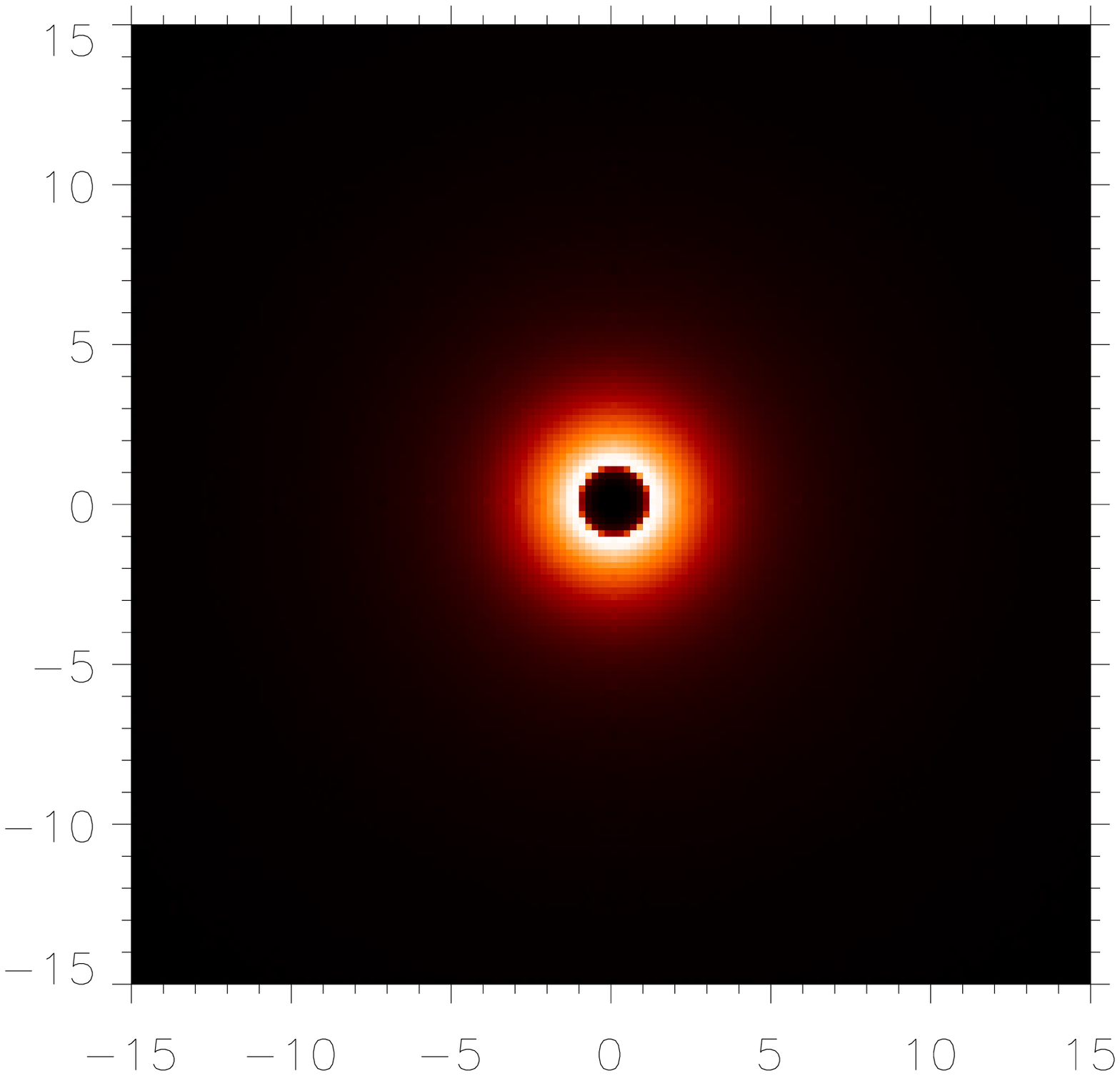} 
   }
   \subfigure{
     \includegraphics[width=4.0cm,angle=90]{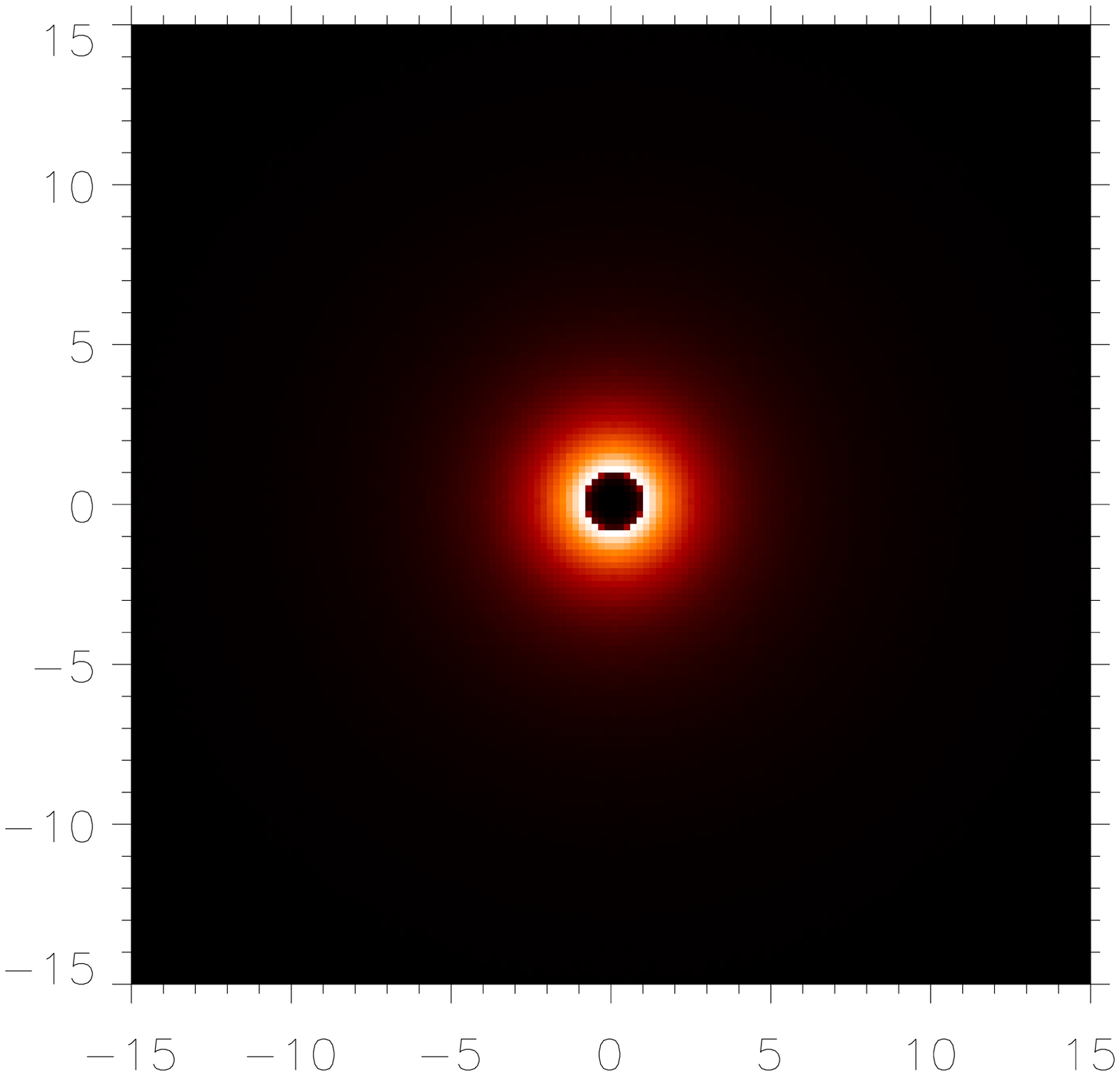} 
   }
}
\mbox{ 
   \subfigure{
     \includegraphics[width=4.0cm,angle=90]{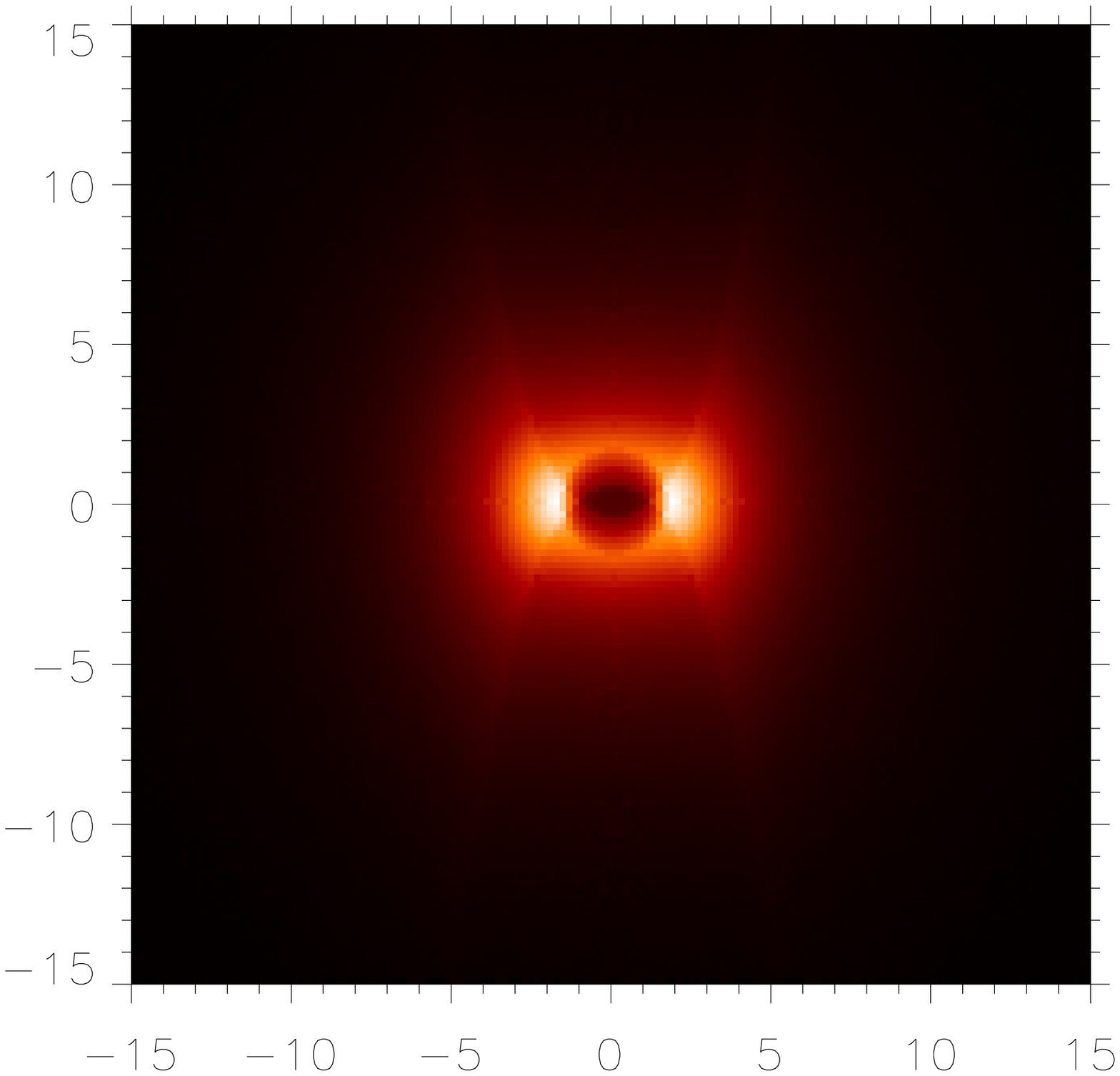}
   }
   \subfigure{
     \includegraphics[width=4.0cm,angle=90]{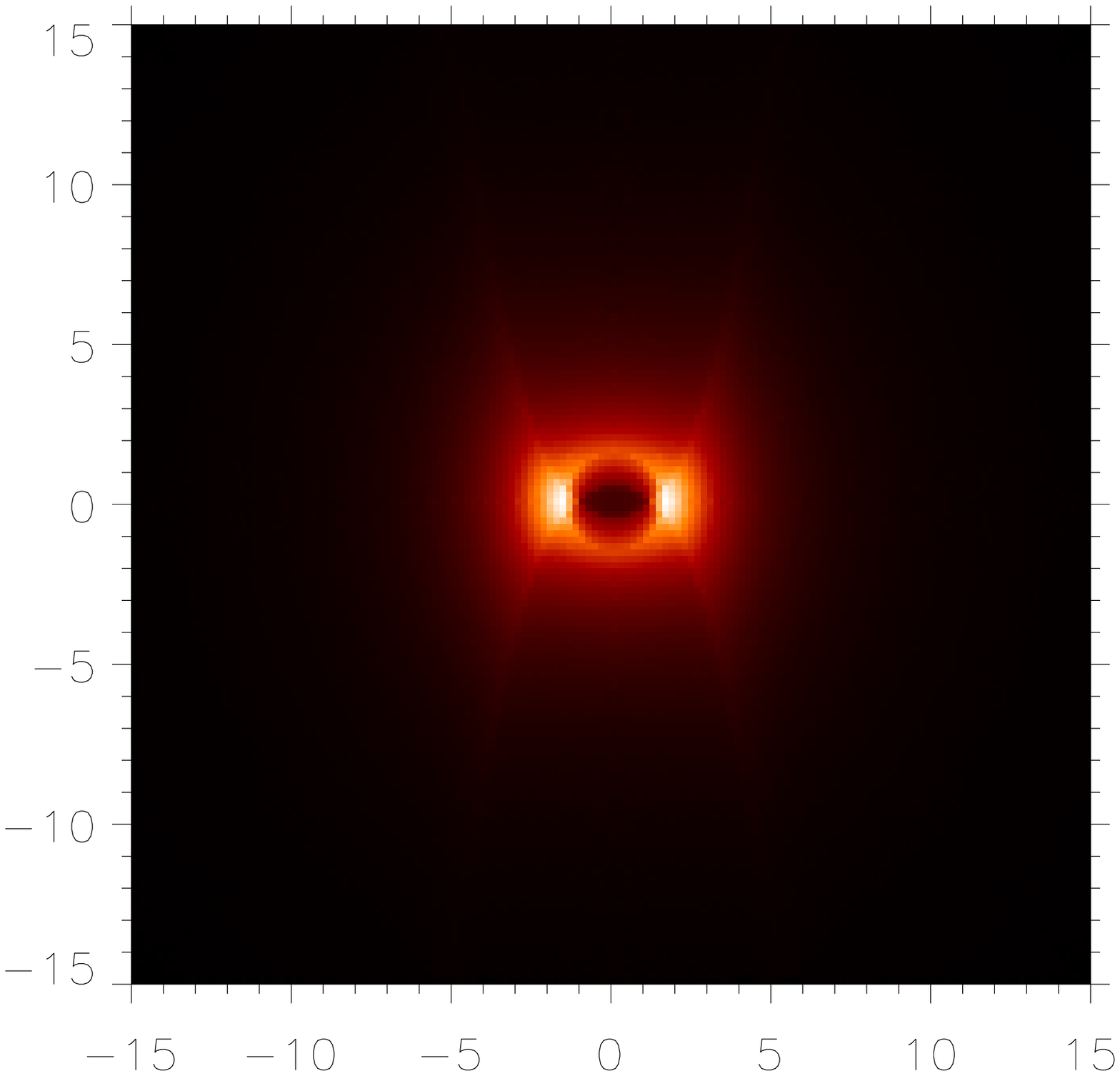} 
   }
   \subfigure{
     \includegraphics[width=4.0cm,angle=90]{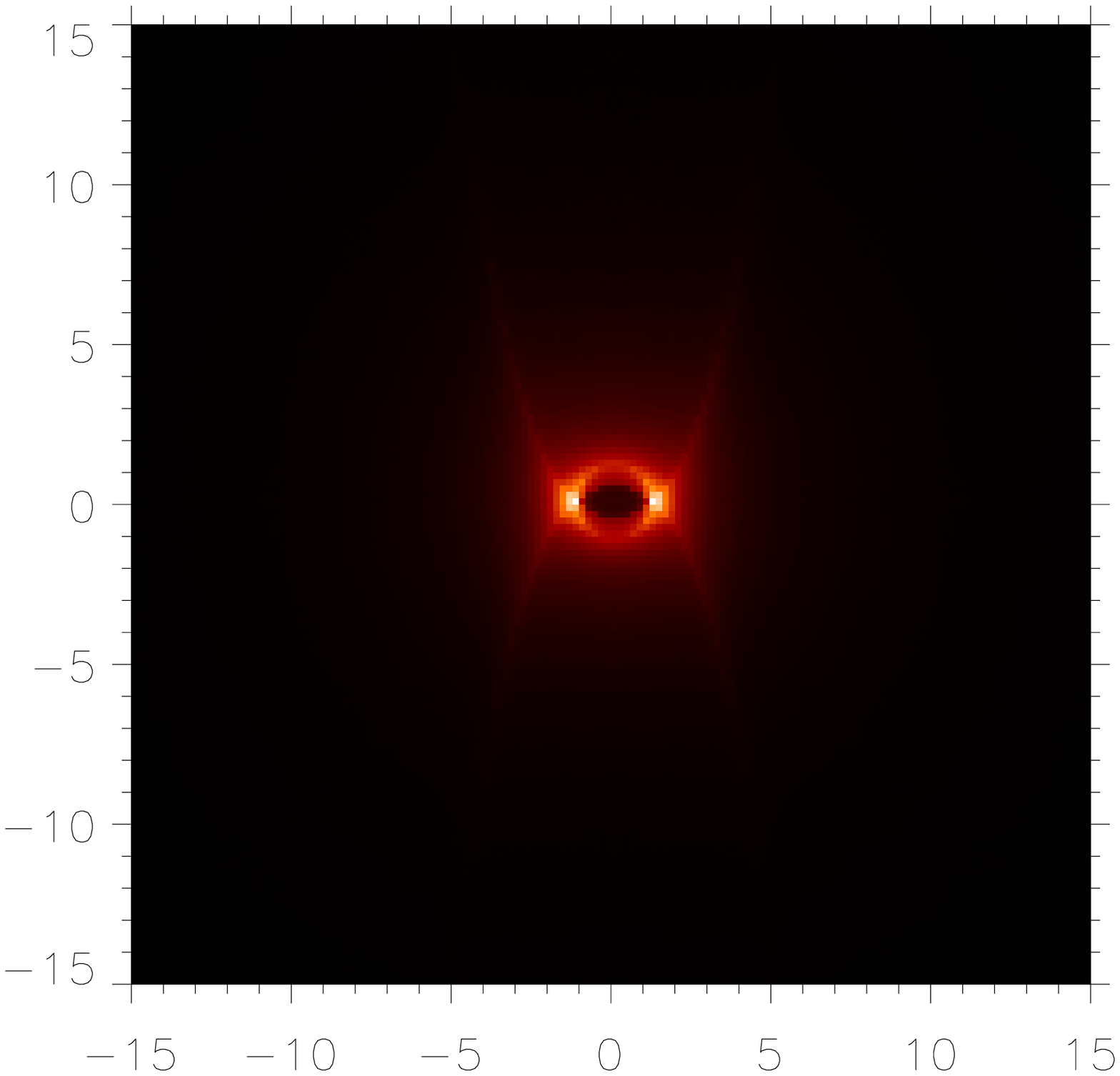} 
   }
   \subfigure{
     \includegraphics[width=4.0cm,angle=90]{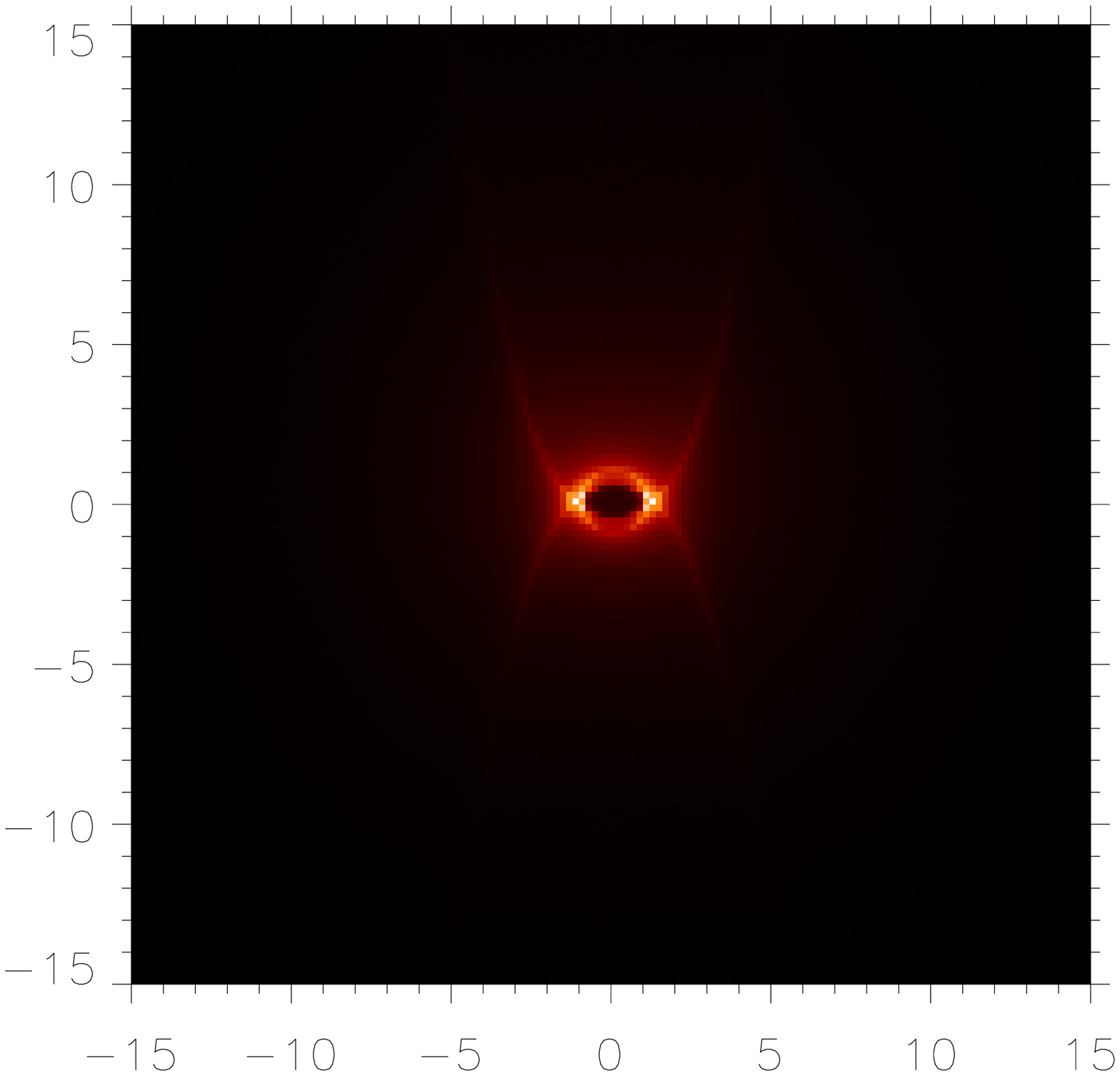} 
   }
}
\mbox{ 
   \subfigure{
     \includegraphics[width=4.0cm,angle=90]{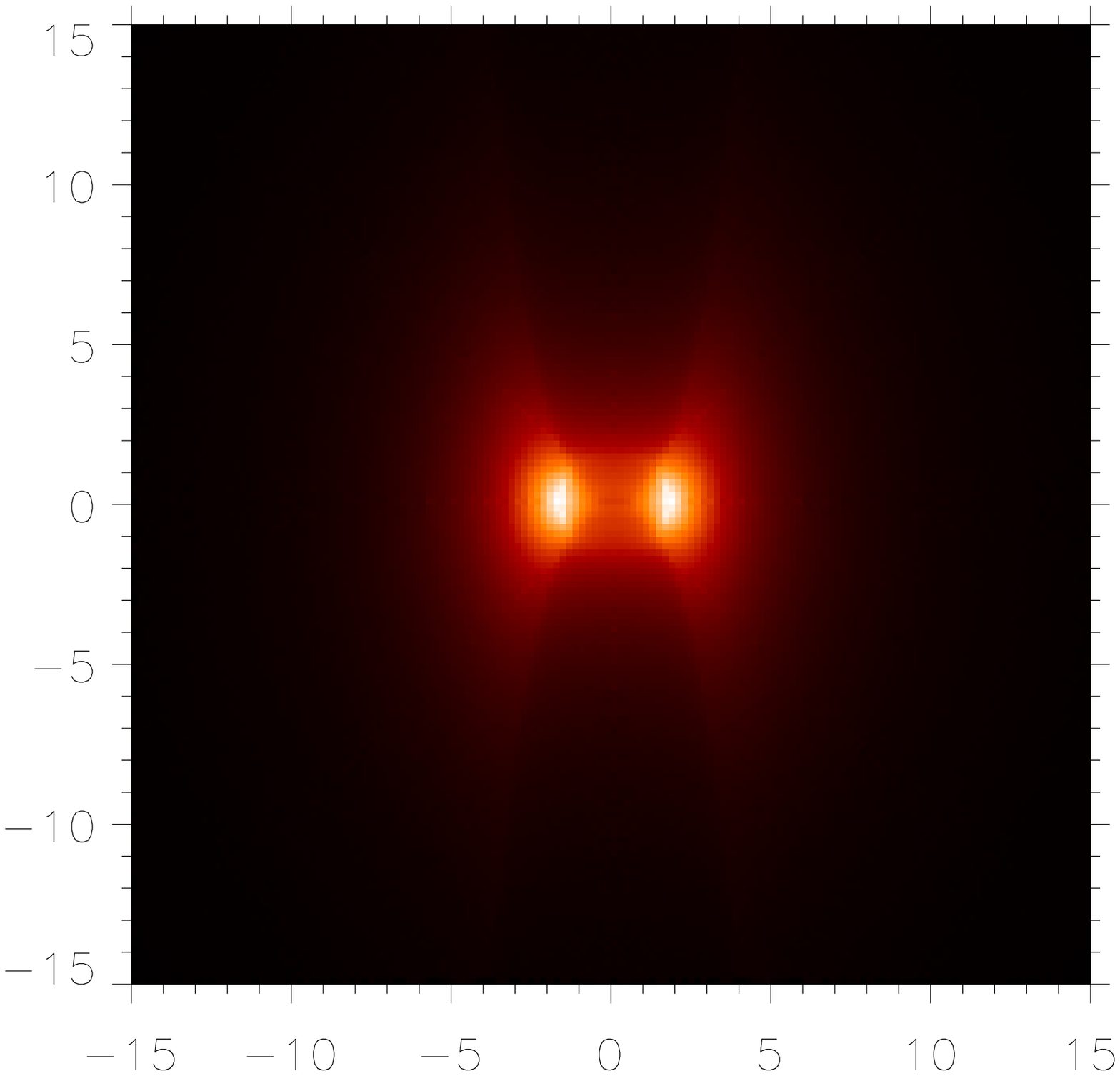}
   }
   \subfigure{
     \includegraphics[width=4.0cm,angle=90]{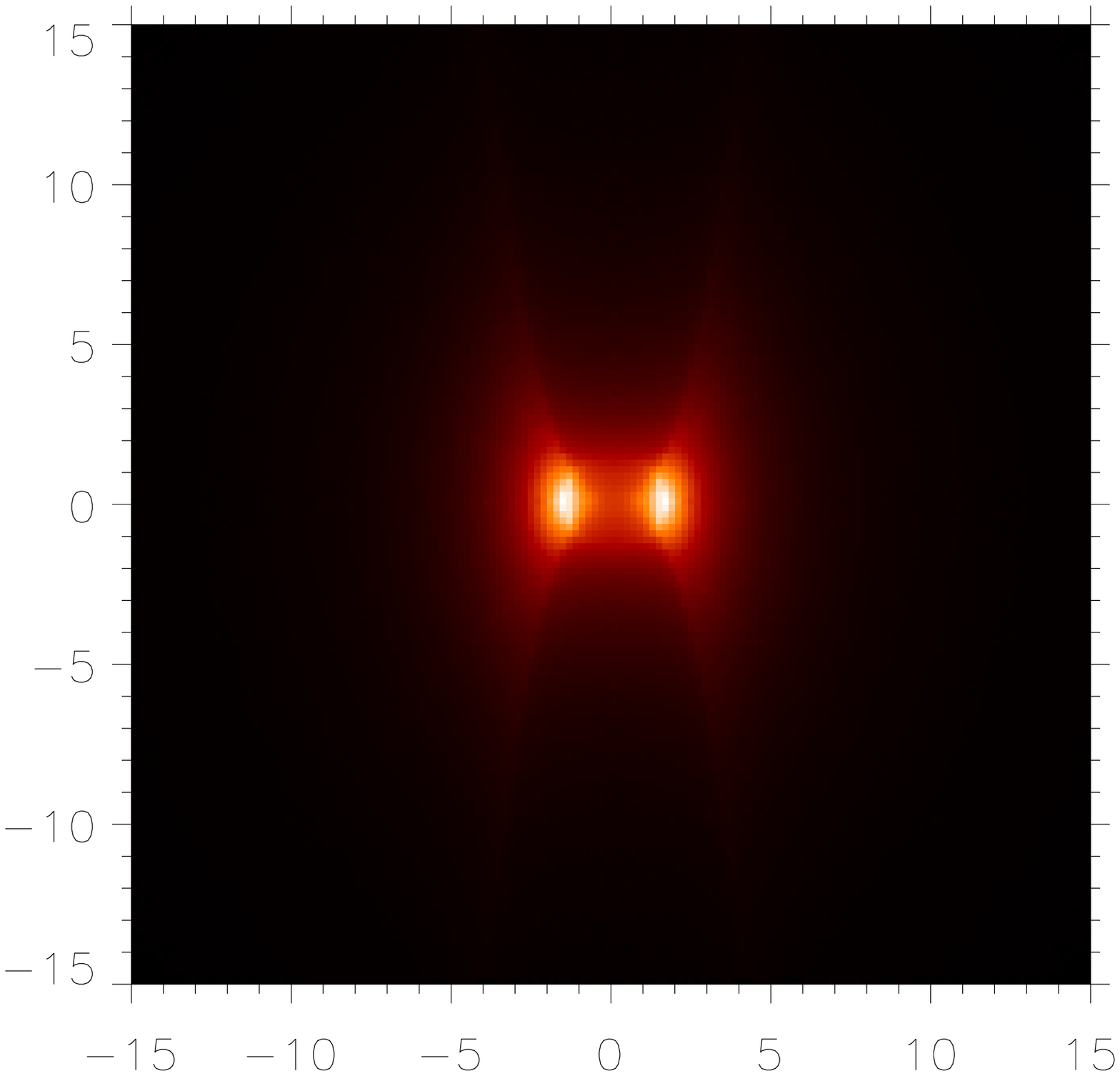} 
   }
   \subfigure{
     \includegraphics[width=4.0cm,angle=90]{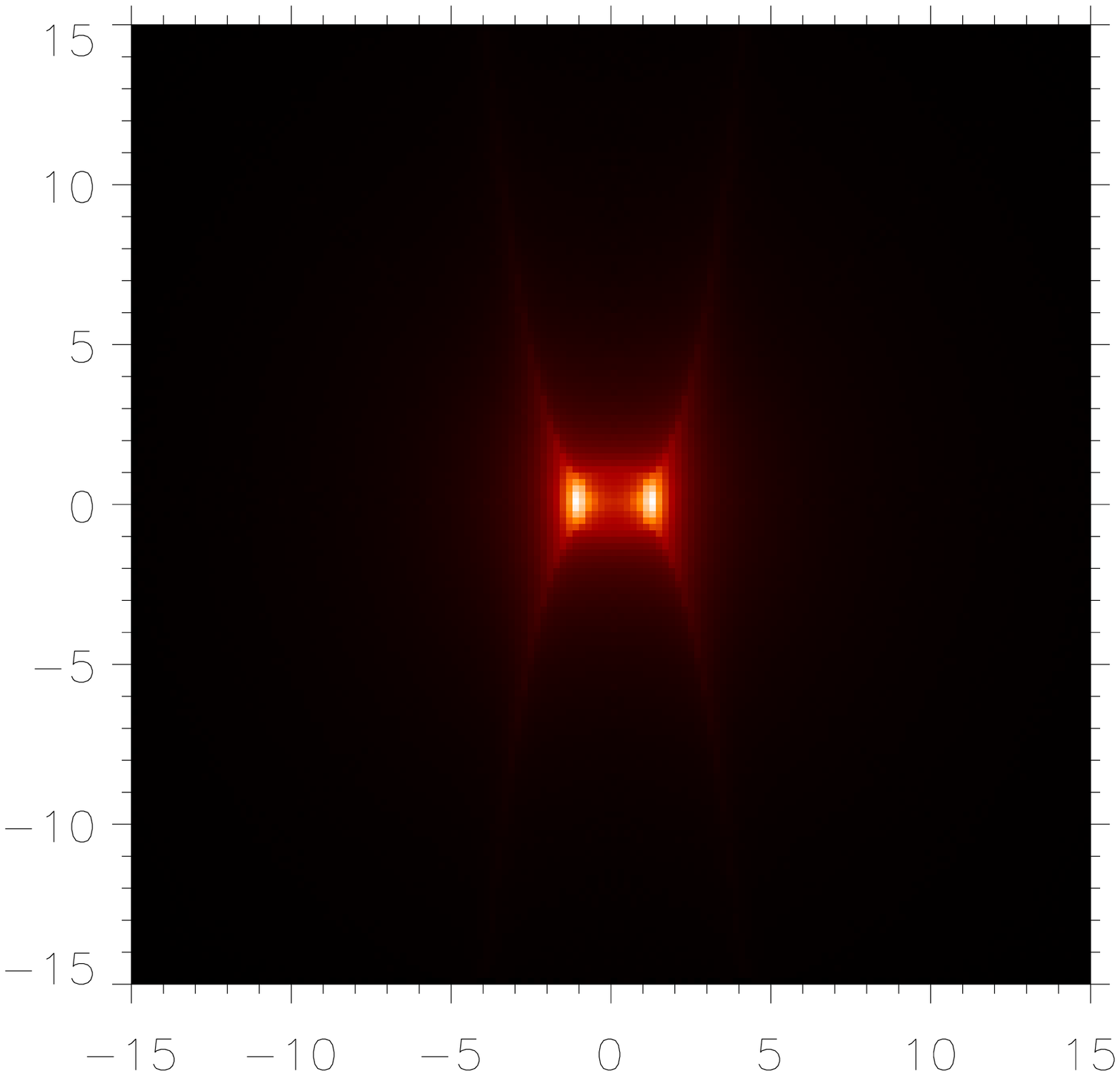} 
   }
   \subfigure{
     \includegraphics[width=4.0cm,angle=90]{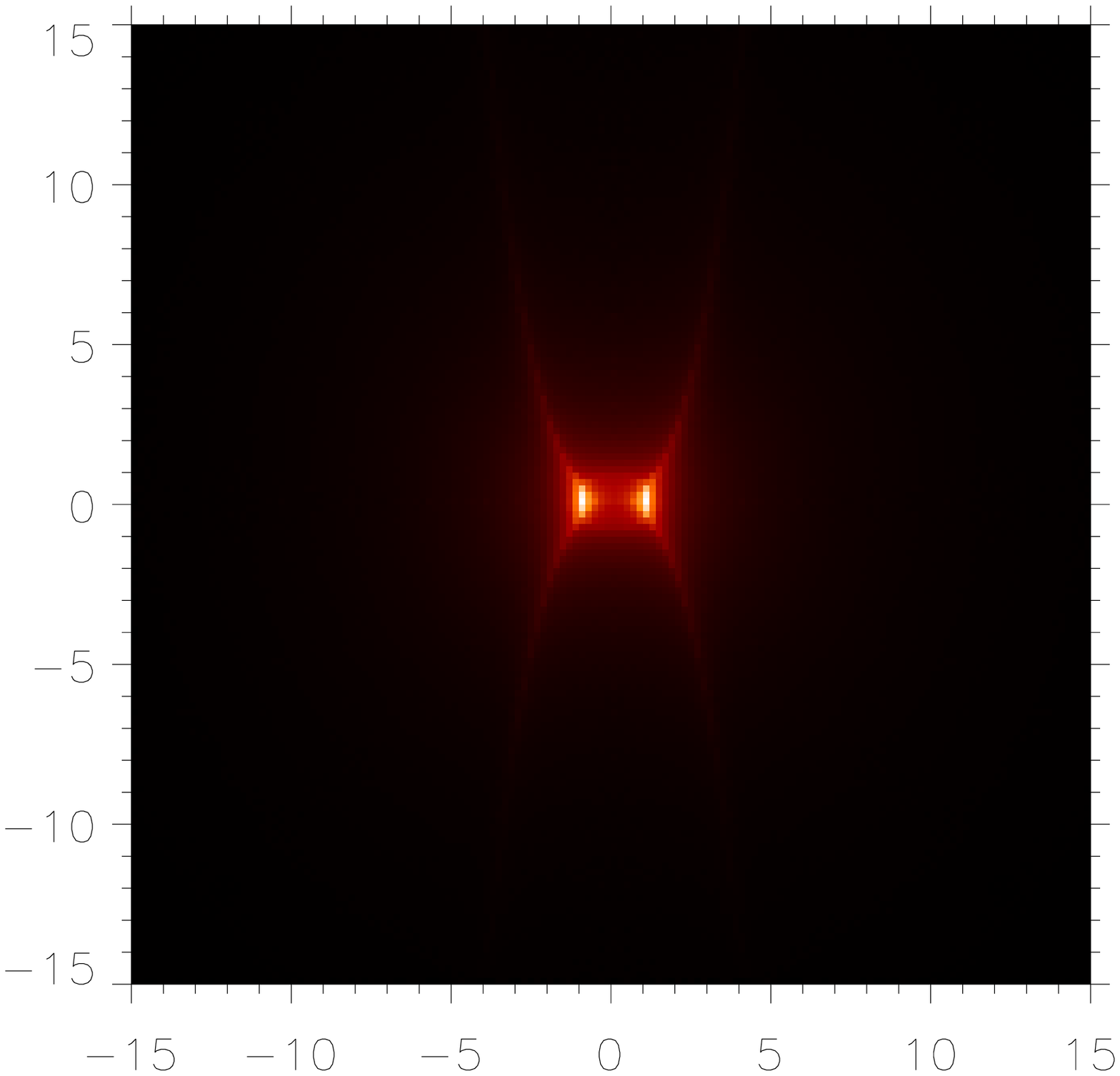} 
   }
}
\addtocounter{subfigure}{-12}
\mbox{ 
   \subfigure[$\tau_{9.7\,\muup \mathrm{m}}$\,=\,$0.5$]{
     \includegraphics[width=4.0cm,angle=90]{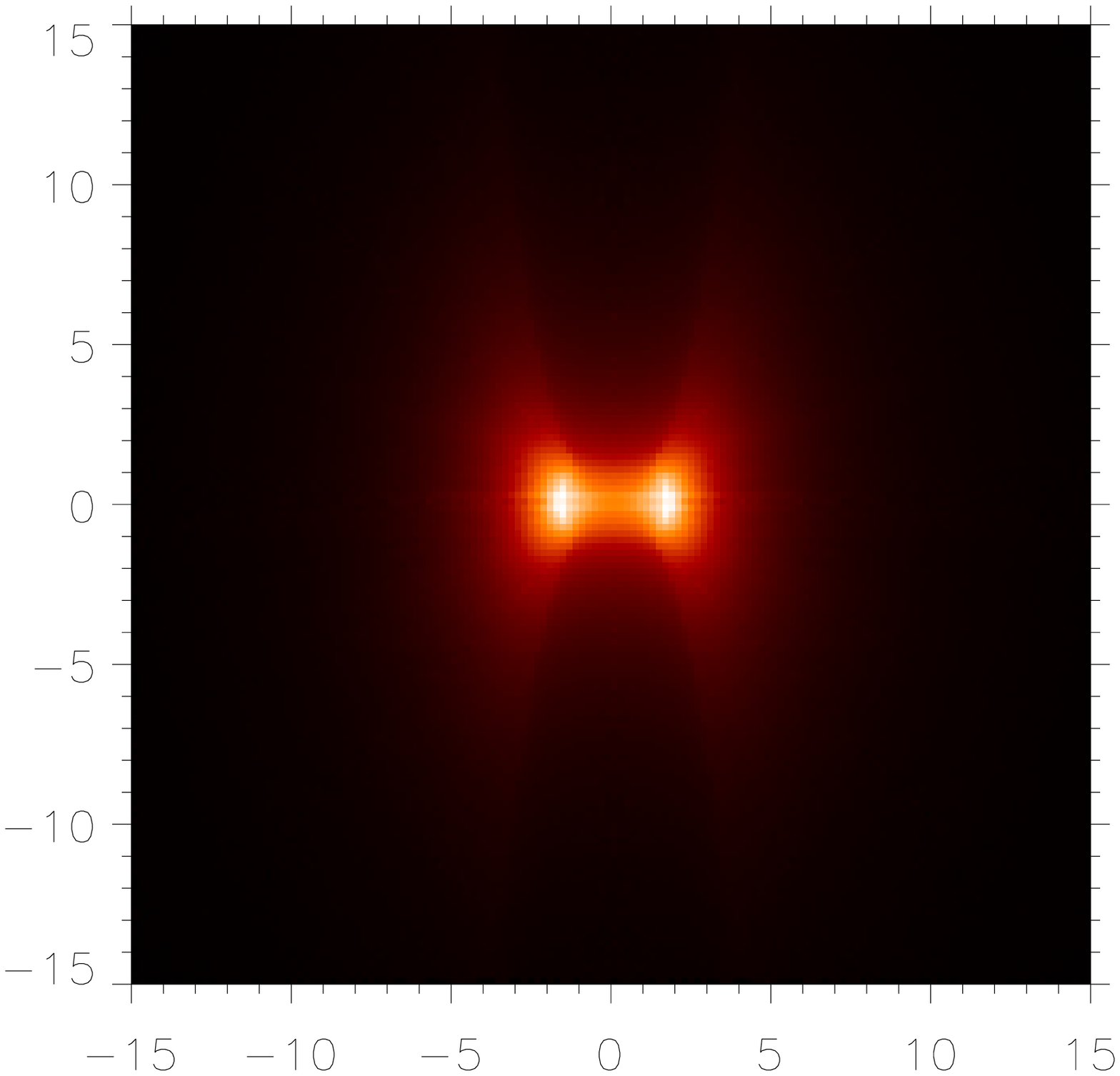}
   }
   \subfigure[$\tau_{9.7\,\muup \mathrm{m}}$\,=\,$1.0$]{
     \includegraphics[width=4.0cm,angle=90]{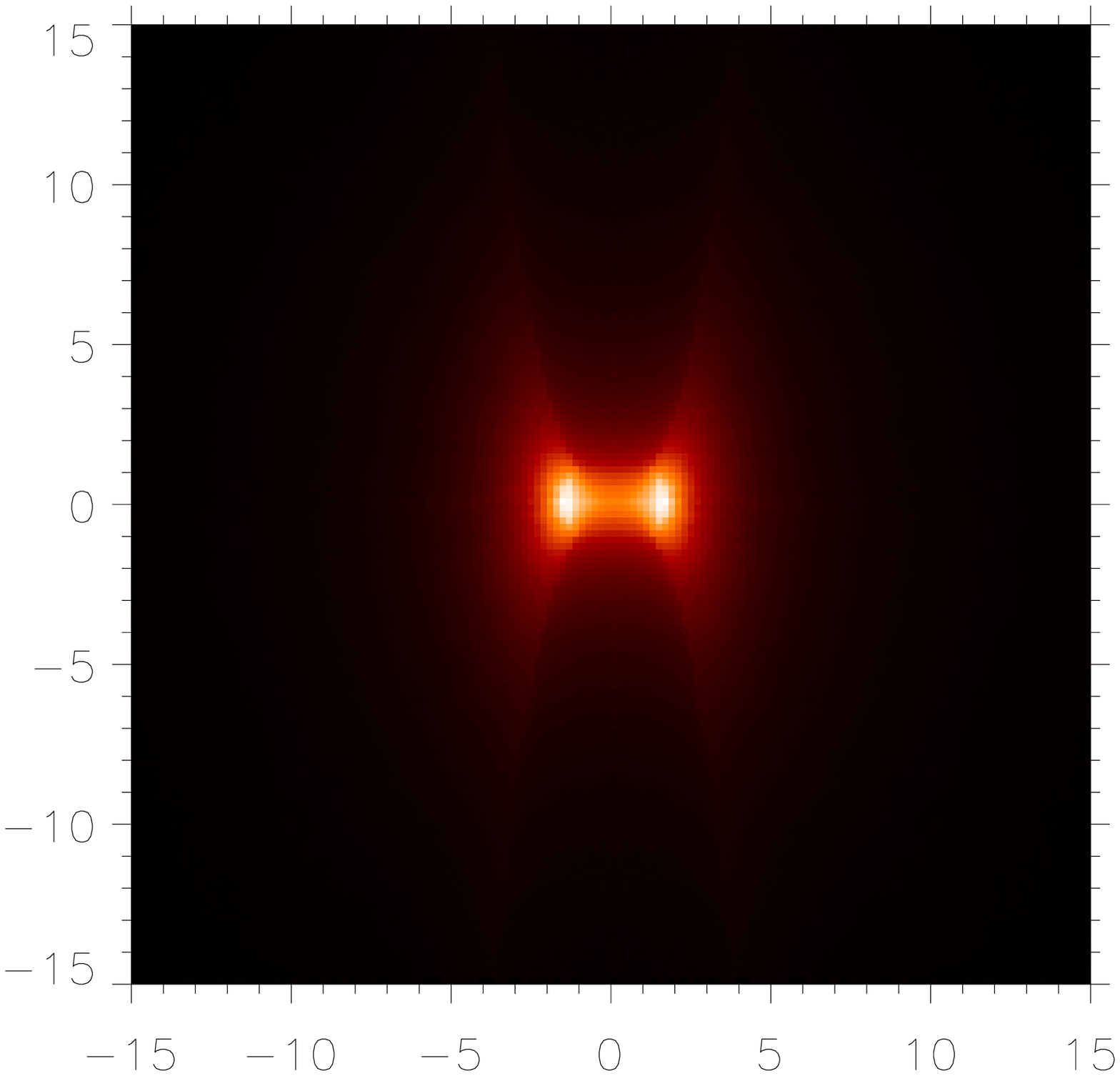} 
   }
   \subfigure[$\tau_{9.7\,\muup \mathrm{m}}$\,=\,$4.0$]{
     \includegraphics[width=4.0cm,angle=90]{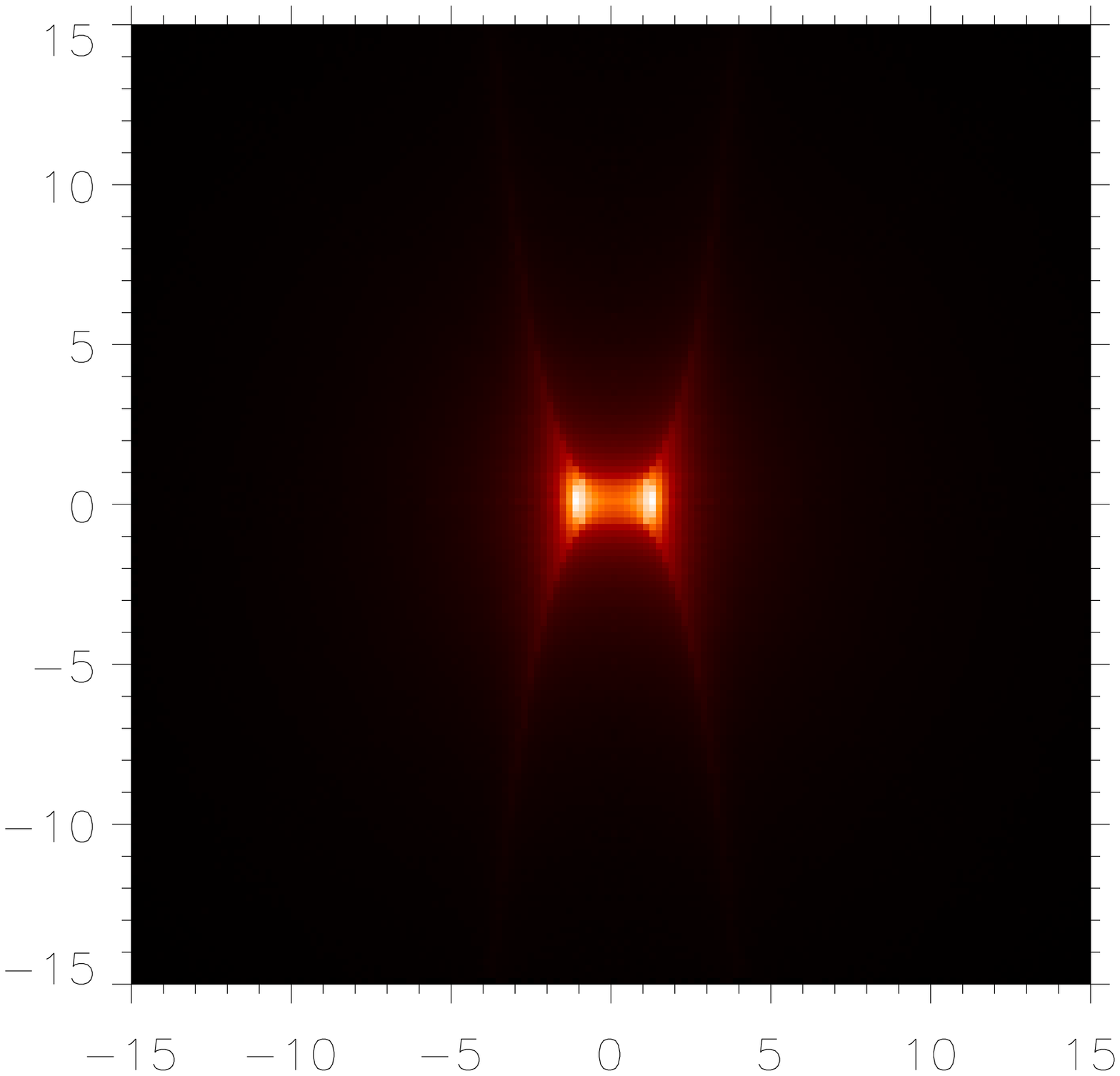} 
   }
   \subfigure[$\tau_{9.7\,\muup \mathrm{m}}$\,=\,$8.0$]{
     \includegraphics[width=4.0cm,angle=90]{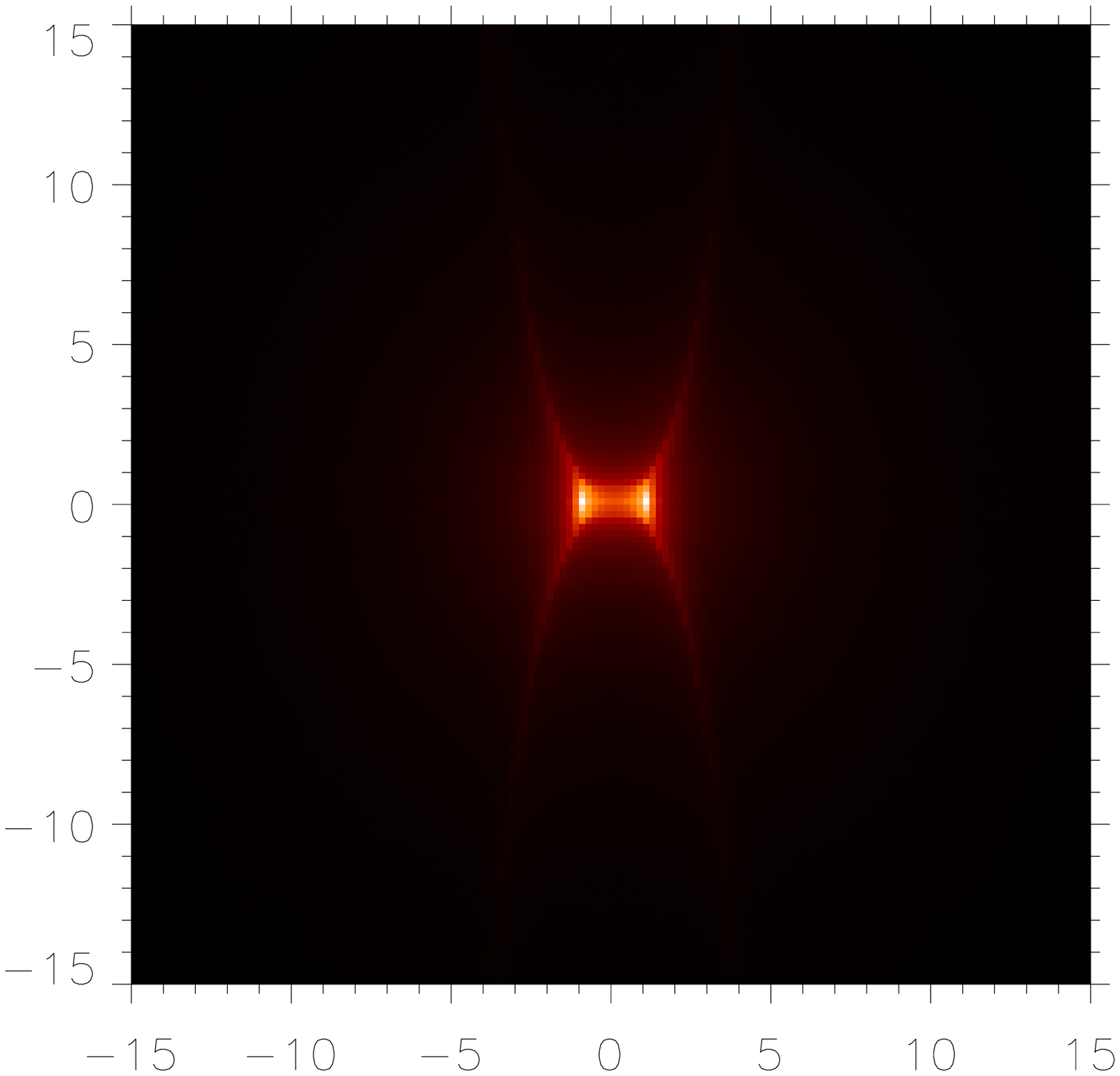} 
   }
}
\caption{Surface brightness distributions at a wavelength of $\lambda$\,=\,$13.18\,\muup$m for models with different optical
  depths. The values for the optical depth are $\tau_{9.7\,\muup \mathrm{m}}$\,=\,$0.5$ (left column), 
  $\tau_{9.7\,\muup \mathrm{m}}$\,=\,$1.0$, 
  $\,\tau_{9.7\,\muup \mathrm{m}}$\,=\,$4.0$ and
  $\tau_{9.7\,\muup \mathrm{m}}$\,=\,$8.0$ (right column)
  and inclination angle changes from 0\,\degr\, (upper row) to 90\,\degr\, (lower
  row) in steps of 30\,\degr. Each step in $\tau$ means doubling the included dust mass. The images with
  $\tau_{9.7\,\muup \mathrm{m}}$\,=\,$2.0$ are omitted (please compare to
  Fig.~\ref{fig:inc_map}). All images are given with a linear 
  colour scale reaching from zero to the
  maximum of the respective image.} 
\label{fig:opdep_map}
\end{figure*}

\begin{figure*} 
\centering
\mbox{ 
   \subfigure{
     \includegraphics[width=8.5cm,angle=90]{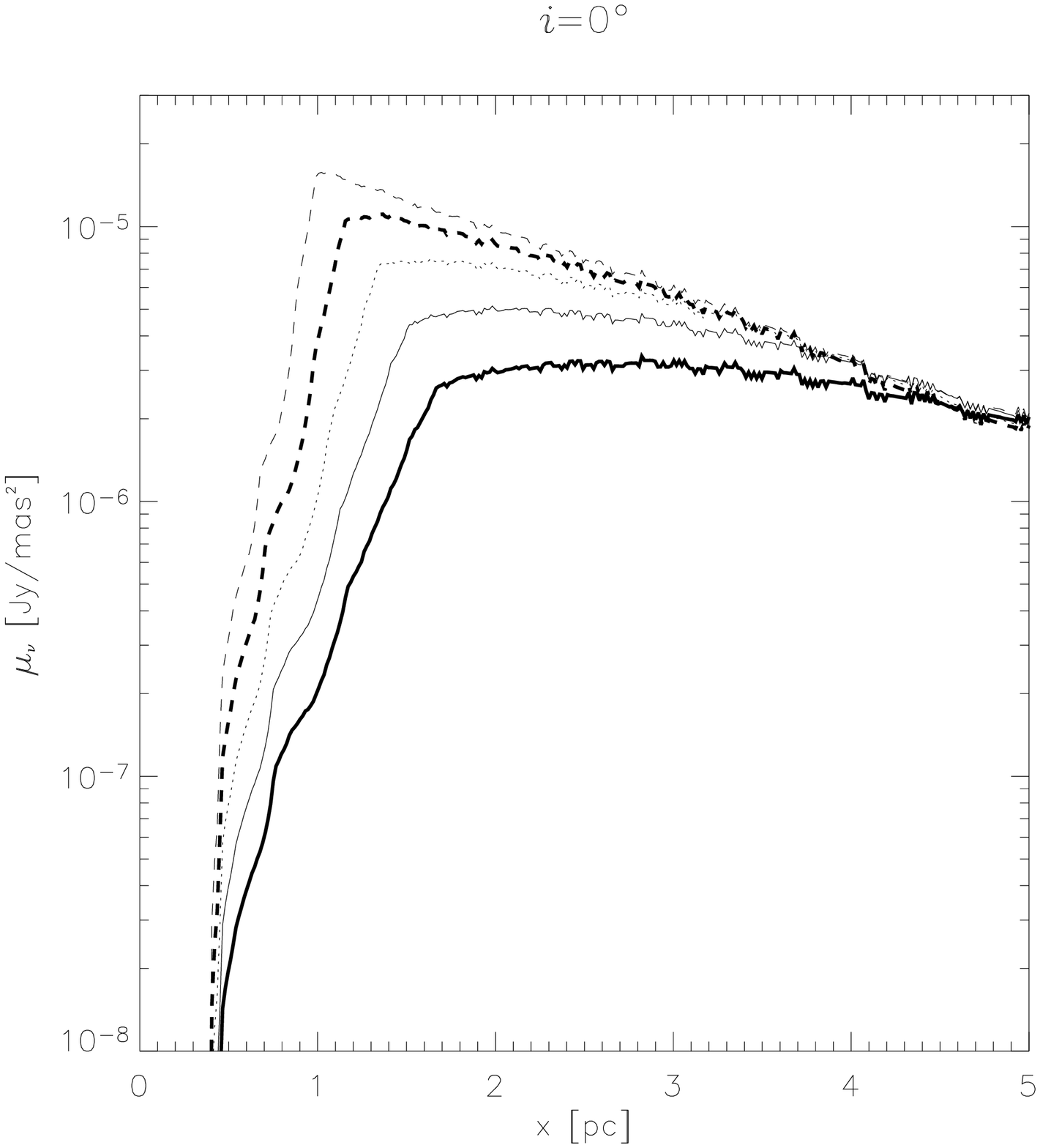}
   }
   \subfigure{
     \includegraphics[width=8.5cm,angle=90]{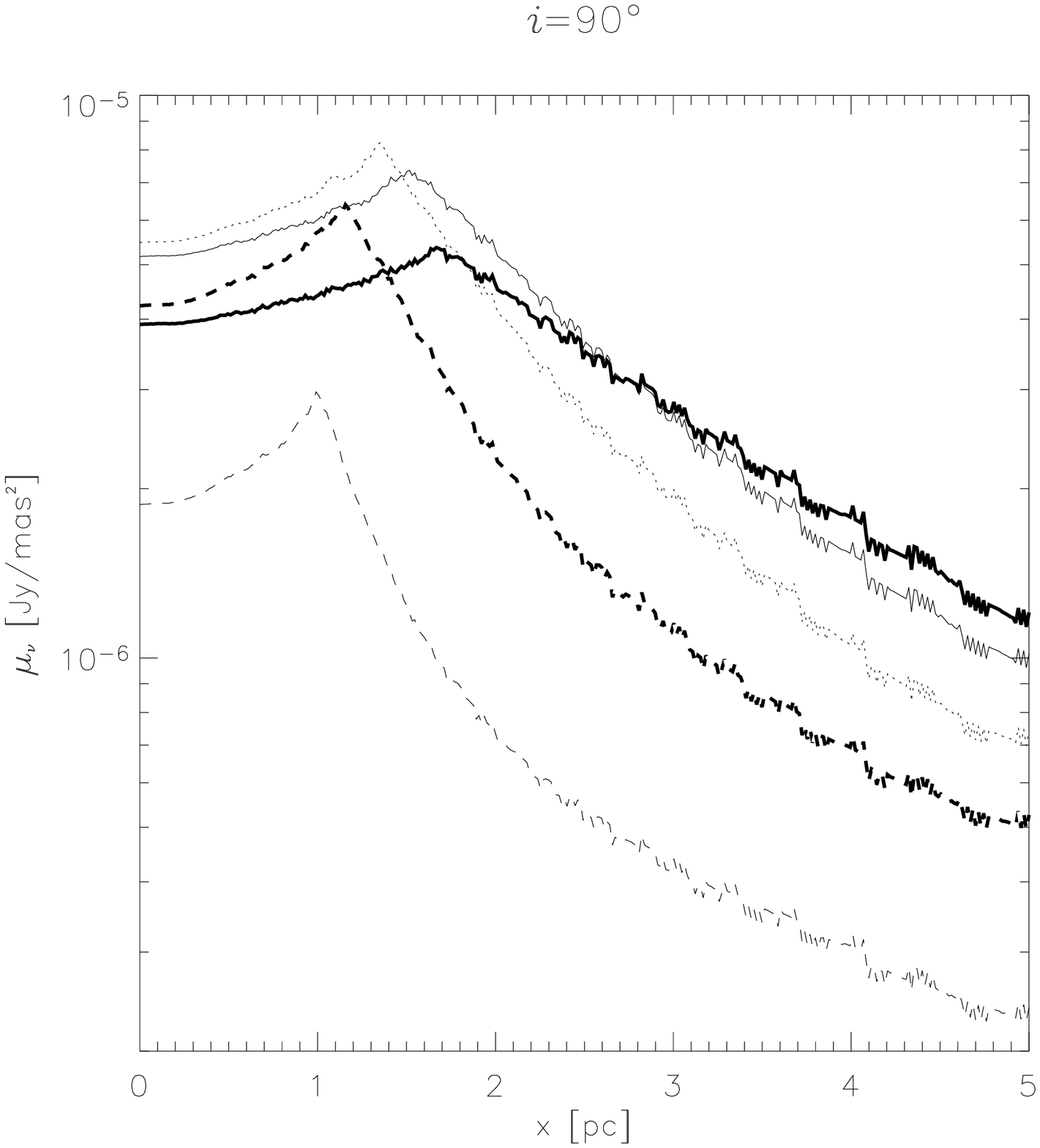} 
   }
}
\caption{Cut through the surface brightness distributions $\mu_{\nu}$ 
  at a wavelength of $\lambda$\,=\,$13.18\,\muup$m along the positive x-axis for inclination angles of 0\,\degr\, and
  90\,\degr. The thick solid line corresponds to the case with $\tau_{9.7\,\muup \mathrm{m}}$\,=\,$0.5$, the thin 
  solid line to $\tau_{9.7\,\muup \mathrm{m}}$\,=\,$1.0$, the dotted line to our standard model with 
  $\tau_{9.7\,\muup \mathrm{m}}$\,=\,$2.0$, the
  thick dashed line to $\tau_{9.7\,\muup \mathrm{m}}$\,=\,$4.0$ and the thin dashed line to $\tau_{9.7\,\muup
  \mathrm{m}}$\,=\,$8.0$. The fluxes shown are given in Jansky per square milliarcsecond for
  an assumed distance of the object of 45\,Mpc.} 
\label{fig:dust_maps_flux}
\end{figure*}

To study the impacts of changing the optical depth along the line of sight on
surface brightness distributions, we
used the same dust masses as in Sect.~\ref{sec:duma} for the case of spectral energy
distributions to gain comparability, but doubled the highest mass again. This means that coming from
our standard model, we doubled the enclosed dust
mass twice and halved it twice.
Again, we chose the same wavelength of $13.18\,\muup \mathrm{m}$. The results are shown in
Fig.~\ref{fig:opdep_map}. The different optical depths ($\tau_{9.7\,\muup
\mathrm{m}}$\,=\,$0.5$, $\tau_{9.7\,\muup \mathrm{m}}$\,=\,$1.0$, $\tau_{9.7\,\muup \mathrm{m}}$\,=\,$4.0$, $\tau_{9.7\,\muup
  \mathrm{m}}$\,=\,$8.0$) are shown in different
columns, while the inclination angle changes are shown in different rows, starting from 0\,\degr\, in the
upper row to 90\,\degr\, in the lower row. Our standard model is omitted here,
please compare to Fig.~\ref{fig:inc_map}. 
All of the images are given in a linear colour scale, with a dynamic range
between zero and the maximum value of the respective map in order to show the
changes in the distribution of maximum surface brightness for the different
optical depths.  
The absolute values for the pixels are given in
Fig.~\ref{fig:dust_maps_flux}, which shows a cut through the maps of the dust mass
variation study from the centre along the positive x-axis for an inclination
of 0\,\degr\, and 90\,\degr. The fluxes are given in Jansky per square
milliarcsecond for an assumed distance of the object of 45\,Mpc.  
For the case of 0\,\degr, all graphs show the same trend: coming from the centre, one can see a
sharp rise of surface brightness with a sharp peak and a flatter decrease in
the outer part. Between 4 and 5\,pc, all of the curves meet into a single
curve, independent of the optical depth. The latter is due to the fact
that our model features very steep funnel walls. As indicated by the
temperature distribution, we expect most of the emission for this wavelength
next to the torus boundary. The intersection of all curves then indicates the
distance from the centre, where the density has decreased that much that a
doubling of it does not change the emitted flux anymore and the expected
surface brightness distribution is
independent of the optical depth within our sample. 
By increasing the optical depth, the maximum is shifted towards the centre and
to higher values and the curves get steeper.     
As already discussed in the previous section, the higher the optical depth is
the steeper the drop of the temperature distribution in radial direction is,
because the photon mean free path length decreases and most of the photon
packages get absorbed within
a reduced volume of the dust torus (steepening of the inner part) and,
therefore, only reemitted photons in the IR
reach the outer region (flattening of the outer part). To estimate the surface
brightness, one has to take into account all emission by dust until $\tau$\,=\,$1$
is reached.
This can also be directly seen in Fig.~\ref{fig:opdep_map} in the decreasing
size of the region of maximum emission. 
The apparent shift of the inner radius of the torus is caused by the increase
of the \emph{atmosphere} around the dense core of the torus or the growth of the
denser core, respectively.

At the other inclination angles one can again see that the area of maximal 
emission changes. At smaller optical depths, it is a diffuse, wide distribution
within the torus volume. Increasing the optical depth leads to a spreading, which is
more located in the direct vicinity of the inner funnel border.
This leads to the sharply emerging inner borders of the
funnel, which are now visible in the typical x-shaped form. These appearances
have been seen in a similar dust configuration around a double system, the
so-called \object{Red Rectangle} \citep{Tuthill_02}, which can be explained by a dusty
torus as well \citep{Menshchikov_02}. It is located in a distance of
only 330\,pc, much closer than Seyfert galaxies and can therefore be resolved. 

At higher inclination angles, two effects have to be taken into
account. With increasing amount of dust along the line of sight, first of all,
we expect a larger surface brightness, but the
extinction along the line of sight increases as well. 
The value of the maximum of the cut along the positive
x-axis of the maps with the highest optical depth is not the brightest
anymore. It decreases with increasing inclinations. In Fig.~\ref{fig:dust_maps_flux}b the case
for an inclination angle of 90\,\degr\, is shown. Here, the model with the
highest dust mass (thin dashed line) has the smallest surface brightness and the curve of the second heaviest
torus (thick dashed line) has also shifted significantly to smaller values. 
The shape of these graphs also changes when changing from the Seyfert\,I ($i$\,=\,$0$\,\degr) case
to the Seyfert\,II ($i$\,=\,$90$\,\degr) case. As the directly illuminated parts of the
funnel are not directly visible anymore, the decreasing curve further away
from the centre decreases faster. With increasing optical depth, the outer
part of the torus gets colder and colder, which can be seen by the decreasing
flux and the steepening of the curves there. In the innermost part, the curves flatten,
which is due to the emission band of the innermost ring of the funnel. 
As less photons reach the outer part, statistic gets poorer there and
therefore the graphs get more noisy.

\begin{figure*}
\centering
\mbox{
  \subfigure{\includegraphics[angle=90,width=8.5cm]{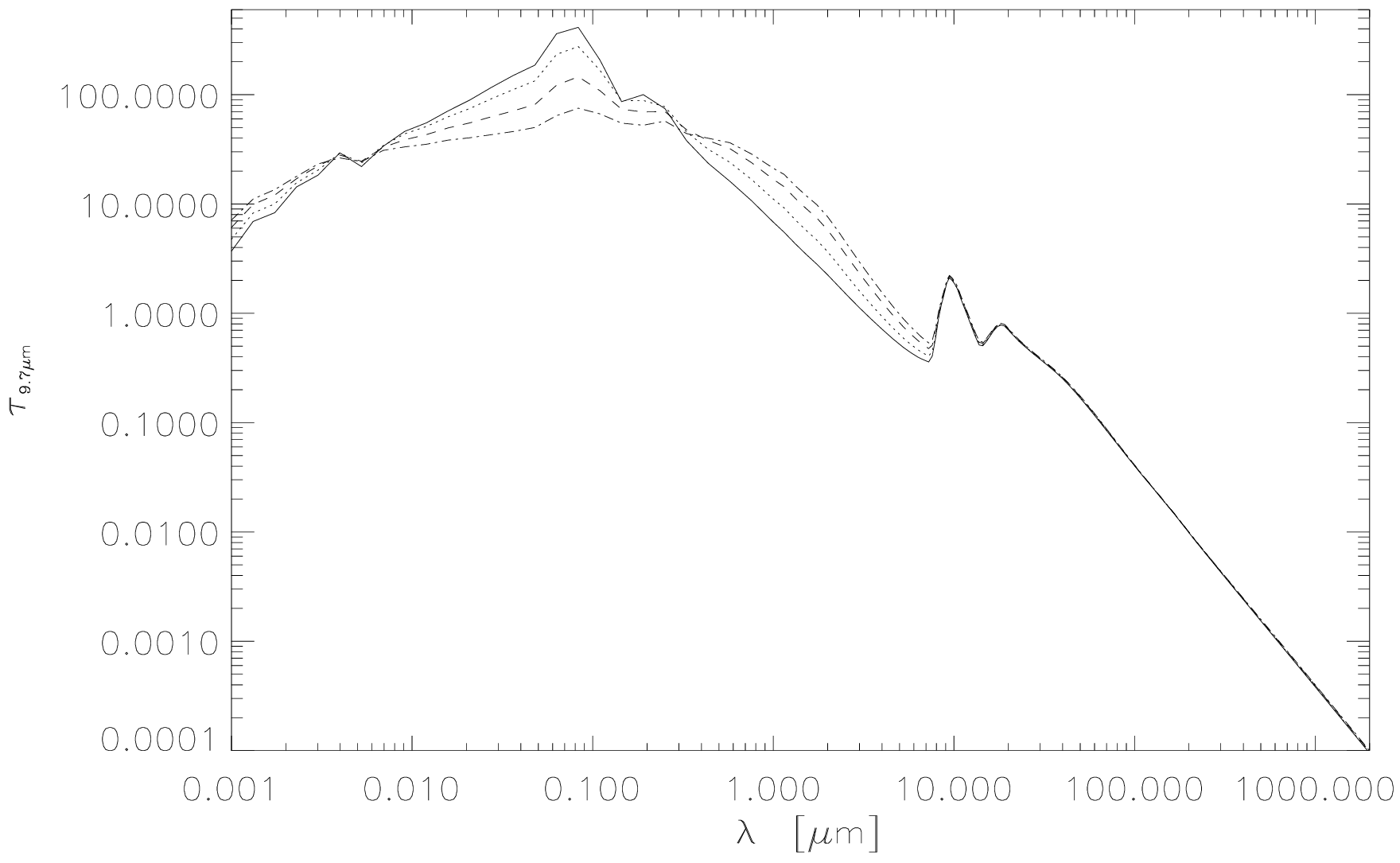}}
  \subfigure{\includegraphics[angle=90,width=8.5cm]{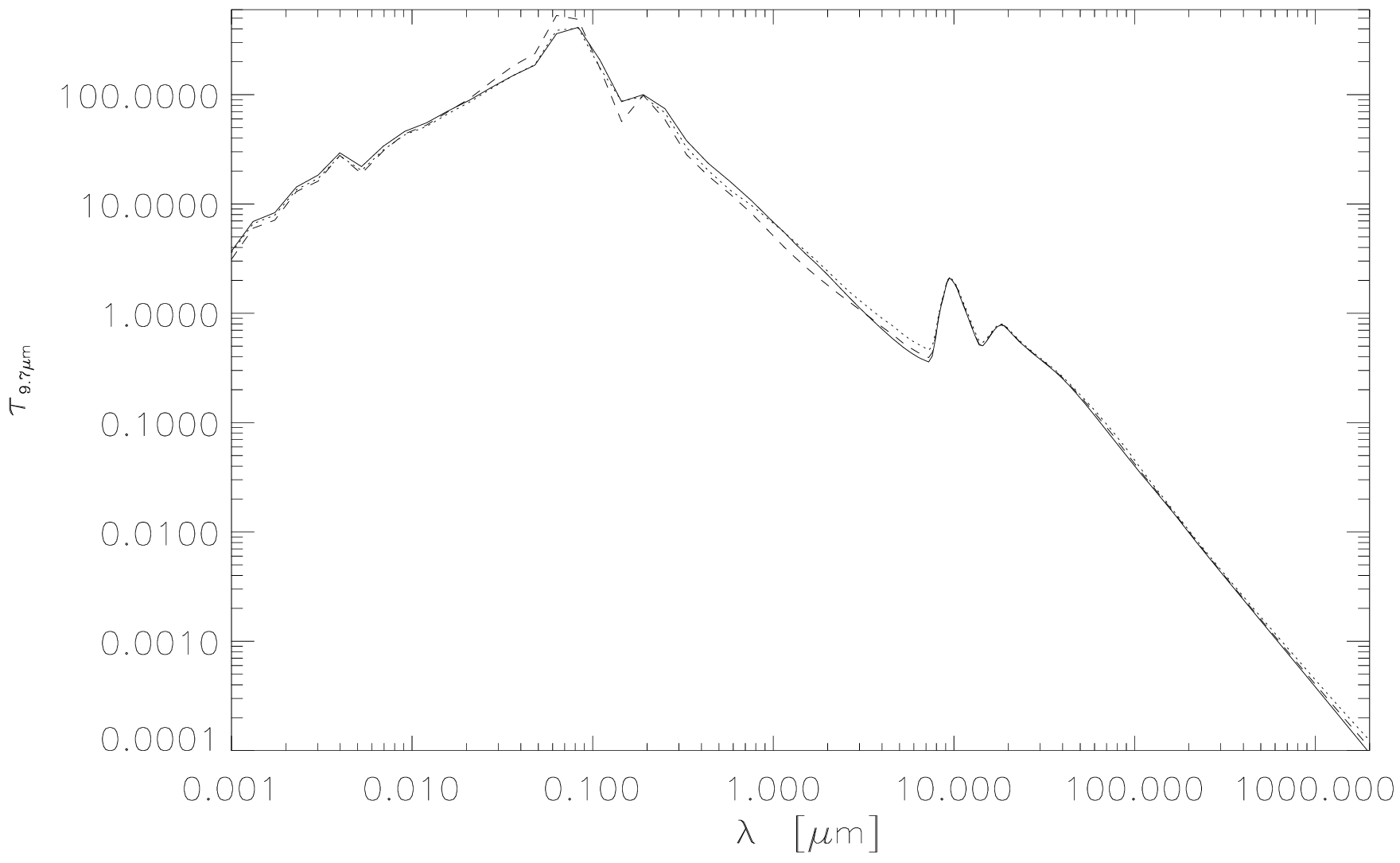}}
}
  \caption{Optical depth within the equatorial plane for different 
           dust models: {\bf a} Increasing the exponent of the number density
	   distribution of grain sizes, from $-3.5$ (our standard model, given by 
	   the solid line) over $-3.0$ (dotted line) and $-2.5$ (dashed line) to 
	   $-2.0$ (dashed-dotted line). {\bf b} Changing the range of grain sizes, from 
	   $0.005 \dots 0.25\,\muup$m (our standard model, given by the solid line)
	   to $0.005 \dots 10\,\muup$m (dotted line) and to $0.001 \dots 10\,\muup$m 
	   (dashed line). \label{fig:dust_opdeps}}
\end{figure*}

\subsection{Variation of dust properties}

As already pointed out in Sect.~\ref{sec:dust}, 
the dust composition in the extreme environment of an AGN 
may be altered compared to interstellar dust in 
our own galaxy. The main effects here are sputtering due to interaction 
of the dust grains with the hot surrounding gas, 
the presence of the strong radiation field of the central source, 
destruction of dust grains due to shocks, etc.

All of the mentioned effects mainly impact the population of small grains.
Therefore, some authors \citep[e.\,g.\,][]{Laor_93,Maiolino_01} claim that within an AGN dust model,
small grains should be depleted. 
In the following section, we test this by 
first of all changing the exponent of the number density distribution 
of grain sizes and later by widening the grain size distribution in both
directions. By doing this, a transition is made between 
the standard MRN galactic dust model -- which we have adopted as 
our standard dust model -- and the dust model proposed by \citet{Bemmel_03}.
In this section, we are especially interested in the behaviour of the $9.7 \muup$m 
silicate feature, as it was claimed that the depletion of smaller 
grains may solve the problem that the feature is never seen in emission in AGN.
On the other hand one needs to check whether other constraints are violated by this change of 
the dust model, e.\,g.\,the appearance of the feature in absorption, as 
observed.

\subsubsection{Changing the exponent of the number distribution of grain sizes}

Different          
exponents of the number density distribution of the grains were
tested by \citet{Bemmel_03}
for their torus model. They finally adopt a somehow more shallow distribution with 
an exponent of $-2$ to be their standard dust model. 

\begin{figure*}
\centering
\mbox{
  \subfigure{\includegraphics[width=8.5cm,angle=90]{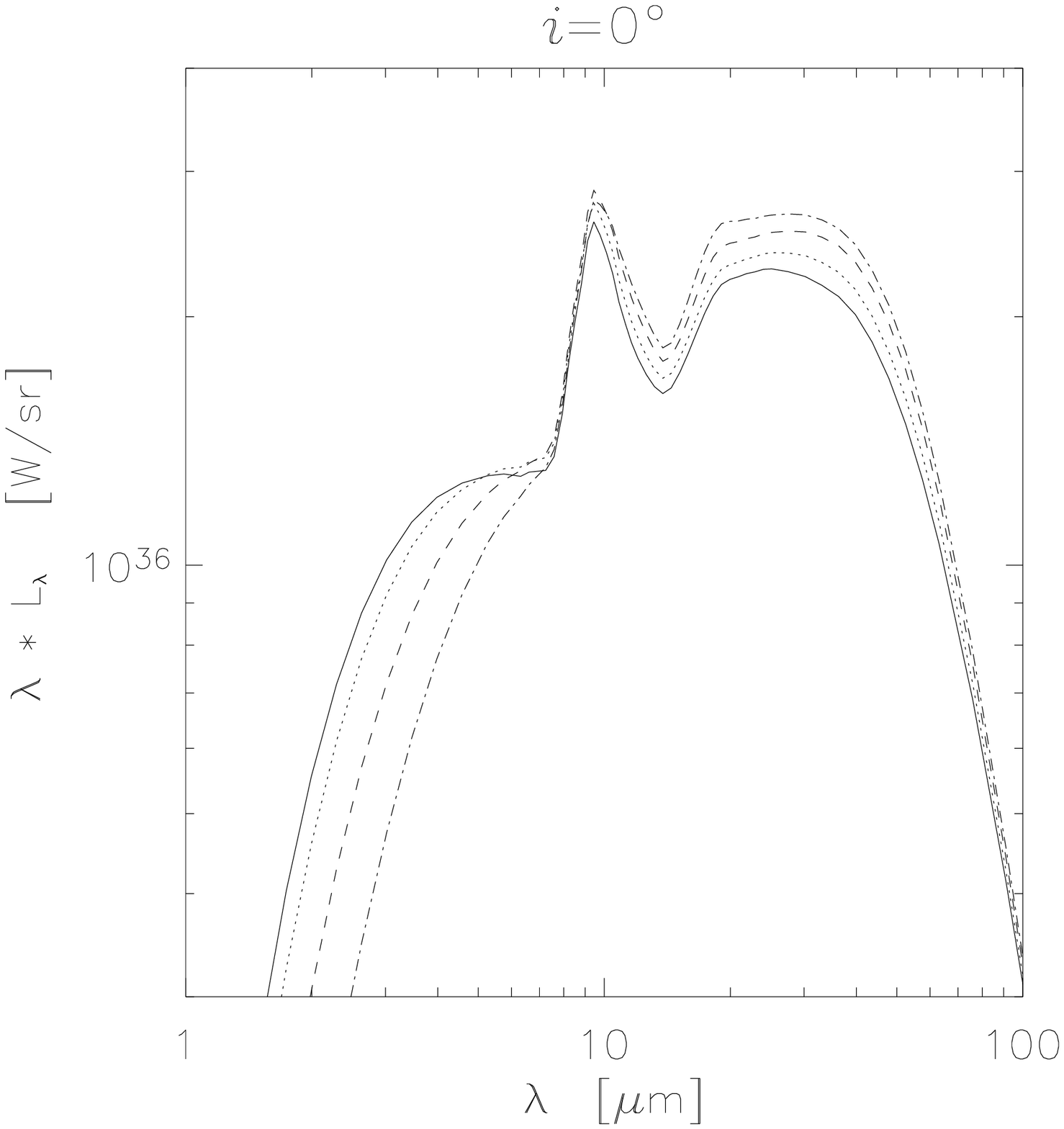}}
  \subfigure{\includegraphics[width=8.5cm,angle=90]{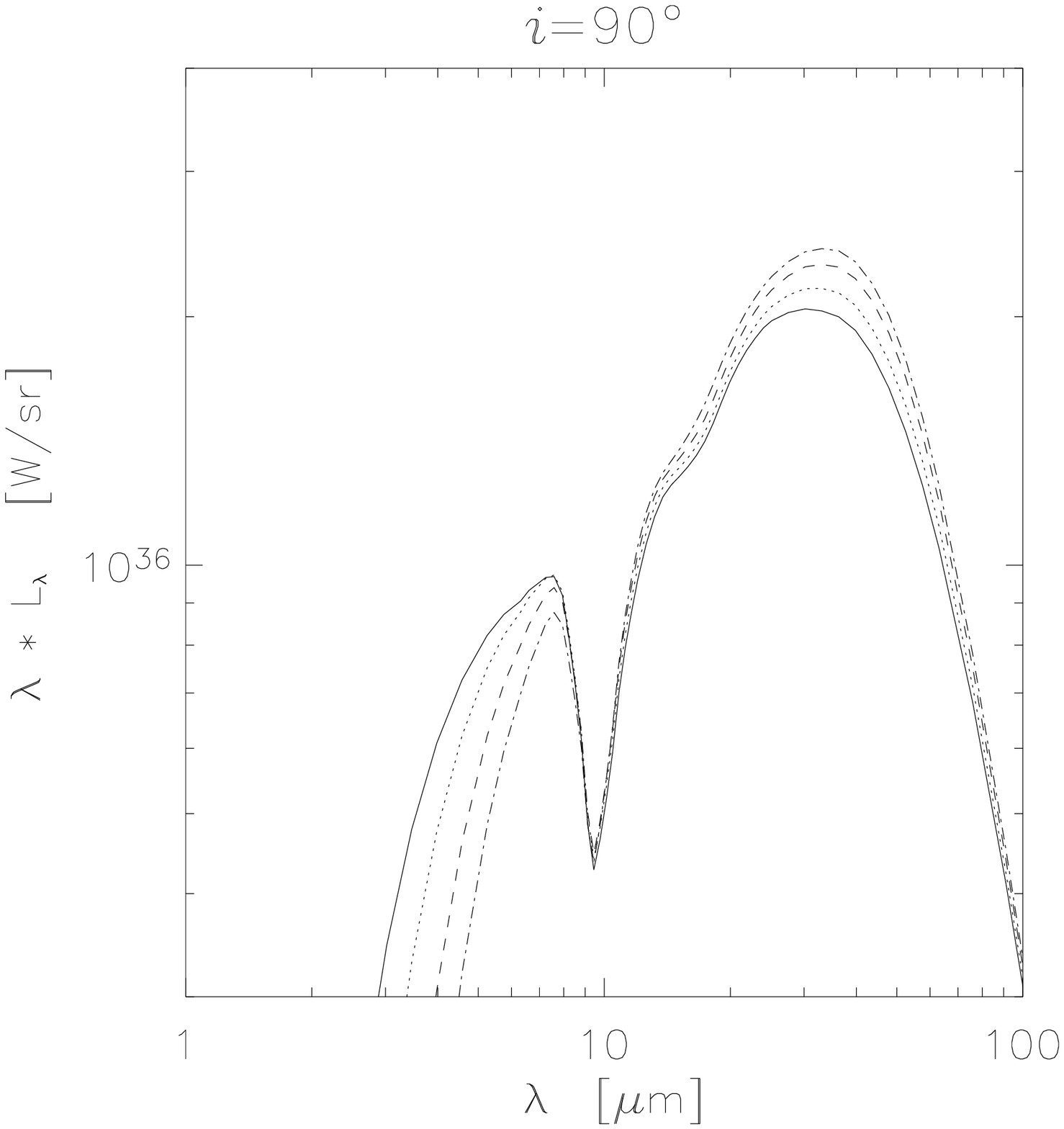}}
}
  \caption{Consequences on the SEDs at inclination angles of $0\degr$ and $90\degr$ of changing the exponent of the dust grain 
           number density distribution. The exponent changes from $-3.5$
           (solid line) in steps of $0.5$ to $-2.0$ (dashed-dotted line). 
	   \label{fig:dust_exponent}} 
\end{figure*}

In this section, we also check different slopes of the distribution for our standard 
torus geometry (see Sect.~\ref{sec:param_model}). 
We start with our standard dust model 
with a slope of $-3.5$ (given by the solid line in Fig.~\ref{fig:dust_opdeps}a and Fig.~\ref{fig:dust_exponent}) and 
test shallower distributions by changing 
the exponent in steps of $0.5$ up to $-2.0$ (dashed-dotted line) used by \citet{Bemmel_03}. Due to renormalisation 
of the dust density after changing the exponent, this means that a flatter distribution leads
to more large grains compared to small grains.
This produces a flatter course of the optical depth (see Fig.~\ref{fig:dust_opdeps}a) at small wavelengths --
compared to our standard model -- 
because of the less efficient heating of larger grains, which dominate the dust composition more 
and more during the variation. Consequently, the temperature 
distributions flatten.

In the SEDs (see Fig.~\ref{fig:dust_exponent}a), this can be seen by the decreasing fluxes 
at the rise of the IR-bump at small wavelengths and the increasing fluxes
at wavelengths around the maximum ($20-30 \muup$m) of the infrared bump (see Fig.~\ref{fig:dust_exponent}a). 
The stratification of dust grains concerning sizes and composition
and the increasing extinction caused by the large grain population (with smaller sublimation radii) lead to
larger sublimation radii for the other components. 
In contradiction to the modelling of 
\citet{Bemmel_03}, a significant change of the silicate feature cannot be 
found in our simulations. For the case of an inclination angle of $90\degr$ (see Fig.~\ref{fig:dust_exponent}b), 
any change is simply due to the fact that there is too little dust left 
that has temperatures higher than about $300$K -- the temperature which is 
required for the emission feature at $9.7 \muup$m.

\begin{figure*}
\centering
\mbox{
  \subfigure{\includegraphics[width=8.5cm,angle=90]{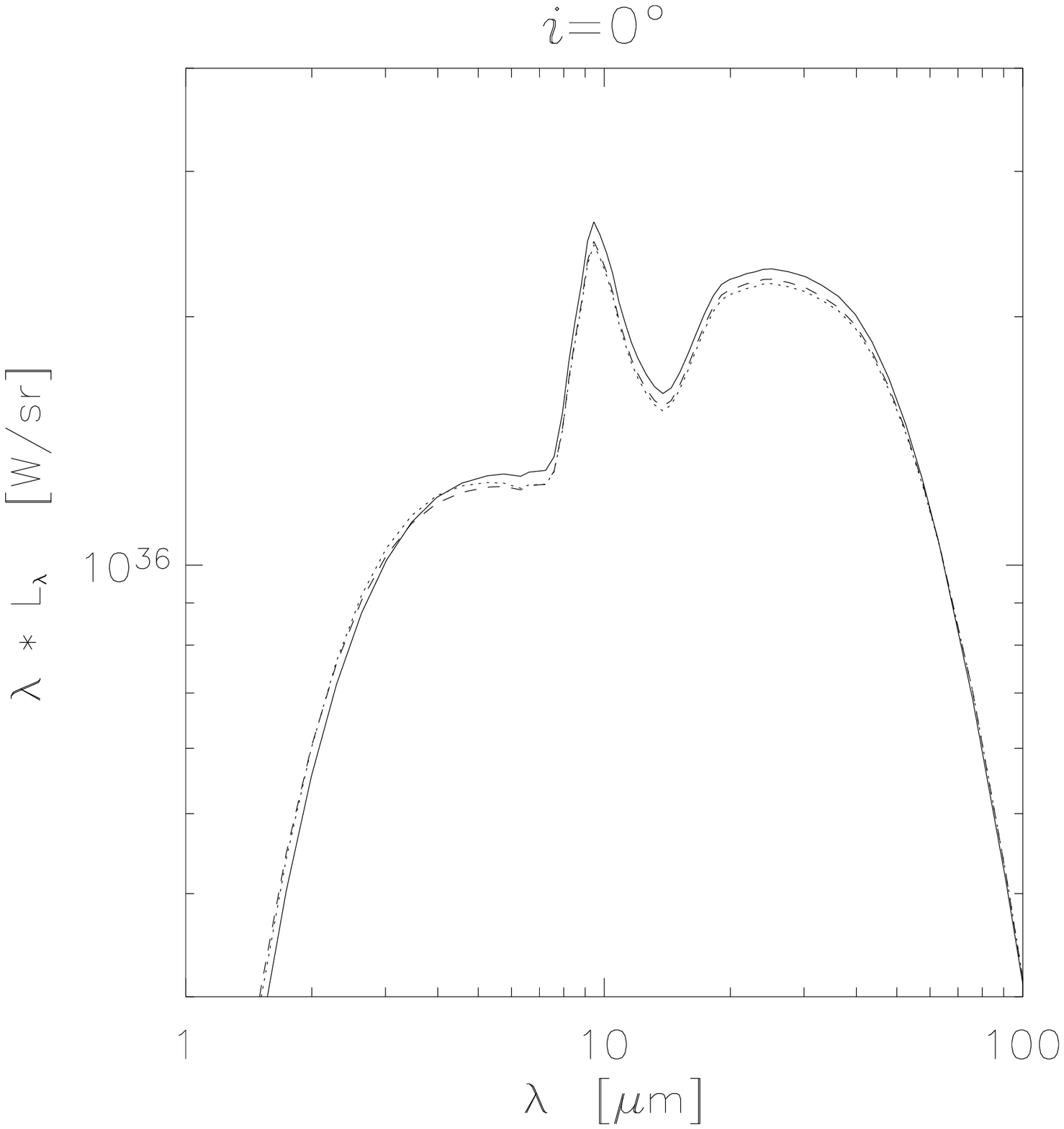}}
  \subfigure{\includegraphics[width=8.5cm,angle=90]{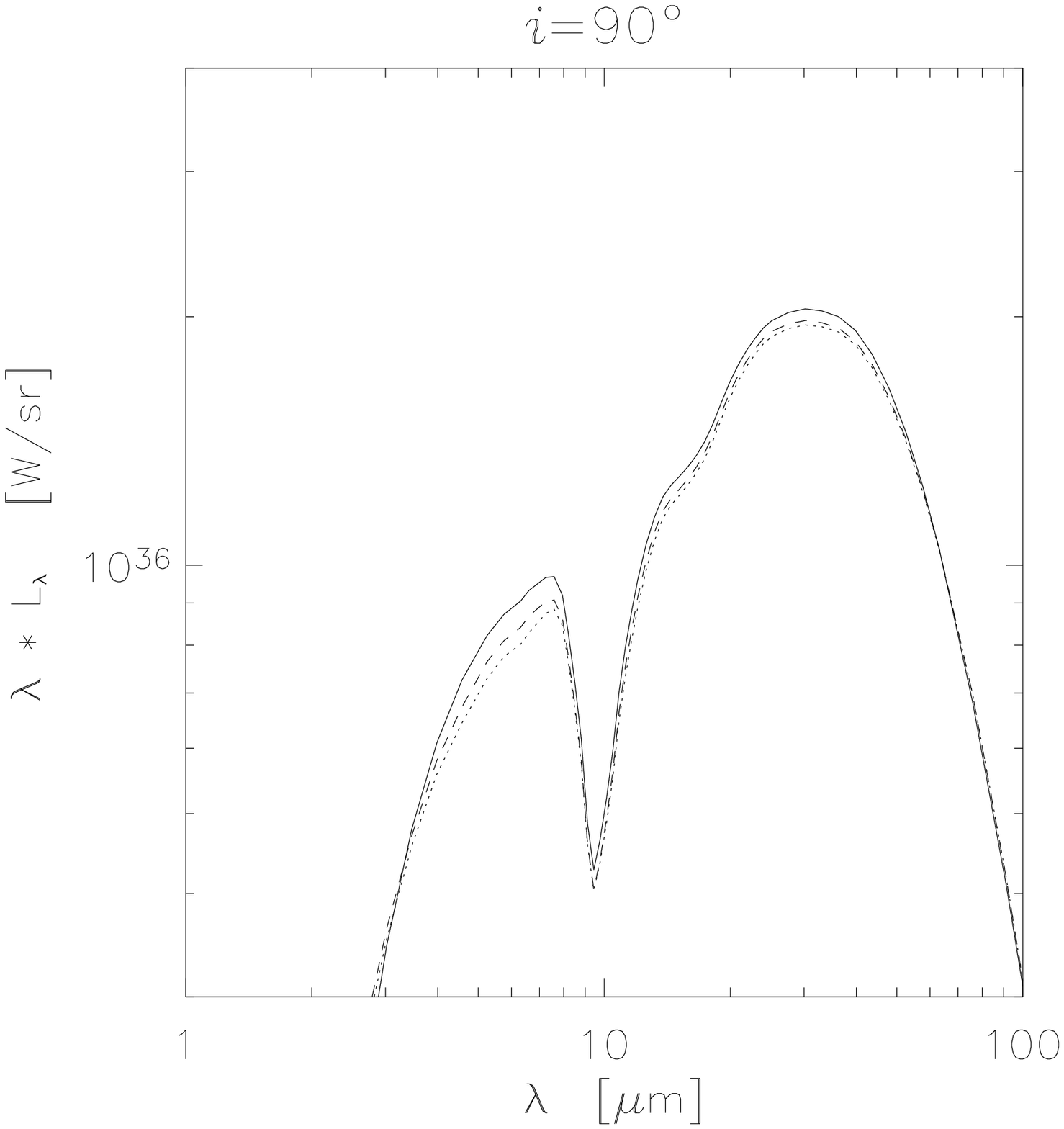}}
}
  \caption{Consequences on the SEDs at inclination angles of $0\degr$ and $90\degr$ of changing the width of the dust grain 
           size distribution from $0.005 \dots 0.25\,\muup$m (solid line) over 
           $0.005 \dots 10\,\muup$m (dotted line) and finally to $0.001 \dots
           10\,\muup$m (dashed line). \label{fig:dust_range}} 
\end{figure*}

\subsubsection{Changing the width of the grain size distribution}

In a further dust parameter study, we test the effects of a broadening 
of the grain size distribution. The corresponding courses of the optical
depths with wavelength within the equatorial plane 
are plotted in Fig.~\ref{fig:dust_opdeps}b and the resulting SEDs in Fig.~\ref{fig:dust_range}b.
Starting from the MRN model (given by the solid line), 
we first extend the grain size range 
towards larger grains (dotted line) up to $10 \muup$m size and later 
additionally towards smaller
grains down to $0.001 \muup$m (dashed line).
By doing this, we finally get the dust model used by \citet{Bemmel_03}. 

As can be seen from Fig.~\ref{fig:dust_opdeps}b, only minor changes of the
optical depth are visible.
Conspicuous is the smaller relative depth of the silicate
feature towards shorter wavelengths after including also large grains
(dotted line). This can also be seen in the
SEDs, displayed in Fig.~\ref{fig:dust_range}b. While the differences for an inclination
angle of $0\degr$ are nearly negligible, the reduced relative depth of the 
silicate feature towards smaller wavelengths is clearly visible in the
SED for edge-on view ($i=90\degr$). 
In the last step of our study -- the inclusion of pronounced slightly smaller
grains -- the effects of extending the size distribution towards
larger grains is partly compensated.

\subsection{Zooming into the torus}
\label{sec:zoomin}

\begin{figure*} 
\centering
\mbox{ 
   \subfigure{
     \includegraphics[width=8.5cm,angle=90]{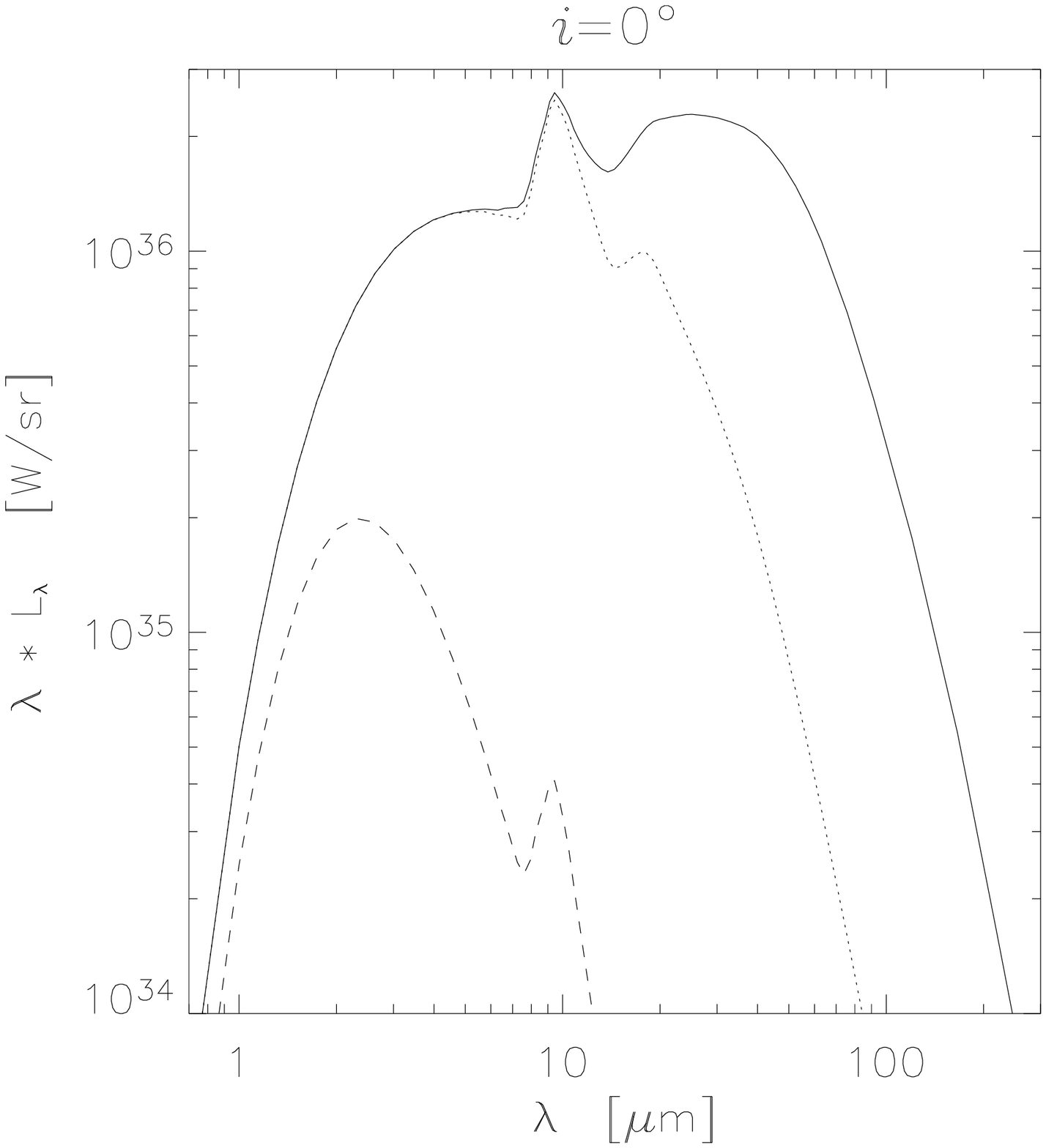}
   }
   \subfigure{
     \includegraphics[width=8.5cm,angle=90]{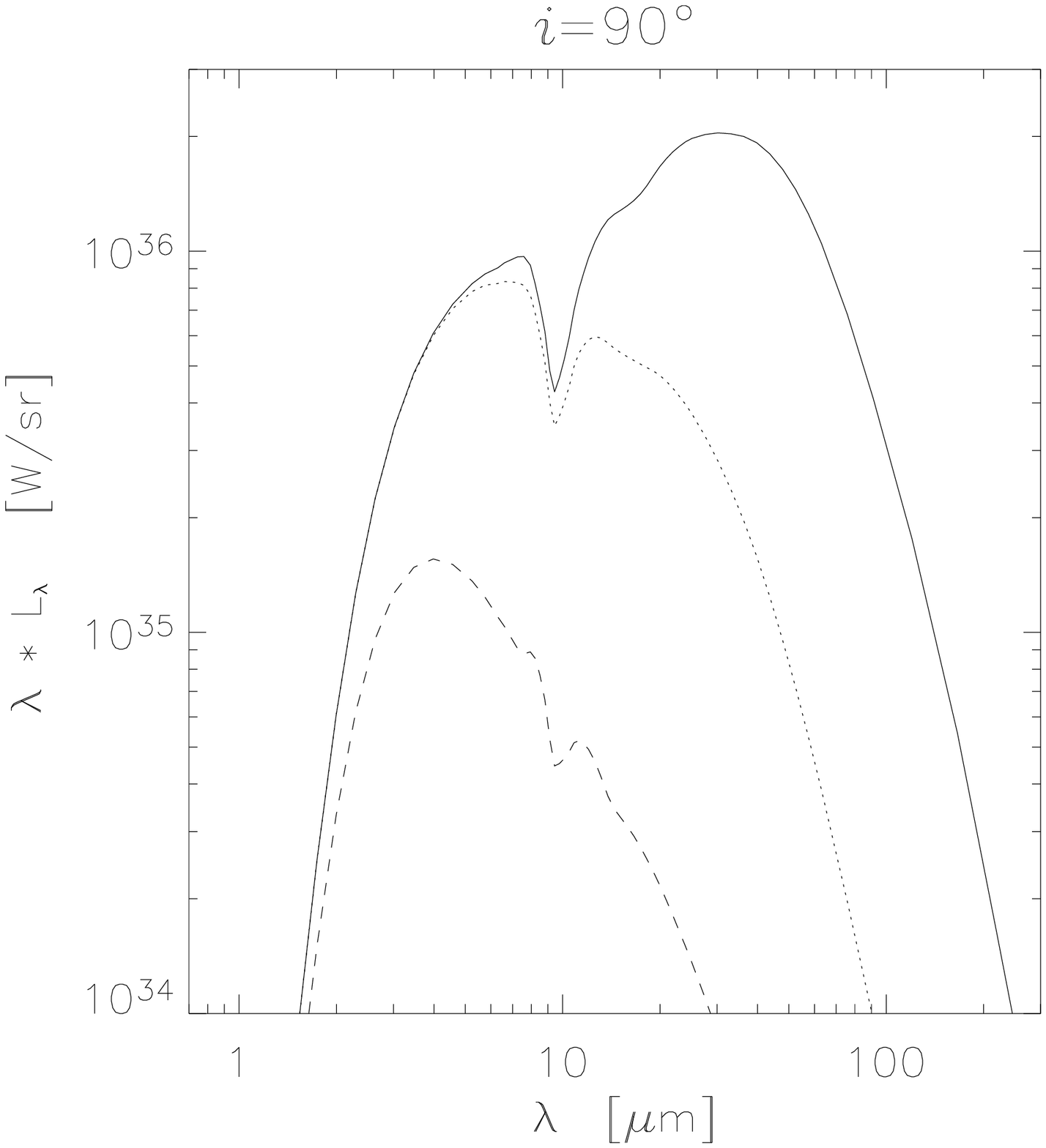} 
   }
}
\caption{Dependence of SEDs on the beamsize of the observing
  device. The solid line corresponds to our standard model, the dotted line to
  the resolution obtainable with a single-dish telescope (100 mas) for a distance to
  the object of 45\,Mpc and the dashed line to the resolution of the   
  MIDI interferometer (10 mas). The left panel shows SEDs for an inclination angle 
  $i=0\,\degr$ and the right panel for $i=90\,\degr$.} 
\label{fig:zoom_SED}
\end{figure*}

Another simulation series we made was to investigate the impact of zooming 
into the torus resembling a certain aperture of
the telescope. To do this, we again started from our standard model
and zoomed into the central 10\,pc and 1\,pc in radius.          
This corresponds to a resolution of
the observing instrument of 100 mas and 10 mas for an object at a distance of
45\,Mpc. These are typical values for the largest single-dish telescopes (100
mas) and the first interferometers in this wavelength range like MIDI (10
mas). The results of this study are shown in Fig.~\ref{fig:zoom_SED}.
In terms of surface brightness distributions compare to Fig.~\ref{fig:inc_map}. 

When we decrease the aperture, we concentrate more and more on the central
parts of the torus emission. As our temperature distribution is very steep, it
is expected that cold dust in the outer part of the torus is excluded from
the field of view. Therefore, the smaller the aperture, the less total flux we
obtain, which is taken from the large wavelengths part of the spectrum. For
the case of 10 mas resolution, only the inner pc in radius is visible.
This is even less than the sublimation radius of the smallest silicate grains
(approximately 1.3\,pc). \\ 

For the silicate feature, one notices a decrease of the relative
depth rather than the expected increase, as seen with MIDI 
(compare to Fig.~2 in \citet{Jaffe_04}). This
might be due to a too steep radial temperature distribution of our
models. Introducing a cloudy structure, which leaves unobscured channels up to
the outer part of the torus, may lead to a direct heating of clouds further
outside and therefore to a broader temperature distribution. Zooming into the
centre then means for the Seyfert\,II case to exclude areas, where emission
features are produced and a deepening of the silicate absorption feature is
expected to take place.

\section{Comparison with observations}
\label{chap:Comparison}

In this section, simulated spectral energy distributions are compared to available 
observational data for the case of large aperture SEDs and high spatial resolution
SEDs. 

\subsection{Comparison with large aperture spectra of type\,I galaxies}

\begin{figure}
  \resizebox{\hsize}{!}{\includegraphics[angle=90]{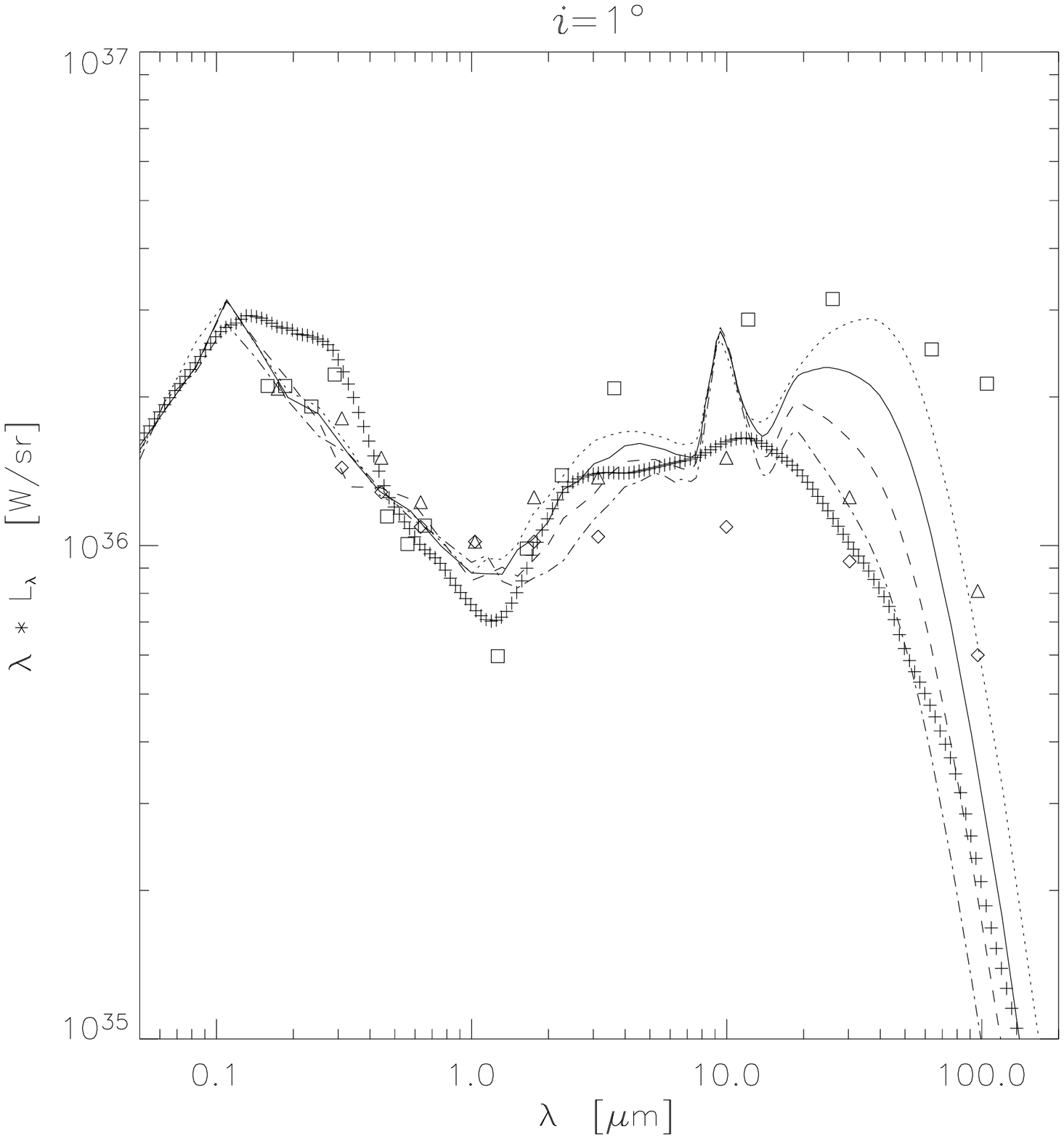}} 
  \caption{Comparison of our standard model (different line styles correspond
  to varying dust masses) with large aperture
  spectra for type\,I objects. Plus signs refer to 29 radio quiet quasars
  \citep{Elvis_1994}, squares to 16 Seyfert\,I galaxies \citep{Granato_94} and
diamonds and triangles 
  to PG Quasars \citep{Sanders_89} with luminosity smaller (diamonds) and  
larger (triangles) than $10^{12}
\,L_{\sun}$. For a typical Seyfert\,I galaxy ($10\,\degr < i < 20\,\degr$), the silicate feature looks much less
pronounced (see Fig.~\ref{fig:inclinationstudy}a).}   
  \label{fig:comp_lowres_1} 
\end{figure}

In Fig.~\ref{fig:comp_lowres_1} we compare the resulting SEDs of our standard
model (different line styles) for an axial view on the torus 
(typically, Seyfert\,I galaxies have inclination angles between 10\,\degr and 20\,\degr)
with varying enclosed dust masses (as described in Sect.~\ref{sec:duma} and shown in
Fig.~\ref{fig:tau_SED}) to different sets of mean large aperture data for type\,I objects.
The plus signs refer to a mean spectrum, extrapolated from 29 radio quiet quasars, provided by
\citet{Elvis_1994}, the squares correspond to 16 Seyfert galaxies, reported by
\citet{Granato_94}. Triangles and diamonds represent radioquiet type\,I PG
quasars with bolometric luminosity larger and smaller than $10^{12}
\,L_{\sun}$ respectively \citep{Sanders_89}. All of these data sets are scaled in order to get
fluxes comparable to our simulations. 

As can be seen from this, our models compare well to the overall shape of the UV to
FIR wavelength range of these mean AGN spectra. An exeption is the pronounced feature at 
$9.7\,\muup$m. As already discussed before, it has never been observed in emission  
in SEDs of Seyfert galaxies -- as well not in SEDs of higher spectral resolution.
The introduction of a clumpy structure of the dust distribution, which is also a more physical solution, 
might help to remove these large emission features of our models. 
A comparison with high spatial resolution SEDs of type\,II sources is given in the next section for some selected
objects.

\subsection{Comparison with special Seyfert\,II galaxies}

To specify as many of our simulation parameters as possible for these special
objects, we started with
a literature study. Some of the remaining variables could be chosen according to
conditional relations as mentioned in Sect.~\ref{sec:param_model}. The value
of the enclosed dust mass was assumed such that the relative depth of the $10\,\muup$m-feature
appears comparable to the observations of the correlated flux with a baseline
of 78\,m done with MIDI for the case of \object{NGC\,1068} and comparable to the depth of
the feature in the MIR spectrum of \citet{Roche_91} for the \object{Circinus} galaxy. 
As zooming into our torus does not reproduce the
deepening of the silicate feature as seen by MIDI, this seems to be a
reasonable procedure and explains the deviations concerning wavelengths around the feature
for the case of \object{NGC\,1068}. 
It results in an optical depth in the equatorial plane at $9.7\,\muup$m of $\tau$\,=\,$3.3$ for 
\object{NGC\,1068} and $\tau$\,=\,$3.9$ for the \object{Circinus} galaxy.  
The free
parameters finally were chosen to obtain a reasonable dust distribution and a 
good comparison with the data.

\subsubsection{\object{NGC\,1068}}

\begin{figure}[t!] 
  \resizebox{\hsize}{!}{\includegraphics[angle=90]{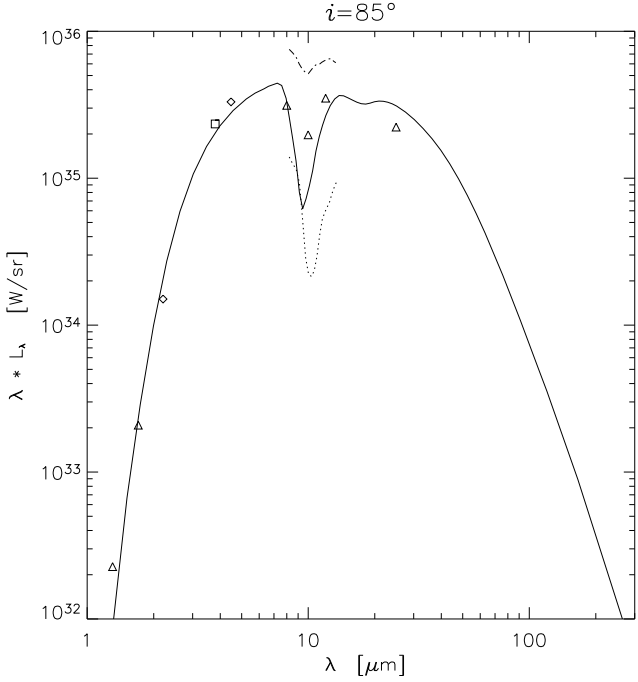}} 
   \caption{Comparison of one of our models (described by the parameters
     given in Table \ref{tab:param_ngc1068} and plotted with a solid line) 
     to a set of data (given in Table 
     \ref{tab:ngc1068_highres_data}) from high spatial
     resolution observations of \object{NGC\,1068}, compiled by M.\,A.\,Prieto. 
     The various symbols correspond to slightly different apertures. The dashed-dotted 
     line gives the total flux spectrum obtained with MIDI and the dotted line corresponds
     to the correlated flux spectrum obtained with MIDI for a baseline of 78\,m 
     \citep{Jaffe_04}.}
\label{fig:comp_highres_NGC1068}
\end{figure}

\object{NGC\,1068} is a nearby Seyfert\,II galaxy at a distance of 14.4\,Mpc \citep{Galliano_03}. Therefore,
it has been studied extensively and we were able to find at least some of the 
needed simulation parameters in literature.   
The procedure described above led to the parameters
summarised in Table \ref{tab:param_ngc1068}.
The inclination angle of $i=85$\,\degr\, was chosen in agreement with the tilt  
of the outflow cone of ionised gas reported by \citet{Crenshaw_00}. 
For our comparison, we use a compilation of high resolution data done by 
M.\,A.\,Prieto, given in Table \ref{tab:ngc1068_highres_data}.
Fig.~\ref{fig:comp_highres_NGC1068} shows the result of the adaptation of
the model given in Table \ref{tab:param_ngc1068} to
this data set. The line corresponds to a zoomed-in view of the torus by a factor of 10, which
resembles the aperture of 0.2\arcsec in diameter of most of the datapoints (given by the triangles) in 
Fig.~\ref{fig:comp_highres_NGC1068}. The diamonds correspond to a slightly larger aperture of 0.27\arcsec. 
The L-band flux, which corresponds to an aperture of 0.7\arcsec in 
diameter is given by the square symbol. 
In addition, we added the total flux (dashed-dotted line) -- corresponding to an 
aperture of $0.6\,\times0.6\arcsec$ -- and the correlated flux spectrum for 
a baseline of 78\,m (dotted line) -- corresponding to an aperture in diameter of 
approximately $30\,mas$ \citep{Jaffe_04}. 
As can be seen in this figure, 
observations are in good agreement with our model except of the wavelength region around the 
$9.7\,\muup$m silicate characteristic. As already mentioned before, zooming into our model torus
does not reproduce the trend of MIDI-observations, which means a deepening of the silicate 
feature. We therefore adapted the depth of our feature to the depth observed in 
correlated flux measurements of MIDI with a baseline of 78\,m. This was also necessary in order 
to describe the datapoints at the smaller wavelengths. 

\begin{table}[!htb]
\centering
\caption{\label{tab:param_ngc1068} Parameters used for the simulation of \object{NGC\,1068}.}
\begin{tabular}{lcc}
\hline
\hline
Parameter & Value & Reference \\
\hline
$M_{\mathrm{BH}}$   & $8.3\cdot10^6 \, M_{\sun}$           &  \citet{Greenhill_1996} \\
$R_{\mathrm{c}}$    &  24\,pc                               &  \citet{Gallimore_03} \\
$M_{*}$             & $6.5\cdot 10^8 \, M_{\sun}$          &  \citet{Thatte_1997}   \\
$L_{\mathrm{disc}}$          & $5.42\cdot 10^{10} \, L_{\sun}$      &  \\
$L_{\mathrm{disc}}/{L_{\mathrm{Edd}}}$ & 20\% & \\
$R_{\mathrm{T}}$    & 3\,pc                                 &  \\
$M_{\mathrm{dust}}$ & $7.8\cdot10^4 \, M_{\sun}$           &  \\ 
$\tau_{9.7\,\muup \mathrm{m}}$ & 3.3 & \\
$R_{\mathrm{out}}$  & 72\,pc                                &  \\
$v_{\mathrm{t}}$    & $164\,\mathrm{km}\,\mathrm{s}^{-1}$ &  \\
$\gamma$            & 0.5                                  &  \\ 
$d$                 & 14.4\,Mpc                             &  \citet{Galliano_03} \\
$i$                 & 85\,\degr                              &  \citet{Crenshaw_00} \\
\hline
\end{tabular}
\end{table}

\begin{table}[!htb]
\centering
\caption{\label{tab:ngc1068_highres_data} High resolution data of \object{NGC\,1068}. The compilation 
         was done by M.\,A.\,Prieto. The references are 
         R98 \citep{Rouan_98}, P04 (Prieto, private communication), M03 \citep{Marco_03} and 
         B00 \citep{Bock_00}.}
\begin{tabular}{ccccc}
\hline
\hline
Wavelength & Band & Flux & Aperture (dia) & Reference \\
$\muup$m & & Jy & arcsec & \\
\hline
1.30  & J &  5e-4     & 0.20  & R98  \\
1.70  & H &  6e-3     & 0.20  & R98  \\
2.20  & K &  5.6e-2   & 0.27  & P04  \\
3.79  & L &  1.5      & 0.70  & M03  \\
4.47  & M &  2.5      & 0.27  & P04  \\
7.99  & N &  4.23     & 0.20  & B00  \\
9.99  & N &  3.33     & 0.20  & B00 \\
11.99 & N &  7.1      & 0.20  & B00 \\
24.98 &   &  9.4      & 0.20  & B00 \\
\hline
\end{tabular}
\end{table}

\subsubsection{\object{Circinus}}

The \object{Circinus} galaxy is one of the nearest spiral galaxies, which harbors an Active Galactic
Nucleus. As it is the second extragalactic object, which has been observed by MIDI and also with
NAOS/CONICA \citep{Prieto_04}, we also carried out a comparison with one of
our models. From the first results of these observations \citep{Prieto_04} and also from
previous modelling of the dusty torus by \citet{Ruiz_01}, we know that it harbors a relatively
small sized toroidal dust distribution. Therefore, we used a scaled down
version of our standard model described above.
The simulation was done with the parameters summarised in Table
\ref{tab:param_Circinus}.

\begin{figure}[t!] 
  \resizebox{\hsize}{!}{\includegraphics[angle=90]{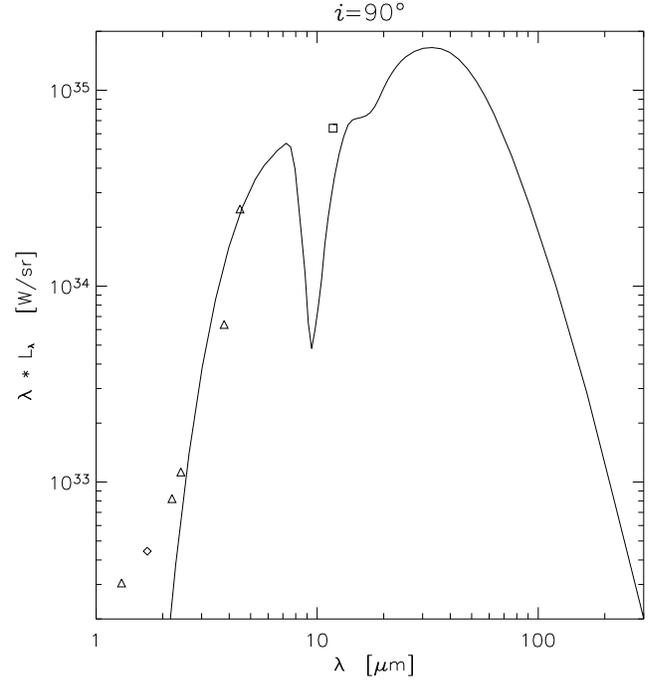}} 
  \caption{Model comparison with high spatial resolution data of the Circinus galaxy 
     taken with
     the NACO camera at the VLT. Data courtesy of \citet{Prieto_04}. The data was 
     corrected for foreground extinction by $A_{\mathrm{V}}$\,=\,$6\,\mathrm{mag}$.
     The various symbols correspond to different apertures (see Table \ref{tab:circinus_highres_data}).}
\label{fig:comp_highres_Circinus}
\end{figure}

\begin{table}[!htb]
\centering
\caption{\label{tab:param_Circinus} Parameters used for the simulation of the
  \object{Circinus} galaxy.}
\begin{tabular}{lcc}
\hline
\hline
Parameter & Value & Reference \\
\hline
$M_{\mathrm{BH}}$   & $1.7\cdot10^6 \, M_{\sun}$           &  \citet{Greenhill_2003} \\
$R_{\mathrm{c}}$    & 10\,pc                               &   \\
$M_{*}$             & $2.0\cdot 10^8 \, M_{\sun}$        &   \\
$L_{\mathrm{disc}}$ & $8.4\cdot 10^{9} \, L_{\sun}$, & \\
$L_{\mathrm{disc}}/{L_{\mathrm{Edd}}}$ & 15\% & \\ 
$R_{\mathrm{T}}$    &  3.5\,pc                                 &  \\
$M_{\mathrm{dust}}$ & $2.0\cdot 10^4 \, M_{\sun}$           &  \\ 
$\tau_{9.7\,\muup \mathrm{m}}$              & 3.9     \\
$R_{\mathrm{out}}$  & 30\,pc                                &  \\
$v_{\mathrm{t}}$    & $140\,\mathrm{km}\,\mathrm{s}^{-1}$ &  \\
$\gamma$            & 0.5                                  &  \\ 
$d$                 & 4.2\,Mpc                              & \citet{Freeman_1977} \\
$i$                 & 90\,\degr                              &  \citet{Crenshaw_00} \\
\hline
\end{tabular}
\end{table}

\begin{table}[!htb]
\centering
\caption{\label{tab:circinus_highres_data} High resolution data of the \object{Circinus} galaxy. 
The data was corrected for foreground extinction by $A_{\mathrm{V}}$\,=\,$6\, \mathrm{mag}$. 
Data courtesy of \citet{Prieto_04}.}
\begin{tabular}{cccc}
\hline
\hline
Wavelength & Band & Flux & Aperture (dia) \\
$\muup$m & & mJy & arcsec \\
\hline
1.30  & J  & $\la$7.9   & 0.38 \\
1.70  & H  & 15.0       & 0.10  \\
2.20  & K  & 36.0       & 0.38 \\
2.42  &    & 54.0       & 0.38 \\
3.79  & L  & 480.0      & 0.38 \\
4.47  & M  & 2200.0     & 0.38 \\
11.80 & N  & 15000.0    & 1.00 \\
\hline
\end{tabular}
\end{table}

Fig.~\ref{fig:comp_highres_Circinus} shows the result of the comparison of this model
(given by the solid line plot)
with data taken with the NACO camera at the VLT \citep{Prieto_04}. Various 
symbols represent different apertures (see Table \ref{tab:circinus_highres_data}). 
Fluxes, corrected for foreground extinction by  
$A_{\mathrm{V}}$\,=\,$6\, \mathrm{mag}$, are used. 

The graph corresponds to a torus model with an inner radius of approximately 0.9\,pc
and a luminosity of the central source of
15\% of the Eddington luminosity.
The modelling leads to a maximum temperature lower than the sublimation temperatures
of the dust grains.   
Models with smaller inner radii and, therefore, higher maximum temperatures can be excluded. For models 
with higher maximum temperature, the rise of the IR bump at small wavelengths is moved towards smaller
wavelengths. In order to avoid this, we would have to increase the amount of dust, leading to a deeper 
silicate feature, in contradiction to available observations.
This shows that sublimation temperatures of the dust grains are only reached in some
of the torus configurations.
Our model fits well to the observations, except of the small wavelength part of the SED. This 
excess is not due to torus emission. The increased fluxes are most probable caused by 
light from the central source, scattered above the torus by material (dust and electrons) 
within the opening of the torus.  

Our attempts to model observational data showed once more the inherent
ambiguity: The same SED can be modelled with different parameter sets or even models.
Therefore, additional 
information from observations is needed in order to be able to distinguish between
different torus models.

\section{Conclusions and outlook}

In this paper we present new radiative transfer simulations for dust tori of 
Active Galactic Nuclei. 
The dust density distribution and the geometrical shape of the
torus model result from a hydrostatic equilibrium in which gravitational and
centrifugal forces of the nuclear stellar
distribution and the central black hole are balanced by pressure
forces due to the turbulent motion of the clouds. Spectral energy distributions (SEDs) and surface brightness distributions are 
obtained by a
fully three-dimensional treatment of the radiative transfer problem.

After fixing as many parameters of the model as possible by observational
constraints, we undertake an extensive parameter study by varying dust
properties, dust masses and other factors.
It shows that our modelling is able to explain
mean observed spectral energy distributions (SEDs) for classes of Active Galactic Nuclei as well as for individual objects. 

Obviously, the innermost part of the tori close to the 
central energy source is crucial
for the determination of SEDs in the mid-infrared wavelength range. Specifically, we found that any realistic model has to incorporate the fact that
sublimation radii for varying dust grain sizes and dust composition differ
considerably. 
Using mean dust characteristics according
to the MRN dust model (see Sect.~\ref{sec:dust})  instead leads to incorrect temperature distributions and SEDs.
Even with this refinement, the model cannot explain the recent interferometric observations 
of the MIDI instrument, as reported by
\citet{Jaffe_04}, 
which show a deepening  of the silicate absorption feature
in the correlated flux -- effectively sampling  the central few parsec
of the dust distribution. In order to reproduce these results, a 
temperature distribution shallower than in our model would be required.
We find that introducing a $|\cos\theta|$ radiation characteristic for the
primary radiation source does not yield significant changes. In a next
step, we will introduce a clumpy structure of our dust configuration. 
Gaps within the
dust distribution should then lead to direct heating of clumps in the outer part of
the torus and thus producing a shallower mean temperature distribution.  
Biasing the grain size distribution of the dust towards larger grains as
proposed by several authors \citep[e.\,g.\,][]{Laor_93,Maiolino_01} does not yield a reduction of the emission
feature of type\,I objects for the case of our dust density distribution. 

With the present  examination of the \emph{Turbulent Torus Model} for dust distributions in AGNs, we made
a first step towards the introduction of more physics into the modelling of
the torus emission. 
%As was already shown by several other authors
%a toroidally shaped dust distribution 
% can account for the emission in mid-infrared bands.
We demonstrated
that a model based on physical assumptions reproduces the SEDs both on kiloparsec and parsec scales as well  as previous models in which the shape and size of the dust distribution are free parameters ({\it  e.g.}
\citet{Pier_93, Granato_94, Manske_98, Nenkova_02, Bemmel_03} ).

As pointed out in Sect.~\ref{sec:TTM-model}, this kind of stationary 
modelling cannot account for all physical effects present in such an 
environment. In a further step, hydrodynamic simulations in combination with
radiative transfer calculations should enable us to take at least some of these
uncertainties
into account. The main effects are continuous energy feedback by supernovae explosions and mass
injection by stellar winds. 

With the first interferometric observations at hand, however, the
problem becomes more challenging: the wide range in observable scales from hundreds of parsec down to {\it one parsec} requires not only to reproduce
global SEDs of the torus but also to describe its shape, structure and
temperature stratification correctly. In addition, these detailed observations
are an essential step to resolve the ambiguities that are present in current
global models of dusty tori in Active Galactic Nuclei.

\begin{acknowledgements}

The authors would like to thank the referee G.-L.\,Granato for providing one of his
torus models in order to do a comparison of the two codes and 
helpful commments to improve the paper. We also wish to thank M.\,A.\,Prieto
for the NACO data and the foreground corrected fluxes as well as fruitful discussion.

\end{acknowledgements}

\bibliographystyle{aa}
\bibliography{astrings,schartmann2005_colour}
  
\end{document}